\documentclass[aps,pra,twocolumn,superscriptaddress,longbibliography]{revtex4-1}
\usepackage{amsmath}
\usepackage{braket}
\usepackage{ulem}
\usepackage{graphicx}
\usepackage[dvipsnames]{xcolor}
\usepackage[colorlinks]{hyperref}

\begin{document}

\title{Unraveling excitation energy transfer assisted by collective behaviors of vibrations}
\author{Zeng-Zhao Li} 
\affiliation{Department of Chemistry, University of California, Berkeley, California 94720, USA}
\author{Liwen Ko} 
\affiliation{Department of Chemistry, University of California, Berkeley, California 94720, USA}
\author{Zhibo Yang} 
\affiliation{Department of Chemistry, University of California, Berkeley, California 94720, USA}
\author{Mohan Sarovar}
\affiliation{Sandia National Laboratories, Livermore, CA, 94551, USA}
\author{K. Birgitta Whaley}
\affiliation{Department of Chemistry, University of California, Berkeley, California 94720, USA}
\affiliation{Berkeley Center for Quantum Information and Computation, Berkeley, California 94720, USA}

\begin{abstract}

We investigate how collective behaviors of vibrations such as cooperativity and interference can enhance energy transfer in a nontrivial way, focusing on an example of  
a donor-bridge-acceptor trimeric chromophore system coupled to two vibrational degrees of freedom. Employing parameters selected to provide an overall uphill energy transfer from donor to acceptor, we  use numerical calculations of dynamics in a coupled exciton-vibration basis, together with perturbation-based analytics and calculation of vibronic spectra, 
we identify clear spectral features of single- and multi-phonon vibrationally-assisted energy transfer (VAET) dynamics, where the latter include up to six-phonon contributions. 
We identify signatures of vibrational cooperation and interference that provide  enhancement of energy transfer relative to that obtained from VAET with a single vibrational mode. We observe a phononic analogue of two-photon absorption, as well as a novel heteroexcitation mechanism in which a single phonon gives rise to simultaneous excitation of both the trimeric system and the vibrational degrees of freedom.
The impact of vibrations and of the one- and two-phonon VAET processes on the energy transfer are seen to be quite different in the weak and strong site-vibration coupling regimes. 
In the weak coupling regime, two-phonon processes dominate, whereas in the strong coupling regime up to six-phonon VAET processes can be induced.
The VAET features are seen to be enhanced with increasing temperature and site-vibration coupling strength, and are reduced in the presence of dissipation.
We analyze the dependence of these phenomena on the explicit form of the chromophore-vibration couplings, with comparison of VAET spectra for local and non-local couplings.
\end{abstract}

\date{\today}
\maketitle

\section{Introduction}

Recent experimental and theoretical studies of the molecular structures present in biological light harvesting complexes have revealed the delicate interplay of electronic and vibrational degrees of freedom, and how these come together to orchestrate efficient transfer of photoexcitations in such systems~\cite{Scholes17nature,WangAllodiEngel19NatRevChem,CinaFleming04jpca,ColliniScholes10nature,AbramaviciusMukamel10jcp,ChristenssonPullerits12jpcb,Plenio13jcp,YehEngelKais18pnas}. Coherent beating patterns in nonlinear spectroscopy signals initially ascribed to long-lived electronic coherence in such systems~\cite{EngelFleming18nature} are now generally agreed to be a due to combination of electronic and vibrational coherence, with a key role played by coupling of the relevant electronic degrees of freedom to long-lived, underdamped vibrational modes of molecules~\cite{Tiwari12pnas,ChinHuelgaPlenio13nphys,HuelgaPlenio13,HuelgaPlenio14nphys,Romero14nphys}. This revelation has brought to light the subtle ways in which vibrational dynamics in molecular complexes can influence electronic and excitonic properties~\cite{FullerOgilvie14nchem,OReilly14ncomms,DeanScholes16chem,DuanThorwartMiller17pnas,bhattacharyya2020role}. 
In this work we draw inspiration from these studies of molecular systems and ask whether vibrational degrees of freedom can exert other subtle influences on energy transfer in such complexes. In particular, can the presence of multiple 
underdamped vibrations in a molecular complex influence energy transfer dynamics in non-trivial ways?

With the recent growth of quantum technologies, controllable artificial quantum simulators have been developed to probe the underlying basic mechanism of the observed long-time coherences~\cite{Gorman18prx,PotocnikWallraff18ncomm,MaierBlattRoos19prl}. 
In Ref.~\cite{Gorman18prx}, an engineered vibrationally assisted energy transfer (VAET) was experimentally demonstrated for an excitonic dimer emulated in a trapped-ion platform. 
In that work, not only was a one-phonon VAET process signified by a peak at the vibrational frequency being equal to the excitonic transition frequency unambiguously reported, but also unresolved peaks at smaller frequencies were found.  It was suggested that the latter were due to multiphonon VAET processes. 
This provides further motivation for the study of multiphonon VAET processes.  In this work we seek to ascertain the extent to which such multiphonon VAET can be resolved, and also the consequences of any collective behavior of the vibrations
for excitonic energy transfer processes. Specifically, it is of interest to explore whether cooperative or  interference effects might play a role in the vibrationally enhanced energy transfer.
In addition, these systems offer the possibility of finding both the phononic analog of the well-known two-photon absorption~\cite{Mayer1931,KaiserGarrett61prl,Abella62prl}, and the inverse phenomenon~\cite{GarzianoNori16prl}, in which 
one phonon might simultaneously excite two excitonic transitions, where the latter could be of different frequencies. The latter inverse situation offers a new twist with phonons relative to atoms, namely that 
in the context of VAET under our consideration, the question that we 
can ask is whether one phonon from a specific vibrational mode can simultaneously cause an excitonic transition and a vibrational transition in a different vibrational mode?

To address these questions, we consider here a donor-bridge-acceptor trimeric chromophore system coupled to two vibrational degrees of freedom. 
 We analyze the dynamics within a single electronic excitation subspace with explicit incorporation of the vibrational states, performing full numerical simulations for the excitation energy transfer probability from donor to acceptor via the bridging chromophore under various conditions. 
We construct a two-dimensional spectral representation of the vibrationally assisted energy transfer (VAET) probabilities by scanning the frequencies of the two vibrations. These 2D VAET spectra allow identification of several mechanisms through which the vibrational modes can influence and/or enhance energy transfer.  These mechanisms include both single mode VAET, and multi-mode VAET in which the two vibrational modes can cooperation and/or interfere. 
Detailed assignment of the VAET features is facilitated by calculation of the vibronic states resulting from the coupling of electronic and vibrational degrees of freedom, which shows that a number of the VAET features are correlated with the presence of avoided crossings in the vibronic energy spectra.
In the weak site-vibration coupling regime we analyze the dynamical results with a perturbative analysis and use of double-sided Feynman diagrams~\cite{Mukamel95}. 
For both weak and strong site-vibration coupling regimes we then investigate the dependence of the VAET features on exciton dissipation and vibrational temperature.

The context of this study is 
quantum emulation of excitonic energy transfer in ion traps and their use in elucidating 
the dynamic consequences of exciton-vibration coupling for energy transport in molecular excitonic systems. 
We focus first on the behavior when the vibrations are coupled locally to individual chromophore sites, as in the case of trapped ions coupled to transverse modes.The model trimeric system of primary interest in this work is generalized from a dimeric system studied previously with an experimental ion trap emulator~\cite{Gorman18prx} and starts from a Hamiltonian in which individual vibrational modes are coupled to Frenkel excitons on specific sites. 
We then extend this to study of the energy transfer dynamics induced by Hamiltonian coupling to vibrations that are correlated between sites, which naturally results from coupling to the longitudinal modes of trapped ions.

The trapped ion internal states in an ion trap emulation of molecular chromophores can be regarded as pseudo-chromophores. The ability to select different states and modulate their energies with external fields allows exploration of the effects of different energetic landscapes on energy transfer.  However, the Hamiltonian for interaction of such internal ionic states with the external vibrational modes of the trapped ions has some subtle differences from the form of Hamiltonian relevant to studies of excitonic energy transfer in molecular aggregates such as natural light harvesting 
 systems~\cite{IshizakiFleming10PCCP} or $J$-aggregates.  
We show here that despite these differences, the effective Hamiltonian that results from projection to the single electronic excitation subspace maps onto the standard form for excitonic energy transfer in the presence of vibrations that are correlated between different chromophores.  We study the influence of these correlations on the effectiveness of the energy transfer, focusing in particular on the role of vibrations in enabling uphill energy transport in the excitonic degrees of freedom. The role of coherence in overcoming energy barriers has motivated discussion of rectification~\cite{IshizakiFleming2009}, while the coherent coupling of vibrational degrees of freedom in a quantum bath to an uphill gradient of excitonic states has been demonstrated to allow quantum ratcheting of energy transfer over long distances~\cite{Hoyer2012spatial}.   Clearly the phenomenon of VAET~\cite{Gorman18prx} is a prime enabler for such uphill energy transport. In this work we shall explore in detail the ways in which VAET processes facilitate and enhance energy transport, with particular emphasis on the role of VAET with correlated vibrational modes.  We shall find that not only do resonant single phonon processes play a key role in enhancing energy transport, but that multi-phonon processes can also be strong facilitators of energy transport.

The remainder of the paper is organized as follows. Sec.~\ref{sec:model} introduces the model of the trimeric chromophore system with excitonic sites coupling locally to individual vibrational modes, e.g., as with a linear array of trapped ions coupled to transverse vibrational modes. In Sec.~\ref{sec:EET}, we summarize the procedure for numerical calculations of the 2D VAET spectra and the perturbation theory for analysis of the excitation energy transfer features. Full details of the latter are presented in the Supplementary Material~\cite{suppl}.
Sections~\ref{sec:weak_kappa} and \ref{sec:strong_kappa} present and analyze results revealing the various VAET signatures in the weak and strong site-vibration coupling regimes, respectively. In Sec.~\ref{sec:others}, we present the vibronic energy spectrum and discuss the insights this offers for the VAET processes. We also discuss here the effects of the vibrational cross coupling terms induced by restriction to the single-exciton subspace of a Hamiltonian suitable for emulation of excitonic energy transfer with trapped ions.
Sec.~\ref{sec:correl} presents 2D VAET spectra and analysis for a trimeric system realizable for linear arrays of trapped ions with coupling only to longitudinal vibrational modes, resulting in a Hamiltonian with explicit correlations in the coupling of individual sites to the vibrational modes.  Sec.~\ref{sec:diss} provides a summary and outlook for observation of the predicted VAET phenomena in trapped ion experiments, together with a discussion of the implications of this VAET study for understanding excitonic energy transfer in molecular systems.

\section{The effective model of a trimeric chromophore system \label{sec:model}}

\begin{figure}
\centering
  \includegraphics[width=.99\columnwidth]{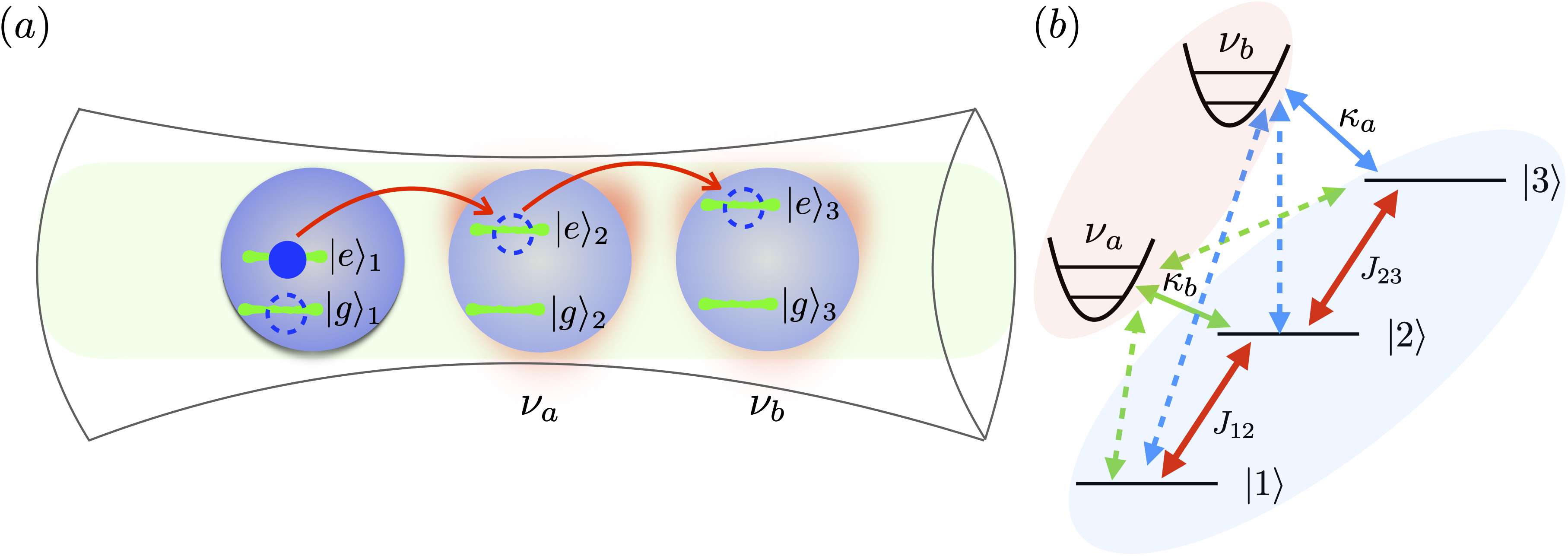} 
\caption{(color online) (a) Schematic of the excitation state transfer in a donor-bridge-acceptor trimeric chromophore system coupled to two vibrations. Each site is modelled by two lowest energy levels $|e\rangle_i$ and $|g\rangle_i$. The backgrounds in light red indicate two vibrations that are individually coupled to the bridge and acceptor sites, respectively, by a $\sigma_z$ type coupling (see text). (b) The effective three-level model of the trimeric chromophore system coupled to two vibrations in the single electronic excitation subspace. In addition to the direct Hamiltonian couplings indicated by the green, blue, and red solid line arrows, there are also cross couplings deriving from the restriction to the single-excitation subspace. These are shown by the green and blue dashed line arrows.}
\label{fig:schematic}
\end{figure}

We consider a donor-bridge-acceptor trimeric chromophore system coupled to two undamped vibrations, shown schematically in Fig.~\ref{fig:schematic}(a). The model can be described by the Hamiltonian 
\begin{eqnarray}
H_{tr}=H_{\rm s}+H_{\rm v}+H_{\rm int} \label{eq:totalH}, 
\end{eqnarray}
where (setting $\hbar=1$)
\begin{eqnarray}
H_{\rm s}&=&\omega_1\sigma_z^{(1)}+\omega_2\sigma_z^{(2)}+\omega_3\sigma_z^{(3)} \notag\\
&&+J_{12}\sigma_x^{(1)}\sigma_x^{(2)}+J_{23}\sigma_x^{(2)}\sigma_x^{(3)}, \label{eq:H_s} \\
H_{\rm v}&=&\nu_a a^{\dagger}a +\nu_b b^{\dagger}b , \label{eq:H_v}\\
H_{\rm int}&=&\kappa_a \sigma_z^{(2)}(a+a^{\dagger}) +\kappa_b \sigma_z^{(3)}(b+b^{\dagger}) . \label{eq:H_int}
\end{eqnarray}
In the electronic Hamiltonian $H_{\rm s}$, the three sites correspond to the donor, bridge, and acceptor, respectively. 
Each site is modelled by a two-level system, with $2\omega_i$ the transition frequency between its ground and excited states, $|g\rangle_i$ and $|e\rangle_i$, respectively.
The Pauli operators $\sigma^{(i)}_{x,z}$ are given by $\sigma^{(i)}_x=|g\rangle_i\langle e| +|e\rangle_i\langle g|$ and $\sigma^{(i)}_z=|e\rangle_i\langle e| -|g\rangle_i\langle g|$. 
$J_{ij}$ is the coupling strength between adjacent $i$th- and $j$th-sites. We assume that the coupling between the first and third sites is vanishingly small. 
The vibrational mode with creation operator $a^{\dagger}$ ($b^{\dagger}$) and frequency $\nu_a$ ($\nu_b$) is coupled to the bridge (acceptor) site with coupling strength $\kappa_a$ ($\kappa_b$). 
In the quantum simulation context, such local coupling of sites to individual vibrations can be realized by coupling to transverse modes of a linear chain of trapped ions, so we denote this Hamiltonian by $H_{tr}$.
In addition to these undamped single-mode vibrations coupled to the electronic sites included in the above Hamiltonian, we will also incorporate dissipation effects by use of non-Hermitian terms in the Hamiltonian. Since the immediate context of this study is the emulation of excitonic energy transfer by trapped ions, these may be due to coupling to optical dephasing or to spontaneous emission, in addition to an additional damped vibrational environment as is usual in natural light harvesting systems.

We focus on the excitation energy transfer in a single electronic excitation subspace. An effective Hamiltonian within this subspace is obtained using the subspace projection operator 
$\Xi \equiv |egg\rangle\langle egg| + |geg\rangle\langle geg| + |gge\rangle\langle gge|$,
using $\tilde{H}_{tr}=\Xi H_{tr} \Xi$. Explicitly, we find 
\begin{eqnarray}
\tilde{H}_{tr} &=& \tilde{\omega}_1 |1\rangle\langle 1|
+\tilde{\omega}_2 |2\rangle\langle 2|
+\tilde{\omega}_3 |3\rangle\langle 3| \notag\\
&&+ J_{12}(|1\rangle\langle 2| + |2\rangle\langle 1|)
+ J_{23}(|2\rangle\langle 3| + |3\rangle\langle 2|) \notag\\
&&+\kappa_a(a^{\dagger}+a)(|2\rangle\langle 2|-|1\rangle\langle 1|-|3\rangle\langle 3|) \notag\\
&&+\kappa_b(b^{\dagger}+b)(|3\rangle\langle 3|-|1\rangle\langle 1|-|2\rangle\langle 2|) \notag\\
&&+\nu_a a^{\dagger}a +\nu_b b^{\dagger}b ,
\label{eq:effec_trimerH}
\end{eqnarray}
where
$|1\rangle \equiv |egg\rangle, |2\rangle\equiv |geg\rangle, |3\rangle\equiv |gge\rangle$, 
$\tilde{\omega}_1 = \omega_1-\omega_2-\omega_3$, 
$\tilde{\omega}_2 = \omega_2-\omega_1-\omega_3$, 
and $\tilde{\omega}_3 = \omega_3-\omega_1-\omega_2$. 
The couplings in the full Hamiltonian are depicted in Fig.~\ref{fig:schematic}(a) and resulting couplings of the effective Hamiltonian in the single excitation subspace are illustrated in Fig.~\ref{fig:schematic}(b).  We note that, just as for a dimer~\cite{Gorman18prx}, assistance from vibrations for energy transport becomes unnecessary with the transition frequencies of three sites are identical, since there are then no energetic difference between spatially separated sites.
Our numerical calculations in this paper will focus on electronically uphill processes, as indicated schematically in Fig.~\ref{fig:schematic}(b).

Two important remarks on the effective model in Eq.~\eqref{eq:effec_trimerH} are in order here. 
First, the counter-rotating terms in the XX-type interaction of Eq.~(\ref{eq:H_s}) do not conserve the number or excitations and therefore do not survive the projection into the single electronic excitation manifold (since $\Xi\sigma_+^{(1)}\sigma_+^{(2)}\Xi = \Xi\sigma_-^{(1)}\sigma_-^{(2)}\Xi =0$).  
Second, it is evident that, in addition to the coupling between an electronic site and its directly connected vibrational mode [see Eq.~(\ref{eq:H_int})], the single-excitation effective model of Eq.~(\ref{eq:effec_trimerH}) contains terms that couple a vibration to excited states of its unconnected sites, 
as indicated by dashed arrows in light green and blue in Fig.~\ref{fig:schematic}(b). 
These terms derive from the projection of the $\sigma_z$ operators through which sites 2 and 3 couple to their individual vibrations, i.e.,
$\Xi\sigma_z^{(2)}\Xi=|2\rangle\langle 2|-|1\rangle\langle 1|-|3\rangle\langle 3|$ and $\Xi\sigma_z^{(3)}\Xi=|3\rangle\langle 3|-|1\rangle\langle 1| -|2\rangle\langle 2|$.

The Hamiltonian Eq.~(\ref{eq:totalH} differs formally 
from the common modelling of vibrational coupling of molecular chromophores, for which only coupling to excited state electonic states is included, i.e., $\kappa_a \sigma_+^{(2)}\sigma_-^{(2)}(a^{\dagger}+a)
 + \kappa_b \sigma_+^{(3)}\sigma_-^{(3)}(b^{\dagger}+b)$~\cite{IshizakiFleming10PCCP}.
 In the Born-Oppenheimer approximation, the ground electronic state of a molecular system is defined to be at the minimum of all relevant vibrational degrees of freedom and therefore there is no linear coupling to the electronic ground state. 
For a single chromophore, this coupling will just shift the overall energy of the system and both pictures are used for single chromophores in the literature.  However, for two or more chromophores, we find that on projection to the single excitation subspace the $\sigma_z$ coupling introduces cross-coupling terms between excitations on different sites that mimic correlations between vibrations at different sites in a molecular system~\cite{Lee2007coherence,Sarovar2011environmental,IshizakiFleming10PCCP,Abramavicius2011correlated,Tiwari12pnas,Rolczynski2018correlated}.
In the following sections we shall see that these cross-couplings can give rise to enhanced collective phenomena in VAET.  
We note that the cross-couplings in Eq.~\eqref{eq:effec_trimerH} show both positive correlations between some sites, eg., between donor and acceptor sites for mode $\nu_a$ and negative correlations between other sites, e.g., between donor and bridge sites for mode $\nu_a$.  Consequently the overall effect of the correlations is not easily rationalized in terms of arguments for correlated dimers~\cite{IshizakiFleming10PCCP}.

\begin{table}[b]
\caption{Comparison of typical parameters for trapped ion emulators used in this work (line 1) with values found in natural photosynthetic systems (line 3). Line 2 shows parameters scaled up from line 1 to the regime for natural systems. Note that the vibrational frequencies on line 1 are for two-phonon processes.}
\centering
\resizebox{\columnwidth}{!}{%
\begin{tabular}{|c|c|c|c|c|c|c|c|c|c|c|c|c|c|}
\hline\hline
Parameters  & $\tilde{\omega}_1$ & $\tilde{\omega}_2 $ & $\tilde{\omega}_3$ & $J_{12}$ & $J_{23}$ & $\nu_a$ & $\nu_b$ & $\kappa_a$ &$\kappa_b$ & $k_BT_a$ & $k_BT_b$ \\ [0.5ex]
\hline
values (kHz)  & $-0.5$ & $0$ & $0.5$ & $0.1$ & $0.1$ & $0.52$ &$0.52$ & $0.1$ & $0.1$ & $0.72$ & $0.72$ \\ [1ex]
\hline
scale-up ($cm^{-1}$)  & $-138.6$ & $0$ & $138.6$ & $27.72$ & $27.72$ & $144$ &$144$ & $27.72$ & $27.72$ & $200$ & $200$ \\ [1ex]
\hline
natural system ($cm^{-1}$)  & $-138.6$ & $0$ & $138.6$ & $-5.9$ & $-13.7$ & $180$ & $180$ & $42.2$ & $42.2$ & $200$ & $200$ \\ [1ex]
\hline\hline
\end{tabular}
}
\label{table:pars_scaleup}
\end{table}

To illustrate the relevance of this study for understanding of natural photosynthetic systems, Table~\ref{table:pars_scaleup} shows typical values for the parameters considered in this work and compares them to the corresponding typical values for the Fenna-Matthews-Olson (FMO) light harvesting complex unit of green sulphur bacteria~\cite{AdolphsRenger06biophys,NalbachThorwart11pre}.  
The first row shows typical values from the parameter ranges employed in this work
and the third row shows corresponding typical values for the FMO system.  The second row shows the result of setting our site energies $\tilde{\omega}_1$, $\tilde{\omega}_2$, and $\tilde{\omega}_3$ to the natural value and scaling up the other parameters accordingly.  
It is evident that the range of parameters available in ion trap emulations scales consistently to the natural system, suggesting that analogs of the phenomena observed here might be present also in some natural light harvesting complexes (for more discussion of this, see the Supplementary Material~\cite{suppl}).  Of particular relevance here are systems that have uphill regions in their landscape of Frenkel (site) exciton energies.  This includes the FMO monomer complex~\cite{AdolphsRenger06biophys}, as well as the purple bacterium {\it Rhodopseudomonas viridis}~\cite{Timpmann1995energy} and the CP43 core antenna of photosystem II~\cite{Muh2012structure}. 

\section{Excitation energy transfer probability \label{sec:EET}}

We calculate the probability of finding an excitation in the acceptor, i.e., the third site, given an initial excitation localized on the donor, i.e., on the first site, as a function time, $P_3(t)$.   Since there is no direct excitonic transfer from donor to acceptor, in the absence of coupling to vibrations the excitation is  transferred via the bridging site.  $P_3(t)$ is given by  
\begin{eqnarray}
P_3(t)&=&Tr [|3\rangle\langle 3| U |1\rangle\langle 1| \rho_a\rho_b U^{\dagger} ] .
\label{eq:P_3}
\end{eqnarray}
Here the unitary time evolution is for the whole system, i.e., $U|1\rangle\langle 1|\rho_a\rho_bU^{\dagger}$, where $|1\rangle\langle1|$, $\rho_a$, and $\rho_b$ are initial states of the chromophoric trimer and two vibrational modes, respectively, and $U=e^{-i\tilde{H}t}$ is an evolution operator with $\tilde{H}$ given by Eq.~(\ref{eq:effec_trimerH}).
Tracing over the vibrational degrees of freedom 
leads to the reduced excitonic system dynamics and then further taking the quantum average of the site number operator $|3\rangle\langle3|$ gives the population at the acceptor site, $P_3(t)$.
Numerical calculations of the transfer probability in Eq.~(\ref{eq:P_3}) are performed assuming thermal initial states for the two vibrations, i.e., $\rho_a=e^{-\nu_a a^{\dagger}a/k_BT_a }/Tr_a[e^{-\nu_aa^{\dagger}a/k_BT_a }]$ and $\rho_b=e^{-\nu_b b^{\dagger}b/k_BT_b }/Tr_b[e^{-\nu_bb^{\dagger}b/k_BT_b }]$, where $k_BT_a$ and $k_BT_b$ are individual temperatures of each vibration. 
We shall use ${\rm Max}[P_3(t)]$ ($0\le t\le t_f$) and $\int_0^{t_f} P_3(t) dt$, which measure the maximum and accumulated population during a given time period $t_f$, respectively, as quantitative measures of the excitonic energy transfer efficiency.

In order to assist in interpreting and understanding the results from these numerical calculations, we also develop an analytic perturbation theory approach.
Our analytical treatment focuses on a symmetric version of the effective three-level model, namely, $J_{12}=J_{23} =J$ and $\tilde{\omega}_{3}-\tilde{\omega}_2 =\tilde{\omega}_2 - \tilde{\omega}_1=\Delta$ in Eq.~(\ref{eq:effec_trimerH}). 
The excitonic Hamiltonian in Eq.~(\ref{eq:effec_trimerH}) then becomes $H_0^{(e)} = \sum_{j=1}^3\lambda_j |e_j\rangle\langle e_j|$, with eigenergies $\lambda_j=0,\pm\Omega$ ($\Omega=\sqrt{2J^2+\Delta^2}$) and eigenstates $|e_j\rangle$.  

Transforming to an the interaction picture with respect to free Hamiltonian [including $H_0^{(e)}$ and the vibrational part (i.e., $H_0^{(\nu)}=\nu_aa^{\dagger}a + \nu_bb^{\dagger}b$)],
the site-vibration Hamiltonian becomes 
\begin{align}
H_I(t)=\sum_{j,k} \kappa_a A_{jk} [a^{\dagger} e^{i(\Delta_{jk}+\nu_a)t} +a e^{i(\Delta_{jk}-\nu_a)t}] |e_j\rangle\langle e_k| \\ \nonumber
+\sum_{j,k} \kappa_b B_{jk} [b^{\dagger} e^{i(\Delta_{jk}+\nu_b)t} +b e^{i(\Delta_{jk}-\nu_b)t}] |e_j\rangle\langle e_k|, 
\end{align}
where $\Delta_{jk}=\lambda_j-\lambda_k$ is the transition frequency between eigenstates and the forms for the coefficients $A_{jk}, B_{jk}$ are given in Appendix~\ref{app:coeff}.
We use fourth-order perturbation theory with respect to the site-vibration coupling to expand the evolution operator as 
\begin{align}
U_I(t)=\mathcal{T} e^{-i\int_0^t ds H_I(s)}\approx1+\sum_{i=1}^4 U_I^{(i)},
\end{align} and correspondingly obtain the transition probability 
\begin{align}
P_3(t)=P_3^{(0)}+P_3^{(1)}+P_3^{(2)} 
\end{align}
with 
\begin{align*}
P_3^{(0)} &= Tr_{a,b}[(\mathcal{A}^{(0)})^{\dagger} \mathcal{A}^{(0)} \rho_a\rho_b], \\ 
P_3^{(1)} &= Tr_{a,b} \{ [ (\mathcal{A}^{(1)})^{\dagger} \mathcal{A}^{(1)} +  (\mathcal{A}^{(0)})^{\dagger} \mathcal{A}^{(2)} + (\mathcal{A}^{(2)})^{\dagger} \mathcal{A}^{(0)} ]  \rho_a\rho_b\}, 
\end{align*}
and 
\begin{align*}
P_3^{(2)} &= Tr_{a,b} \{[ (\mathcal{A}^{(2)})^{\dagger} \mathcal{A}^{(2)} 
+ (\mathcal{A}^{(1)})^{\dagger} \mathcal{A}^{(3)} \\ 
&\,+(\mathcal{A}^{(3)})^{\dagger} \mathcal{A}^{(1)} 
+(\mathcal{A}^{(0)})^{\dagger} \mathcal{A}^{(4)} +(\mathcal{A}^{(4)})^{\dagger} \mathcal{A}^{(0)}]\rho_a\rho_b \}, 
\end{align*}
where the transition amplitudes 
are given by $\mathcal{A}^{(0)} = \langle 3|U_0 |1\rangle$ and $\mathcal{A}^{(i)} = \langle 3|U_0 U_I^{(i)} |1\rangle$, with $U_0=e^{-i(H_0^{(e)} + H_0^{(\nu)})t}$.  

Unless otherwise stated, in the numerical calculations and perturbative analysis of these in the following two sections we shall study the maximum probability ${\rm Max}[P_3(t)]$ during a given time period ($0\le t\le t_f=400$ms), for a symmetric trimeric system with $J_{12}=J_{23}=J$ and $\tilde{\omega}_{3}-\tilde{\omega}_2=\tilde{\omega}_2-\tilde{\omega}_1=\Delta$ in Eq.~(\ref{eq:effec_trimerH}), coupled to two vibrations at a temperature larger (by a factor of $\simeq 10$) than these energies.
This choice of parameters represents an energetically uphill process in the single excitation subspace.  
We shall present two-dimensional (2D) VAET spectra obtained by evaluating  ${\rm Max}[P_3(t)]$ over the given time period as a function of the two vibrational frequencies $\nu_a$ and $\nu_b$. These frequencies are given in units of the energy difference $\Delta_{31}$ between the eigenstates $|e_3\rangle$ and $|e_1\rangle$ of the electronic Hamiltonian $H_0^{(e)}|e_j\rangle=\lambda_j|e_j\rangle$.
A full description of the perturbative analysis is given together with explicit expressions for the eigenstates of the symmetric model in the Supplementary Material~\cite{suppl}.
The numerical calculations typically employ a vibrational basis of $N=15$ Fock states, which is sufficient for convergence of the VAET spectra over a broad range of parameters, as described in Appendix~\ref{app:converge}.

\section{VAET signatures in the weak site-vibration coupling regime \label{sec:weak_kappa}}

When the site-vibration coupling is weak, e.g., $\kappa_a,\kappa_b<\Delta, J$, the electronic states of the trimeric chromophore system are only weakly perturbed and do not gain substantial vibronic character.  In this situation the energy transfer processes are primarily excitonic in origin but assistance by vibrations that are coupled to the excitonic states can be still expected. 
Fig.~\ref{fig:maxPop_N15_kappa_0d01_kBT_1d5_1d5_amplify}(a), presents a 2D VAET spectrum  in this regime and
Fig.~\ref{fig:maxPop_N15_kappa_0d01_kBT_1d5_1d5_amplify}(b) presents a corresponding schematic diagram summarizing the energy transfer processes responsible for each of the main features of the 2D VAET spectrum.  The perturbative analysis of these features is then summarized in Fig.~\ref{fig:vaet}. We now discuss the 2D VAET spectral features, starting with those due to single-mode VAET processes and then proceeding to the multi-mode VAET processes.

\subsection{Single-mode VAET \label{sec:singlemode}}

The main features due to single modes in the weak site-vibration coupling regime are represented by the three vertical lines at $\nu_a/\Delta_{31}=1,0.5,0.25$ and the horizontal lines at $\nu_b/\Delta_{31}=1,0.5,0.25$.  These signify resonant 
one-, two-, and four-phonon VAET processes, respectively, assisted by either the vibration $\nu_a$ which is coupled to the bridging site 2 (vertical lines) or the vibration $\nu_b$ which is coupled to the terminal, acceptor site (horizontal lines).
For a given number of phonons, e.g., one, two or four, the vertical line is more intense than the corresponding horizontal line, indicating a stronger impact of the bridging vibration $\nu_a$ in assisting the energy transfer. 

\begin{figure}
\centering
  \includegraphics[width=.86\columnwidth]{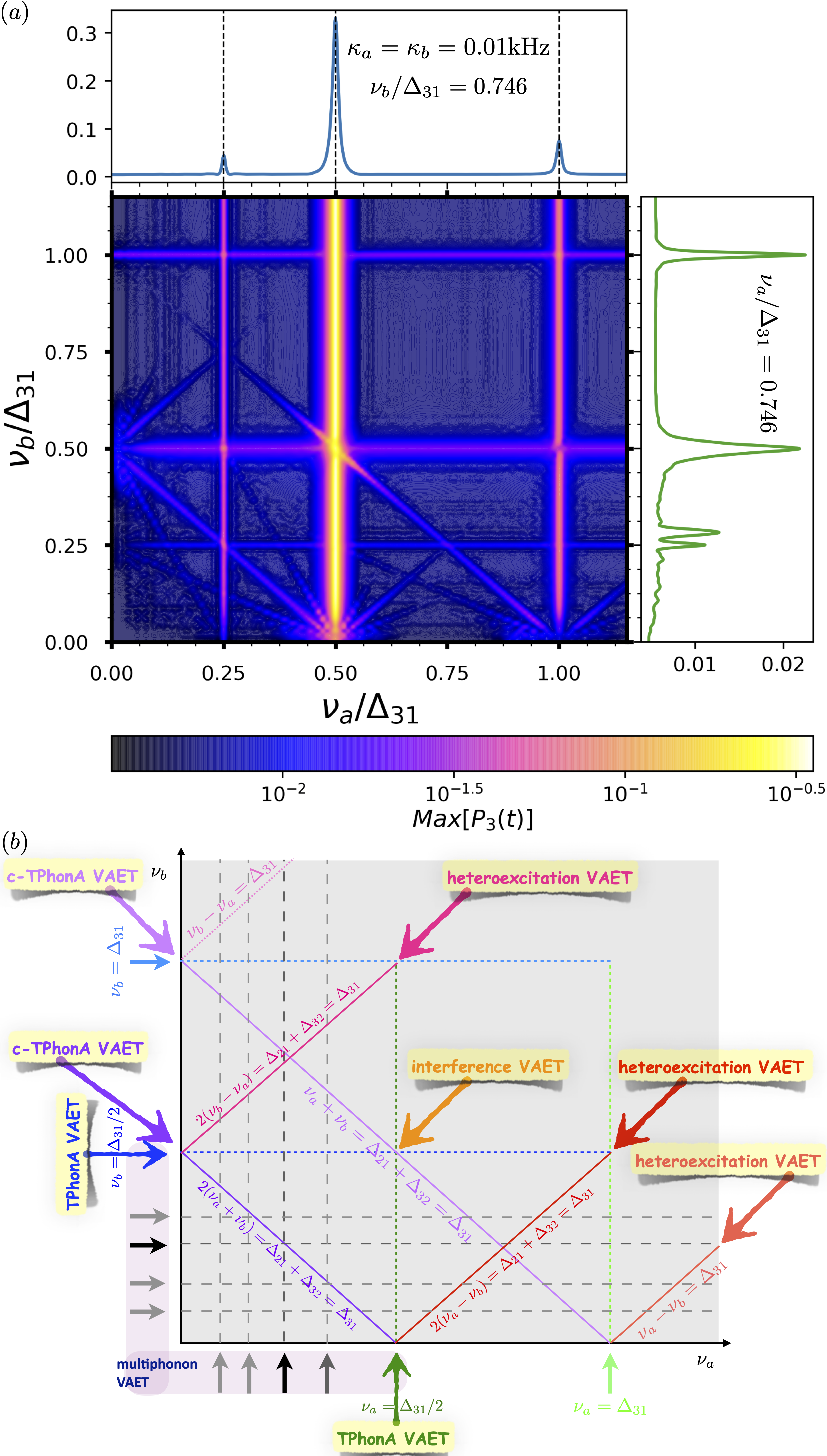} 
\caption{(color online) (a) 2D VAET spectrum of a symmetric trimeric chromophore system coupled to two non-interacting vibrations in the weak site-vibration coupling regime $\kappa_a=\kappa_b=0.01$kHz. The maximum transfer probability ${\rm Max}[P_3(t)]$ is taken during a time period $t\in[0,400]$ms. 
$\Delta_{ij}$ is the energy difference between eigenstates $|e_i\rangle$ and $|e_j\rangle$ of the electronic part in Eq.~(\ref{eq:effec_trimerH}) with $\{\tilde{\omega}_1,\tilde{\omega}_2,\tilde{\omega}_3,J_{12},J_{23}\}=\{-0.5,0,0.5,0.1,0.1\}$kHz, satisfying  $\Delta_{31}=2\Delta_{21}$~\cite{suppl}. 
We consider two vibrations with identical temperatures $k_BT_a=k_BT_b=1.5$kHz. The truncation number of each vibrational Fock space is $N=15$. The 1-dimensional slices on the top and right of the 2-dimensional contour plot are taken at $\nu_b/\Delta_{31}=0.746$ and $\nu_a/\Delta_{31}=0.746$, respectively. 
(b) Schematic diagram identifying the different single- and multi-mode VAET features of the 2D VAET plot for the trimeric chromophore system shown in (a).}
\label{fig:maxPop_N15_kappa_0d01_kBT_1d5_1d5_amplify}
\end{figure}

\begin{figure*}
\centering
  \includegraphics[width=1.6\columnwidth]{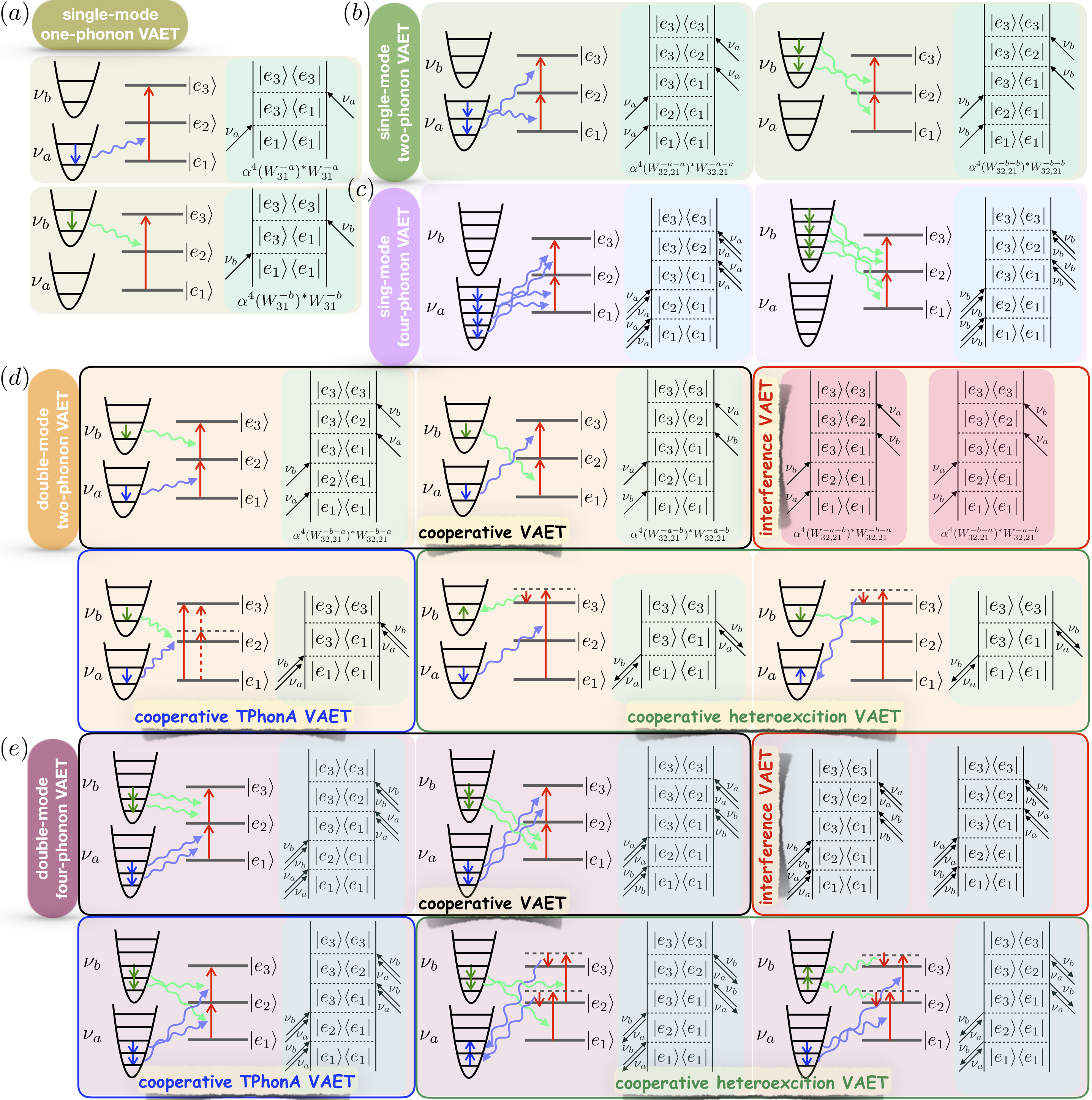} 
\caption{(color online)  
Schematics of the various VAET processes contributing to the 2D VAET specrum of Fig.~\ref{fig:maxPop_N15_kappa_0d01_kBT_1d5_1d5_amplify}. Panels (a) - (c) show single-mode VAET and panels (d) - (e) show multimode VAET.  Depicted are schematics of one-, two-, and four-phonon transfer processes, together with the corresponding Feynman diagrams connecting the initial state $|e_1\rangle$ and final state $|e_3\rangle$. 
The interaction amplitudes $W^{q_1x_1}_{jk}$ and $W^{q_1x_1,q_2x_2}_{jk,kk'}$ with $x_i\in\{ a,b\}$, and $q_i\in\{+,-\}$ appearing below some of the Feynman diagrams are defined in the Supplementary Material~\cite{suppl}.
}
\label{fig:vaet}
\end{figure*}

This can be confirmed by our perturbative analysis as follows. For the resonant one-phonon VAET, 
perturbative expansion of the transfer probability (see Ref.~\cite{suppl} for full details) shows that the first order term, which is second order in the interaction Hamiltonian, is proportional to $\alpha^4(W_{31}^{-a})^*W_{31}^{-a}\sim t^2\kappa_a^2A_{13}^2$ for transitions along the line $\nu_a/\Delta_{31}=1$, and to $\alpha^4(W_{31}^{-b})^*W_{31}^{-b}\sim t^2\kappa_b^2B_{13}^2$ along the line $\nu_b/\Delta_{31}=1$. Here $\alpha \sim 1$ for the weak coupling regime and the coefficients $A_{13}$, $B_{13}$ are given by Eqs.~(\ref{eq:A_13}) and (\ref{eq:B_13}) of Appendix~\ref{app:coeff}, respectively.
This VAET process is illustrated schematically by the Feynman diagrams in Fig.~\ref{fig:vaet}(a). 
Averaging this transfer probability over the thermal distributions of vibrational states for modes $\nu_a$ and $\nu_b$
\cite{suppl}, shows that even under conditions of identical site-vibration couplings ($\kappa_a=\kappa_b$) and identical vibrational temperatures ($k_BT_a=k_BT_b$), there will nevertheless be a higher probability for excitations along the vertical line than along the horizontal line, confirming the stronger impact of the vibration $\nu_a$ that is coupled to the bridge site.
This effect is clearly visible in the 1-dimensional top and right slices of Fig.~\ref{fig:maxPop_N15_kappa_0d01_kBT_1d5_1d5_amplify}(a), where it can be seen that the probability at the one-phonon VAET peak $\nu_a/\Delta_{31}=1$ (top slice) is almost four times as large as that of the one-phonon VAET $\nu_b/\Delta_{31}=1$ (right slice).
Closer examination of the perturbative couplings shows that difference arises from the factor of 2 in $A_{13}$ relative to $B_{13}$ (see Eqs.~(\ref{eq:A_13}) and (\ref{eq:B_13})). $A_{13}$ is the matrix element of the electronic coupling to the bridge vibrational mode $\nu_a$, between the lowest excitonic state $\ket{e_1}$ and the highest excitonic state $\ket{e_3}$~\cite{suppl}, i.e.,  $A_{13}=\langle e_1|(|2\rangle\langle 2| -|1\rangle\langle 1| -|3\rangle\langle 3|) |e_3\rangle$. $B_{13}$ is the corresponding matrix element of the terminal vibrationa mode, i.e., $B_{13}=\langle e_1|(|3\rangle\langle 3| -|1\rangle\langle 1| -|2\rangle\langle 2|)|e_3\rangle$. For the symmetric Hamiltonian we have $\langle e_1|1\rangle\langle 1|e_3\rangle=-\langle e_1|2\rangle\langle 2|e_3\rangle= J^2/2\Omega^2$ and $\langle e_1|3\rangle\langle 3|e_3\rangle=J^2/\Omega^2$, resulting in a factor of 2 larger coupling for the electronic coupling to the bridge vibrational mode.

When the vibrational frequency, either $\nu_a$ or $\nu_b$, is equal to half of an excitonic transition frequency, the transition from $|e_1\rangle$ to $|e_3\rangle$ is still accessible but only via an intermediate state $|e_2\rangle$. The excitonic transition is then accompanied by an absorption of two phonons. 
This single-mode two-phonon VAET appears in Fig.~\ref{fig:maxPop_N15_kappa_0d01_kBT_1d5_1d5_amplify}(a) at the vertical line $\nu_a/\Delta_{31}=0.5$ and at the horizontal line $\nu_b/\Delta_{31}=0.5$.  We denote these processes as single mode two-phonon absorption (TPhonA) in Fig.~\ref{fig:vaet}(b). The corresponding transfer probabilities in the weak site-site coupling limit ($J<\Delta$) are given by $\alpha^4(W_{32,21}^{-a-a})^*W_{32,21}^{-a-a}\sim \alpha^4t^4\kappa_a^4A_{12}A_{23}A_{32}A_{21}$ 
and $\alpha^4(W_{32,21}^{-b-b})^*W_{32,21}^{-b-b}\sim \alpha^4t^4\kappa_b^4B_{12}B_{23}B_{32}B_{21}$, with $\alpha\sim 1$ for weak coupling as before, and $A_{ij}$ and $B_{ij}$ given in Eqs.~(\ref{eq:A_12})-(\ref{eq:B_23}).
Further perturbative analysis with respect to $J/\Delta$ reveals that the transition $|e_1\rangle\rightarrow|e_2\rangle\rightarrow|e_3\rangle$ is a second-order process when the absorbed phonons are both borrowed from the bridging vibration $\nu_a$ (since $A_{12}\propto\frac{J}{\Delta}$ and $A_{23}\propto \frac{J}{\Delta}$, see Appendix~\ref{app:coeff} and Ref.~\onlinecite{suppl}), 
but becomes a third-order process when the two phonons are provided by the terminal mode $\nu_b$ (since $B_{21}\propto(\frac{J}{\Delta})^3$ and $B_{23}\propto \frac{J}{\Delta}$).
Therefore, by a similar argument to the one-phonon VAET above, i.e., multiplying the above transfer probability by a prefactor $2n_a^2$ or $2n_b^2$~\cite{suppl} to account for thermal averaging over the vibrational modes,
it is then evident that just as in the one-phonon VAET process, the vibration ($\nu_a$) coupled to the bridge site has a stronger impact on the two-phonon VAET processes than the vibration $\nu_b$ connected to the terminal, i.e., acceptor site. 

Another observable feature of the single-mode VAET in Fig.~\ref{fig:maxPop_N15_kappa_0d01_kBT_1d5_1d5_amplify}(a) is the four-phonon process at $\nu_a/\Delta_{31}=0.25$ or $\nu_b/\Delta_{31}=0.25$, which is described by the Feynman diagrams in Fig.~\ref{fig:vaet}(c). Comparison between the corresponding vertical and horizontal lines in Fig.~\ref{fig:maxPop_N15_kappa_0d01_kBT_1d5_1d5_amplify}(a) supports the conclusion above that the vibration $\nu_a$ which is coupled to the bridge site has a stronger impact on the energy transfer than the vibration coupled to the terminal site.
Since this four-phonon VAET is a higher order process than that considered in our perturbation theory, we do not provide an analytical expression for the transfer probability here. 

We also find that the two-phonon VAET at $\nu_{a(b)}/\Delta_{31}=0.5$ is dominant over the one-phonon VAET at $\nu_{a(b)}/\Delta_{31}=1$). This is particularly marked for the single-mode VAET enabled by the vibration $\nu_a$ that is coupled to the bridge site. This is evident from the vertical lines in Fig.~\ref{fig:maxPop_N15_kappa_0d01_kBT_1d5_1d5_amplify}(a) and the 1-dimensional slice located above this. 
While the dominance of TPhonA VAET is clearly visible for the bridging site vibration, it is also manifested to a lesser degree for the terminal site vibration (see 1-dimensional slice to the right of Fig.~\ref{fig:maxPop_N15_kappa_0d01_kBT_1d5_1d5_amplify}(a) where the integrated strength of TPhonA VAET is clearly stronger~\cite{suppl}, despite a slightly lower maximal value).
This dominance of the two-phonon VAET over one-phonon VAET is particularly marked in the weak coupling regime (recall $\kappa_a=\kappa_b=0.01$kHz for Fig.~\ref{fig:maxPop_N15_kappa_0d01_kBT_1d5_1d5_amplify}), but will also be evident in the strong coupling regime results presented in Sec.~\ref{sec:strong_kappa} below.
The reason for this dominance is the relatively high temperature considered here, i.e., $k_BT_{a(b)}=1.5$kHz, which ensures availability of the required number of phonons for both one- and two-phonon VAET, while the probability of energy transfer increases with average phonon number. 
Thus, for the one-phonon VAET, the average phonon numbers are $n_{a(b)}\sim 1$ when $\nu_{a(b)}/\Delta_{31}=1$, while for the two-phonon VAET, the average phonon numbers are $n_{a(b)}\sim 2.4$ when $\nu_{a(b)}/\Delta_{31}=0.5$. In the latter situation there is a higher than required average phonon number for $\nu_a(b)$, implying that the two-phonon processes are more likely ($1<2.4/2=120\%$). 
This analysis also holds when the temperature becomes so low that the necessary number of phonons cannot be taken from the thermal state, e.g., for $k_BT_{a(b)}=0.5$kHz, where the two-phonon VAET is still dominant relative to one-phonon VAET. 
Here the average phonon number is $n_{a(b)}\sim 0.143$ when $\nu_{a(b)}/\Delta_{31}=1$, and $n_{a(b)}\sim 0.548$ when $\nu_{a(b)}/\Delta_{31}=0.5$. Since $0.143\%<0.548/2=0.274\%$, the two-phonon VAET is still dominant at this lower temperature (see also Section~\ref{subsec:vibrationaltemp} below).

Finally, we note that the single-mode two- and four-phonon VAET features seen in Fig.~\ref{fig:maxPop_N15_kappa_0d01_kBT_1d5_1d5_amplify} are in good agreement with the additional partially-resolved peaks observed in the recent experimental study of single-mode VAET in a dimer system emulated with trapped ions~\cite{Gorman18prx}.

\subsection{Multimode VAET}

The collective behaviors of multiple vibrations enable unique signatures of VAET arising in our trimeric system, relative to those due to individual ones of single vibrations presented above. 
Such signatures are represented by the diagonal and anti-diagonal lines in Figs.~\ref{fig:maxPop_N15_kappa_0d01_kBT_1d5_1d5_amplify}(a) and (b).  They include  cooperativity and interference of the two vibrations. Of particular interest for the former is the manifestation of the phononic analog of Mayer's two-photon absorption~\cite{Mayer1931}, constituting the anti-diagonal lines, and the inverse of this that combines vibrational and excitonic transitions, which we refer to as heteroexcitation, and which constitute the diagonal lines.

\subsubsection{Cooperative Two-Phonon Absorption VAET}

The anti-diagonal 
lines in Fig.~\ref{fig:maxPop_N15_kappa_0d01_kBT_1d5_1d5_amplify}(a) signify cooperative processes in which phonons from both modes are involved in a VAET process.  For example, the line $\nu_a/\Delta_{31}+\nu_b/\Delta_{31}=1$ signifies a double-mode two-phonon cooperative VAET process.  We designate this as a cooperative TPhonA process (c-TPhonA) in Fig.~\ref{fig:maxPop_N15_kappa_0d01_kBT_1d5_1d5_amplify}(b).
This cooperative process with a simultaneous absorption of phonons from two vibrations in assisting the energy transfer constitutes a phononic analogue of two-photon absorption~\cite{Mayer1931}.

\begin{figure}
\centering
  \includegraphics[width=.85\columnwidth]{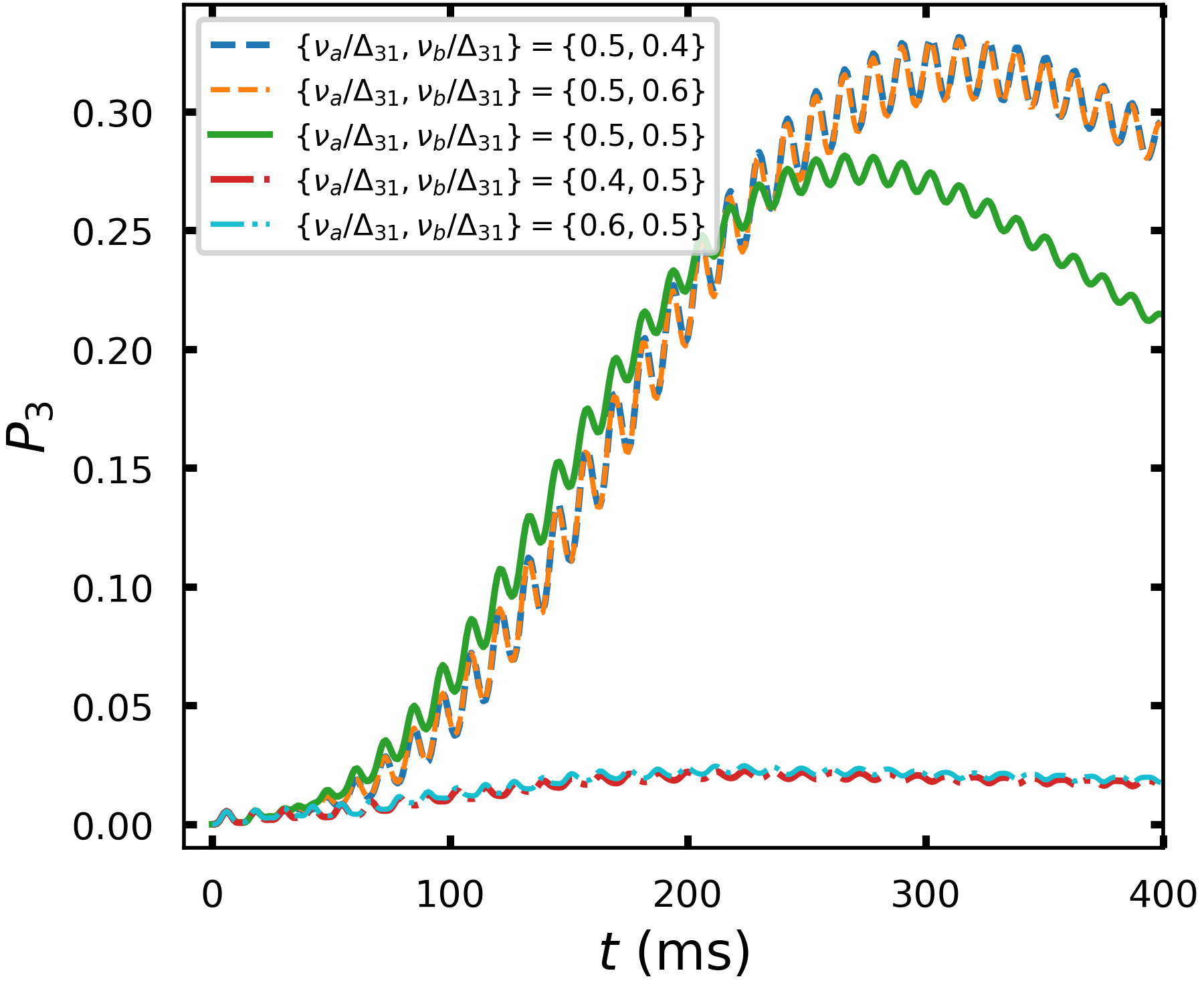}
\caption{(color online) Time evolution of the probability $P_3$ of the acceptor at the symmetric point $\{\nu_a/\Delta_{31},\nu_b/\Delta_{31}\}=\{0.5,0.5\}$ and nearby points along the $\nu_a/\Delta{31}=0.5$ and $\nu_a/\Delta{31}=0.5$ lines in Fig.~\ref{fig:maxPop_N15_kappa_0d01_kBT_1d5_1d5_amplify}. All other parameters are the same as in Fig.~\ref{fig:maxPop_N15_kappa_0d01_kBT_1d5_1d5_amplify}(a).}
\label{fig:interference}
\end{figure}

At the symmetric lattice point $\{\nu_a/\Delta_{31},\nu_b/\Delta_{31}\}=\{0.5,0.5\}$ satisfying the resonance condition $\nu_a=\nu_b=\Delta_{21}=\Delta_{32}$, the transition from $|e_1\rangle$ to $|e_3\rangle$ proceeds 
via the bridge state $|e_2\rangle$ and is assisted by two 
strongly cooperative processes consisting of absorption of a single phonon from one mode, followed by a second phonon from the other mode, with perturbative transfer probability $\alpha^4(W_{32,21}^{-b-a})^*W_{32,21}^{-b-a}$ or $\alpha^4(W_{32,21}^{-a-b})^*W_{32,21}^{-a-b}$~\cite{suppl}, 
as illustrated in Fig.~\ref{fig:vaet}(d).
The two transfer processes having absorption of two phonons from distinct vibrational modes in different orders can interfere with each other, giving rise to a double-mode two-phonon interference VAET. The Feynman pathways for this interference are shown in the right hand (red) subpanel of Fig.~\ref{fig:vaet}(d) and the associated  perturbative expression for the probability is $\alpha^4(W_{32,21}^{-a-b})^*W_{32,21}^{-b-a}+\alpha^4(W_{32,21}^{-b-a})^*W_{32,21}^{-a-b}$~\cite{suppl}. 
Fig.~\ref{fig:interference} shows the time dependence of $P_3(t)$ at $(0.5,0.5)$ (solid green line) together with corresponding time traces for nearby points along the $(0.5, \nu_b /\Delta_{31})$ (dashed blue and yellow lines). This shows that the symmetric point (green line) is a point of destructive interference along the $\nu_b$ axis, since it has a smaller maximal probability than the time traces of the nearby points.
Similarly comparing the time traces of $P_3(t)$ for points along the  $(\nu_a /\Delta_{31}, 0.5)$ line shows that the point $(0.5,0.5)$ is a point of constructive interference along the $\nu_a$ axis, since here the green curve has a higher maximal value than those of nearby points (red and cyan dot-dashed curves). 

At all points away from the symmetric lattice point, i.e., along the rest of the anti-diagonal line $\nu_a/\Delta_{31}+\nu_b/\Delta_{31}=1$ with $\nu_a\neq\nu_b$, the transition from $|e_1\rangle$ to $|e_3\rangle$ involves a virtual intermediate state,  i.e., $|e_2'\rangle$ rather than $|e_2\rangle$ as illustrated in the lower left subpanel of Fig.~\ref{fig:vaet}(d). These points therefore show relatively small transfer probability in Fig. ~\ref{fig:maxPop_N15_kappa_0d01_kBT_1d5_1d5_amplify}.
We point out that the upper portion of the anti-diagonal with $\nu_b<\nu_a$ is noticeably more intense than the lower part where $\nu_b<\nu_a$. This is once again a consequence of the stronger impact of the vibration $\nu_a$ coupled to the bridge site in assisting the energy transfer, relative to that of the vibration coupled to the terminal site.  
In situations for which the impact of the vibration $\nu_b$ becomes dominant in the energy transfer, for example the Hamiltonian with site-correlated vibrational modes analyzed in Sec.~\ref{sec:correl}, this effect is reversed.

Similarly, the second anti-diagonal line visible in Fig. ~\ref{fig:maxPop_N15_kappa_0d01_kBT_1d5_1d5_amplify}(a), i.e., $\nu_a/\Delta_{31}+\nu_b/\Delta_{31}=0.5$ at the bottom left, signifies a double-mode, four-phonon cooperative VAET involving two processes of two-phonon absorption each.
Here the lattice point, i.e., $\{\nu_a/\Delta_{31},\nu_b/\Delta_{31}\}=\{0.25,0.25\}$, 
can host an interference between two transfer pathways involving different ordering of the two-phonon absorptions, while
the rest of this anti-diagonal line, i.e., $\nu_a\neq\nu_b$, shows a relatively low transfer probability due to the off-resonant nature of the intermediate virtual state.  This constitutes a higher-order phononic analogue of  two-photon absorption, which is  illustrated in Fig.~\ref{fig:vaet}(e). 

\subsubsection{Heteroexcitation VAET}

In addition to these cooperative VAET processes for which both vibrations contribute phonons to assist excitonic energy transfer, we also find evidence of VAET processes in which a phonon from one vibrational mode can simultaneously excite 
not only the electronic system but also the other vibrational mode. 
This is a new kind of cooperative mechanism of VAET, which we shall refer to as  ``heteroexcitation". It is evidenced by the diagonal lines in Fig.~\ref{fig:maxPop_N15_kappa_0d01_kBT_1d5_1d5_amplify}(a), which are also shown with their assignments as solid diagonal lines in Fig.~\ref{fig:maxPop_N15_kappa_0d01_kBT_1d5_1d5_amplify}(b). For example, the line $\nu_a/\Delta_{31}-\nu_b/\Delta_{31}=1$ indicates that one phonon of vibrational mode $\nu_a$ generates an electronic transition from site 1 to 3, together with absorption of a single phonon in mode $\nu_b$. The line $\nu_a/\Delta_{31}-\nu_b/\Delta_{31}=0.5$ represents processes in which two phonons of vibrational mode $\nu_a$ generate the same electronic transition, but now together with absorption of two phonons in mode $\nu_b$. An analogous interpretation applies to the line $\nu_b/\Delta_{31}-\nu_a/\Delta_{31}=0.5$.

We note that the alternative heteroexcitation associated with the diagonal line $\nu_b/\Delta_{31}-\nu_a/\Delta_{31}=1$, shown as the dotted diagonal line at the top left of Fig.~\ref{fig:maxPop_N15_kappa_0d01_kBT_1d5_1d5_amplify}(b), in which one phonon from vibrational mode $\nu_b$ generates an excitonic transition together with absorption of a phonon of vibrational mode $\nu_a$ is not observable in Fig.~\ref{fig:maxPop_N15_kappa_0d01_kBT_1d5_1d5_amplify}(a).  
This is because of the weaker impact of the vibration coupled to the terminal site. This vibration now has to provide one phonon to be absorbed by both the timer and the other vibration, which a significantly weaker process at this temperature.
However this transition would appear on further increasing the temperature of mode $b$ ($k_BT_b$), so we also show the relevant Feynman diagrams for this process in Fig.~\ref{fig:vaet}(e).

We note that while the reverse of two photon absorption, namely one photon exciting two atoms, has been discussed previously, it was assumed there that atoms of identical frequency are excited by a single  photon~\cite{GarzianoNori16prl}, consistent with the larger wavelength of optical photons relative to atoms. The heteroexcitations seen here constitute a generalized phononic analog of that optical phenomenon, where now not only the energies, but also the nature of the two degrees of freedom being excited can be different.
Another interesting interpretation of this process is that of redistribution of energy from one phonon reservoir to another, mediated by the electronic degrees of freedom. 

\begin{figure}
\centering
  \includegraphics[width=.99\columnwidth]{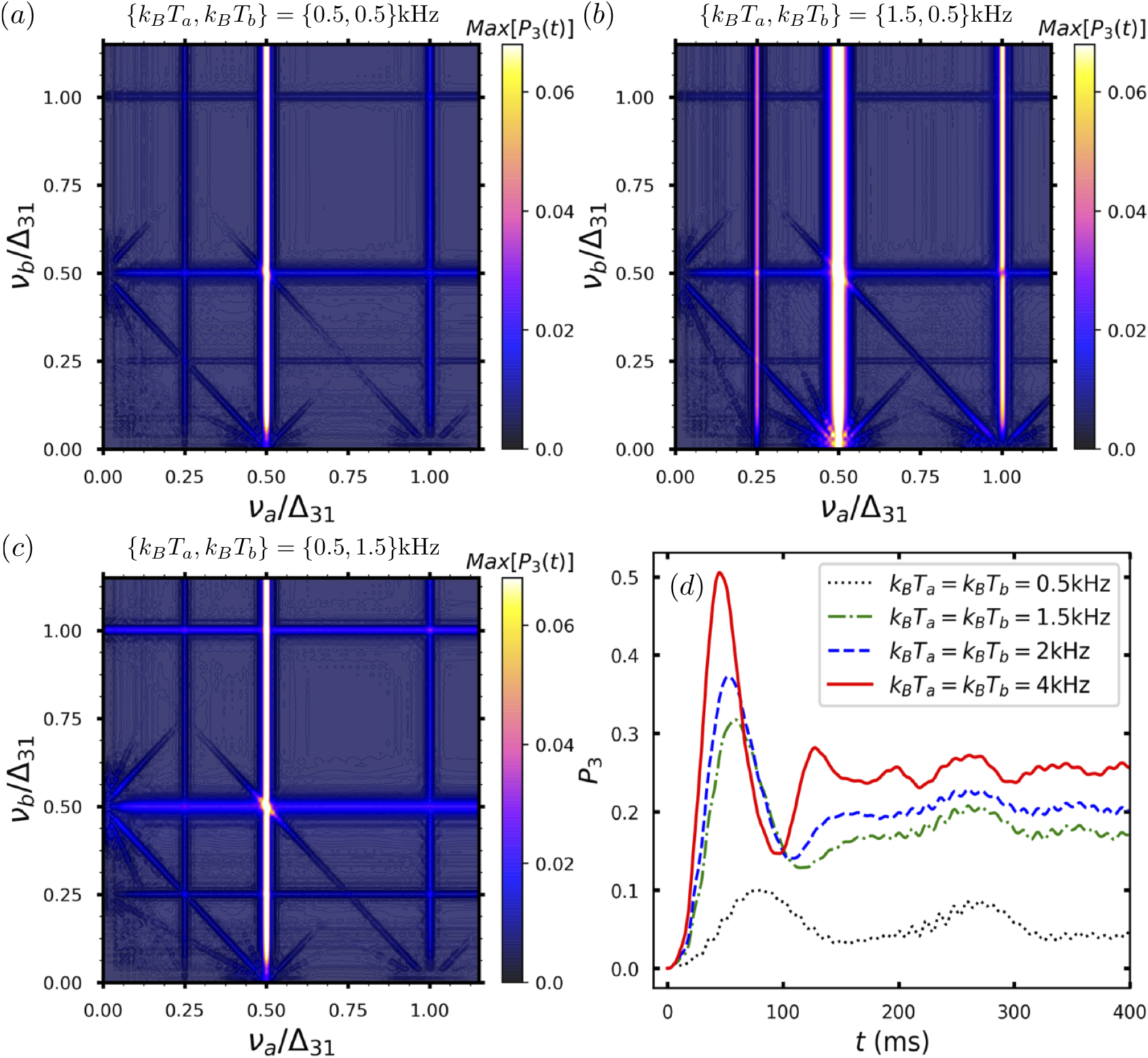} 
\caption{(color online) 2D VAET spectra showing ${\rm Max}[P_3(t)]$ in a time period $t\in[0,400]$ms for different vibrational temperatures $\{k_BT_a,k_BT_b\}$ in the weak coupling regime. (a) $\{k_BT_a,k_BT_b\}= \{0.5,0.5\}$, (b) $\{k_BT_a,k_BT_b\}= \{1.5,0.5\}$, and (c) $\{k_BT_a,k_BT_b\}= \{0.5,1.5\}$. All calculations employed phonon truncation at $N=15$ and site-vibration coupling strength $\kappa_a=\kappa_b=0.01$kHz. 
(d) Time trace of the transfer probability $P_3(t)$ at the resonance points $(\nu_a/\Delta_{31},\nu_b/\Delta_{31})=(0.5,0.5)$ and $(\nu_a,\nu_b)=(0.52,0.52)$ kHz, for temperatures $k_BT_a=k_BT_b=0.5$kHz, $1.5$kHz, $2$kHz,  
$4$kHz, with $N=40$ phonon truncation and site-vibration couplings $\kappa_a=\kappa_b=0.05$kHz. 
All other parameters are the same as in Fig.~\ref{fig:maxPop_N15_kappa_0d01_kBT_1d5_1d5_amplify}.} 
\label{fig:maxPop_N15_kappa_0d01_kBT_1d5_0d5_amplify}
\end{figure}

\subsection{Vibrational temperature effects}
\label{subsec:vibrationaltemp}

One interesting capability of artificial energy transport as studied with emulations using e.g., trapped ions, that is not possible in real molecular systems, is the ability to individually vary the effective temperature of different vibrational modes. Here we assess the effects of these temperatures on VAET features.  Fig.~\ref{fig:maxPop_N15_kappa_0d01_kBT_1d5_0d5_amplify} presents 2D VAET spectra at three  different temperatures from that in Fig.~\ref{fig:maxPop_N15_kappa_0d01_kBT_1d5_1d5_amplify}, including also the presence of a temperature bias between the two 
vibrations in panels (b) and (c).

In the absence of a temperature bias, i.e, when the vibrational temperatures are equal (panel (a)), comparison 
with the higher vibrational temperature spectrum of Fig.~\ref{fig:maxPop_N15_kappa_0d01_kBT_1d5_1d5_amplify} ($k_BT_a=k_BT_b=1.5$kHz) shows that collective VAET features such as the cooperative behavior evidenced by the anti-diagonal and diagonal lines in the 2D spectrum become weaker as the vibrational temperature decreases, indicating a suppression of vibrationally assisted energy transfer processes. However the two-phonon VAET is still dominant over the one-phonon VAET, as discussed in Section~\ref{sec:singlemode}.

When a temperature bias between the two vibrations is present, as in panels (b) and (c), we find that increasing either $k_BT_a$ [Fig.~\ref{fig:maxPop_N15_kappa_0d01_kBT_1d5_0d5_amplify}(b)] or $k_BT_b$ [Fig.~\ref{fig:maxPop_N15_kappa_0d01_kBT_1d5_0d5_amplify}(c)] will enhance the transfer processes assisted by either of the vibrations coupled to the bridge or to the acceptor. This suggests that the weaker impact of the vibration coupled to the acceptor seen above could be enhanced by selectively raising the temperature of this mode in an emulation experiment. 

Panels (b) and (c) of Fig.~\ref{fig:maxPop_N15_kappa_0d01_kBT_1d5_0d5_amplify} show that the presence of a temperature bias across the two vibrations can also enhance heteroexcitations at $|\nu_a-\nu_b|=\Delta_{31}, \Delta_{31}/2$, relative to that seen for equal temperatures in Fig.~\ref{fig:maxPop_N15_kappa_0d01_kBT_1d5_0d5_amplify}(a).   
Such enhancement of heteroexcitations would increase with further increase of the temperature bias. 

The time dependence of energy transfer probability in VAET processes is also strongly dependent on the vibrational temperatures.
Fig.~\ref{fig:maxPop_N15_kappa_0d01_kBT_1d5_0d5_amplify}(d) shows the time evolution of the transfer probability at the resonance points $(\nu_a/\Delta_{31},\nu_b/\Delta_{31})=(0.5,0.5)$ and $(\nu_a,\nu_b)=(0.52,0.52)$kHz in Figs.~\ref{fig:maxPop_N15_kappa_0d01_kBT_1d5_1d5_amplify} and \ref{fig:maxPop_N15_kappa_0d01_kBT_1d5_0d5_amplify}(a). We see that increasing the temperature generically enhances the energy transfer, with the maximum probability reaching values up to  $\sim0.5$ when $k_BT_a=k_BT_b=4$kHz.

\begin{figure}
\centering
  \includegraphics[width=.99\columnwidth]{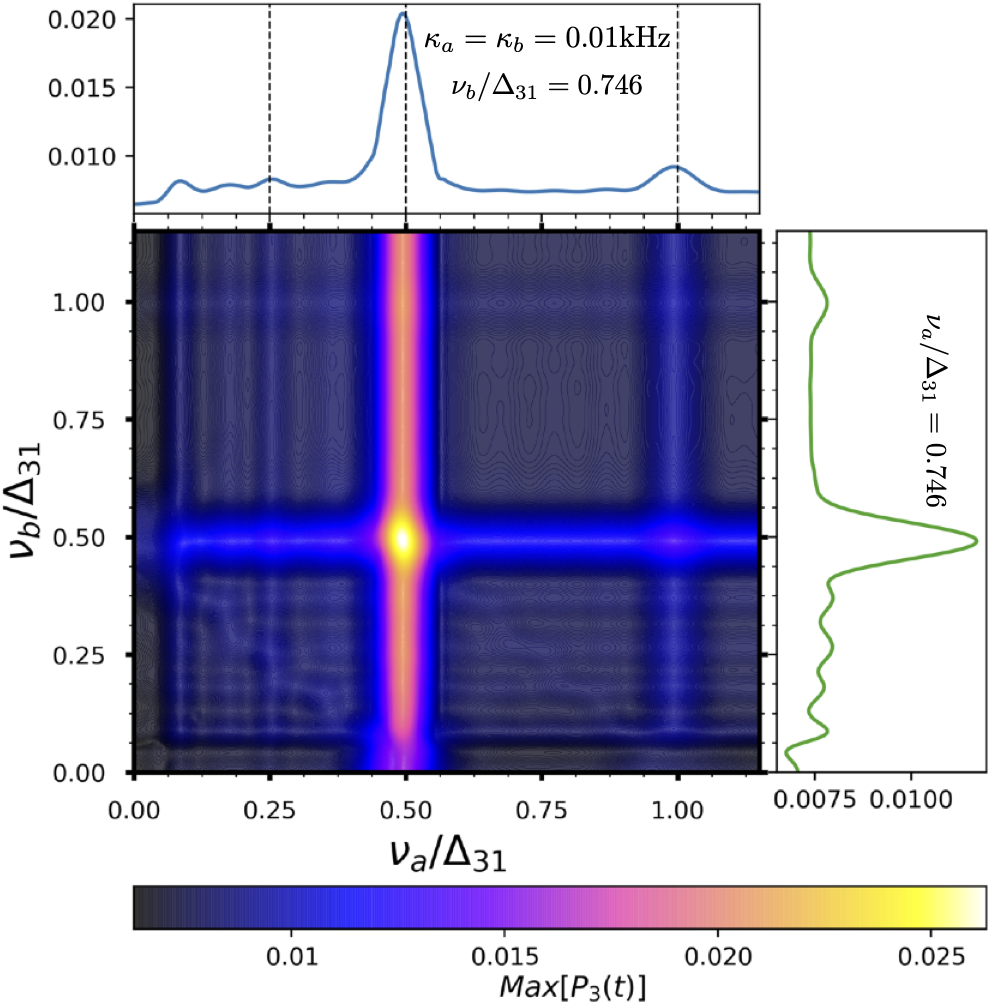} 
\caption{(color online) 2D VAET spectrum for a trimeric chromophore system coupled to two non-interacting vibrations in the weak site-vibration coupling regime, $\kappa_a=\kappa_b=0.01$kHz, with dissipative parameters $\gamma_1=\gamma_2=\gamma_3=0.001$kHz. Here the truncation number of each vibrational Fock space is $N=10$ and other parameters are same as in Fig.~\ref{fig:maxPop_N15_kappa_0d01_kBT_1d5_1d5_amplify}. }
\label{fig:maxPop_N10_gamma_0d001}
\end{figure}

\begin{figure}
\centering
  \includegraphics[width=1.\columnwidth]{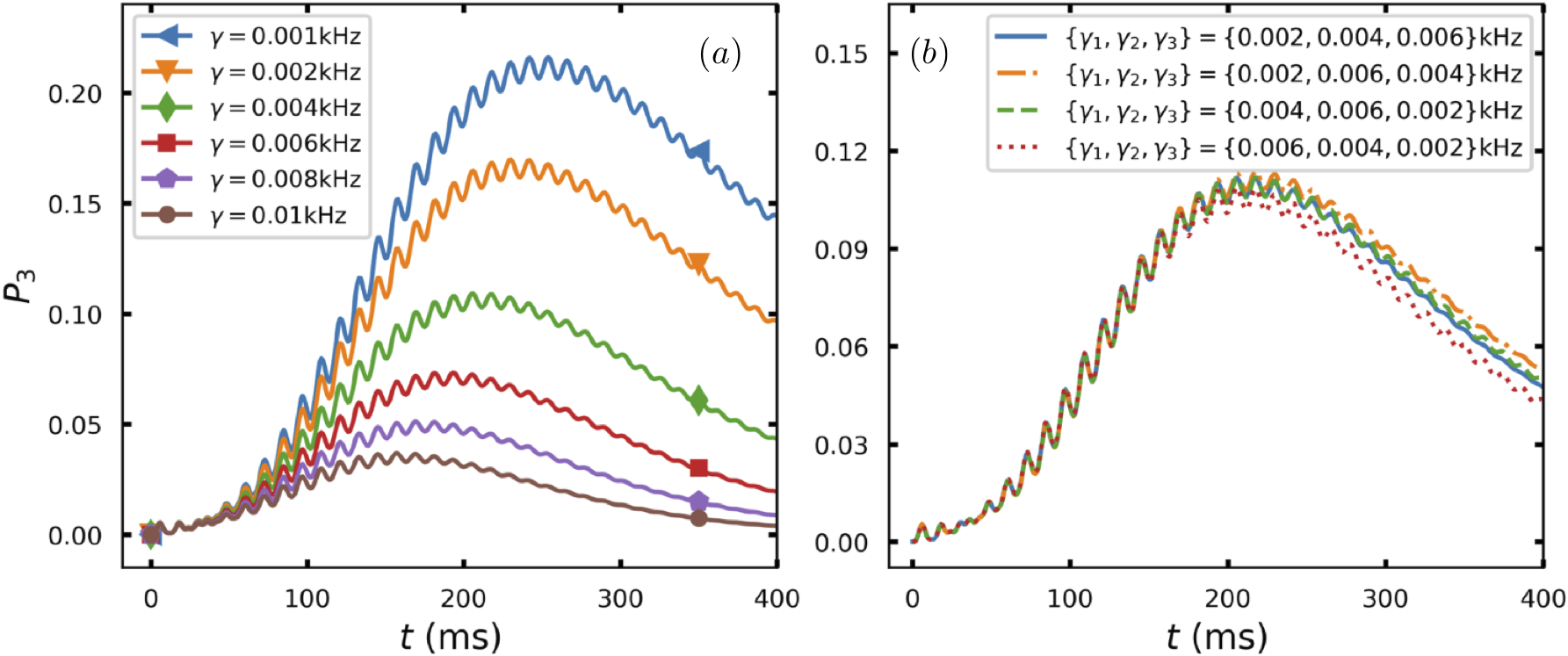}
\caption{(color online) Energy transfer probability $P_3(t)$ as a function of time for (a) identical ($\gamma_1=\gamma_2=\gamma_3=\gamma$) and (b) different values of site dissipation rates. Here we take the site-vibrational coupling $\kappa_a=\kappa_b=0.01$kHz, the vibrational frequency $\nu_a/\Delta_{31}=\nu_b/\Delta_{31}=0.5$, and the temperature $k_BT_a=k_BT_b=1.5$kHz. All other parameters are the same as in Fig.~\ref{fig:maxPop_N15_kappa_0d01_kBT_1d5_1d5_amplify}.}
\label{fig:dissipation}
\end{figure}

\subsection{Dissipative effects}

In contrast to the assistance provided by undamped vibrations for excitation energy transfer discussed above, the decay of an electronic excitation at each site, resulting from e.g., spontaneous emission or coupling to damped vibrational environments, is expected to suppress energy transfer processes. Here we study the effect of such relaxation processes, using a non-Hermitian approach that in the single excitation subspace is equivalent to use of the Lindblad master equation with a relaxation operator~\cite{Bertlmann06pra} (see detailed analysis in the Supplementary Material~\cite{suppl}).

Including a non-Hermitian Hamiltonian term, the system under such dissipation is described by $H_{\rm eff}=H-\frac{i}{2}\sum_{j=1}^3 \gamma_j |e_j\rangle\langle e_j|$ where $H=H_{\rm s}+H_{\rm v}+H_{\rm int}$ given by Eq.~(\ref{eq:totalH}). 
The effective Hamiltonian in the single electronic excitation subspace is then obtained as $\tilde{H}_{\rm eff}=\Xi H_{\rm eff}\Xi=\tilde{H}-\frac{i}{2}(\gamma_1|1\rangle\langle 1| +\gamma_2|2\rangle\langle 2| +\gamma_3|3\rangle\langle 3|)$, where $\tilde{H}$ is given by Eq.~(\ref{eq:effec_trimerH})~\cite{suppl}. 
The average effect of dissipation on the excitation energy transfer is then obtained by repeating the 2D VAET spectral calculations with $\tilde{H}$ replaced by $\tilde{H}_{\rm eff}$. 

Fig.~\ref{fig:maxPop_N10_gamma_0d001} shows the 2D VAET spectrum with dissipation given by parameters $\gamma_1=\gamma_2=\gamma_3=\gamma=0.001$kHz.  We see suppression of all energy transfer processes, particularly those along the anti-diagonal and diagonal lines, relative to the no dissipation results in 
Fig.~\ref{fig:maxPop_N15_kappa_0d01_kBT_1d5_1d5_amplify}(a).
As expected, the single-mode two-phonon VAET is the most pronounced VAET process in Fig.~\ref{fig:maxPop_N10_gamma_0d001}. 

Fig.~\ref{fig:dissipation}(a) shows that for a specific VAET transfer process, e.g., the single mode TPhonA VAET at the resonant position  $\nu_a/\Delta_{31}=\nu_b/\Delta{31}=0.5$, the time-dependent probability of finding an excitation at the acceptor, $P_3(t)$, is increasingly suppressed for all $t$ as $\gamma$ increases. 
To analyze which sites contribute to this suppression, Fig.~\ref{fig:dissipation}(b) shows calculations with different dissipative parameters $\gamma_i$ at each site.  Only small variations are seen, within the general trend that a strong dissipation at the donor site provides the greatest suppression (red dotted line), followed by having the strongest dissipation at the acceptor site (blue solid line). Interestingly, when the strongest dissipation is at the bridge site, the energy transport is most robust to the dissipation (yellow and green dashed lines).

\section{VAET signatures in the presence of strong site-vibration coupling \label{sec:strong_kappa}}

\begin{figure}
\centering
  \includegraphics[width=0.8\columnwidth]{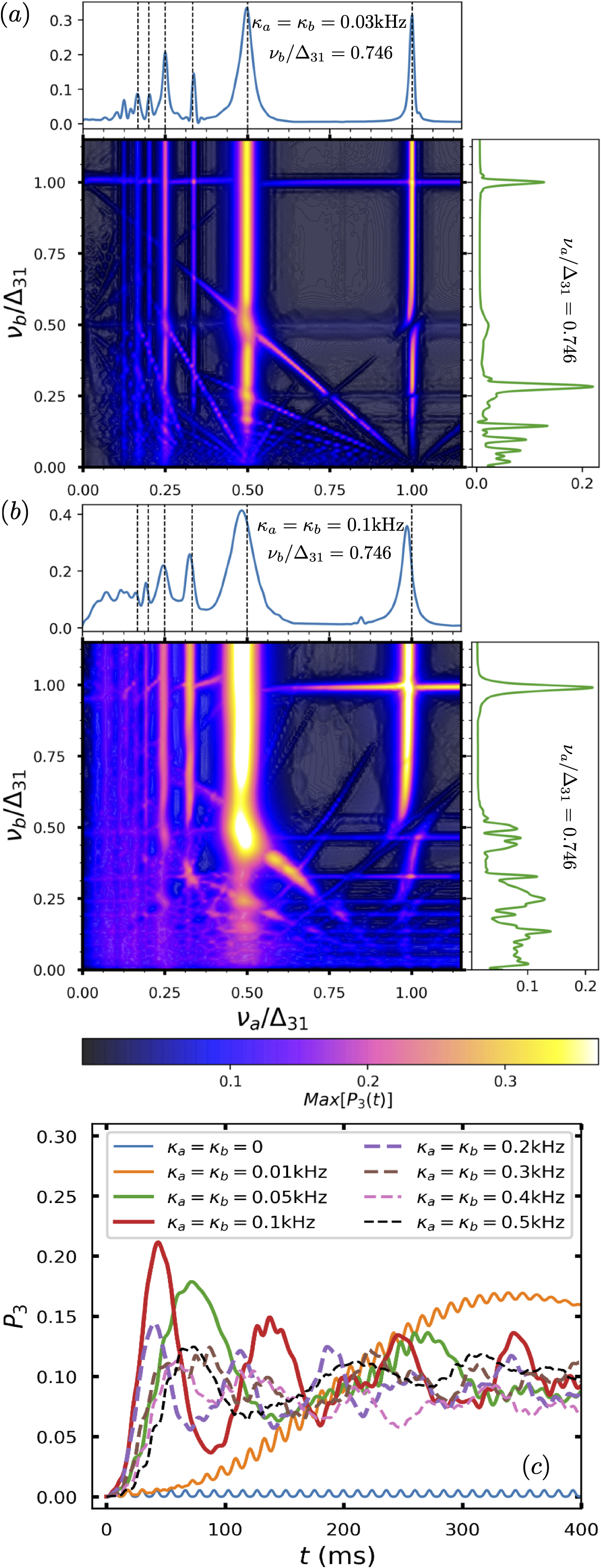} 
\caption{(color online) 2D VAET spectra for the trimeric chromophore system coupled to two non-interacting vibrations in the strong site-vibration coupling regime.  Panels (a) - (b) show the maximum transfer probability ${\rm Max}[P_3(t)]$ in a time period $t\in[0,400]$ms for two values of the site-vibration coupling $\kappa_a=\kappa_b=\kappa$: (a) $\kappa=0.03$kHz and (b) $\kappa=0.1$kHz. The slices on the top and right side of each contour plot are taken at $\nu_b/\Delta_{31}=0.746$ and $\nu_a/\Delta_{31}=0.746$, respectively. (c) Time trace of the energy transfer probability $P_3(t)$ for several combinations of $\kappa_a,\kappa_b$ at the resonance points $(\nu_a/\Delta_{31},\nu_b/\Delta_{31})=(0.5,0.5)$ and $(\nu_a,\nu_b)=(0.52,0.52)$kHz). The vibrational temperatures are  $k_BT_a=k_BT_b=1.5$kHz. All other parameters are the same as in Fig.~\ref{fig:maxPop_N15_kappa_0d01_kBT_1d5_1d5_amplify}.}
\label{fig:maxPop_N15_kBT_1d5_kappa}
\end{figure}

As the site-vibration coupling strength increases, different VAET features emerge and the balance between single- and multi-phonon VAET processes changes.
We explore these changes by considering larger coupling strengths $\kappa_a=\kappa_b=0.03$kHz and $0.1$kHz, summarized in Figs.~\ref{fig:maxPop_N15_kBT_1d5_kappa}(a) and~\ref{fig:maxPop_N15_kBT_1d5_kappa}(b), respectively. The top and right side slices in each of these plots are taken at $\nu_a/\Delta_{31}=0.746$ (right slice) and $\nu_b/\Delta_{31}=0.746$ (top slice). 

In addition to the basic VAET features from Fig.~\ref{fig:maxPop_N15_kappa_0d01_kBT_1d5_1d5_amplify}(a) (where $\kappa_a=\kappa_b=0.01$kHz), 
we now observe additional multiphonon VAET processes in  Fig.~\ref{fig:maxPop_N15_kBT_1d5_kappa}(a) that involve three, five, and six phonons, indicated by vertical lines at $\nu_a/\Delta_{31}=1/3$, $1/5$, and $1/6$, respectively. 
For the larger site-vibration coupling strength $\kappa_a=\kappa_b=0.1$kHz shown Fig.~\ref{fig:maxPop_N15_kBT_1d5_kappa}(b), these vertical lines become more distinct and also start to shift noticeably away from the excitonic resonant transition frequencies. 
We also see that in the strong coupling regime, not only do the one-phonon (e.g., $\nu_a/\Delta_{31}=1$) and two-phonon (e.g., $\nu_a/\Delta_{31}=0.5$) VAET processes become more comparable in intensity, 
but also the impact of the vibration coupled to the acceptor, $\nu_b$, becomes comparable to that of the bridging vibration $\nu_a$.  Thus, we now see local maxima at $\nu_a/\Delta_{31},\nu_b/\Delta_{31}=1$ and  $0.25$, in both the right and top slices of Fig.~\ref{fig:maxPop_N15_kBT_1d5_kappa}(a) and (b).  
The greater structure in these intensity patterns contrasts with the simpler structure obtained for weak site-vibration coupling in Fig.~\ref{fig:maxPop_N15_kappa_0d01_kBT_1d5_1d5_amplify}(a).

To further demonstrate the effects of strong site-vibration coupling on the VAET,  we plot in Fig.~\ref{fig:maxPop_N15_kBT_1d5_kappa}(c) the time evolution of the $P_3(t)$ at the resonance point $(\nu_a/\Delta_{31},\nu_b/\Delta_{31})=(0.5,0.5)$ for a two-phonon VAET process
at various values of the coupling strength $\kappa = \kappa_a=\kappa_b$.
When the electronic sites are decoupled from the vibrations, i.e., $\kappa_a=\kappa_b=0$, the blue reference curve in Fig.~\ref{fig:maxPop_N15_kBT_1d5_kappa}(c) shows Rabi oscillations characterized by the transition frequency $\nu_a=\Omega=\sqrt{\Delta^2+2J^2}=0.52$kHz, with corresponding oscillatory period $2\pi/\nu_a\sim 12$ms (approx. eight cycles in each period of $100$ms). 
When the coupling is nonzero, we observe modulated Rabi-like oscillations that show slow oscillations superimposed on the fast oscillations with the frequency $\Omega$.  See, for example, the solid orange curve in Fig.~\ref{fig:maxPop_N15_kBT_1d5_kappa}(c), for which $\kappa_a=\kappa_b=0.01$kHz.
As $\kappa$ increases to $0.05$kHz (green curve) and beyond to $0.1$kHz (red curve), the initial rise of $P_3(t)$ is faster and the first maximum higher. 
However, further increase of the site-vibration coupling strength beyond $0.1$kHz reverses this trend. In the next section we shall see that this is a result of the formation of vibronic states with strong mixing of excitonic and vibrational degrees of freedom, giving rise to very different transition  frequencies. 

\begin{figure}
\centering
  \includegraphics[width=.9\columnwidth]{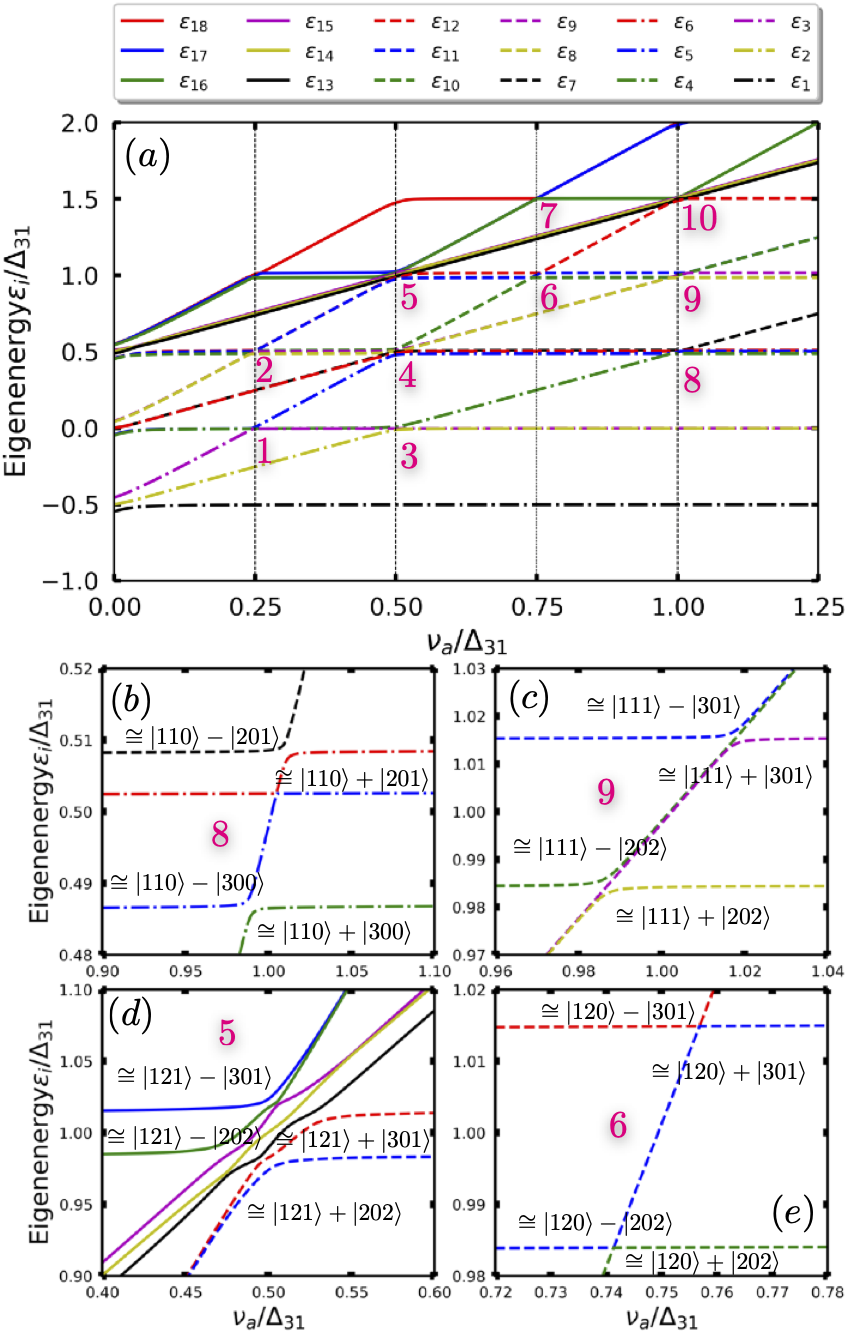} 
\caption{(color online) Vibronic energy spectrum of the effective excitonic levels $|e_i\rangle$ coupled to two vibrational modes $|j\rangle$ and $|k\rangle$ with $N=3$ levels each, as a function of $\nu_a/\Delta_{31}$. Here only eighteen of the twenty seven lowest energy levels are shown. Panels (b)-(k) show enlarged views of each avoided crossing in panel (a). The parameters are $\{\tilde{\omega}_1,\tilde{\omega}_2,\tilde{\omega}_3,J_{12},J_{23}\}=\{-0.5,0,0.5,0.1,0.1\}$kHz and $\nu_b=\Delta_{21}=\Delta_{32}=0.52$kHz and $\kappa_a=\kappa_b=0.03$kHz.}
\label{fig:vibronicSpectrum}
\end{figure}

\section{Vibronic spectral analysis of VAET and role of cross-coupling terms in the effective Hamiltonian}
\label{sec:others}

\subsection{Vibronic states}

In order to better understand the origin of the VAET features, 
we have calculated the energy spectrum for the trimer excitonic system coupled to the two vibrational modes including three vibrational levels [Eq.~(\ref{eq:effec_trimerH})] as a function of the scaled frequencies $(\nu_a/\Delta_{31},\nu_b/\Delta_{31})$, for specific values of the coupling strengths. This reveals the energies of the vibronic states formed as a consequence of the two exciton-vibration couplings. Fig.~\ref{fig:vibronicSpectrum} shows the corresponding two dimensional vibronic spectrum for the case of coupling strengths $\kappa_a=\kappa_b=0.03$kHz. 
The figure clearly shows the presence of avoided crossings that derive from the exciton-vibration coupling.  For example,
along the horizontal line $\nu_b/\Delta_{31}=0.5$, 
whenever $\nu_a/\Delta_{31}$ approaches a resonant transition frequency of the trimer excitonic system (i.e., zero detuning at $\nu_a/\Delta_{31}=0.25,0.5,1$ as shown in Fig.~\ref{fig:maxPop_N15_kappa_0d01_kBT_1d5_1d5_amplify}), gives rise to an avoided crossing due to the site-vibration coupling.
Each avoided crossing in the spectrum shown in Fig.~\ref{fig:vibronicSpectrum}
indicates a vibronic state, i.e., a mixing of the electronic and vibrational degrees of freedom~\cite{WangAllodiEngel19NatRevChem}. 

Some of the avoided crossings in the vibronic energy spectrum correspond to VAET features discussed above.
A perturbative analysis of the vibronic energies predicts the presence of avoided crossings at the degenerate states.
Thus the specific avoided crossings magnified in Figs.~\ref{fig:vibronicSpectrum}(b) and (c) indicate the hybridized vibronic states ($\cong|110\rangle\pm|300\rangle$ and $|111\rangle\pm|301\rangle$  ($i$, $j$, $k$ in $|ijk\rangle$ represent the excitonic state $|e_i\rangle$, and vibrational occupation states $|j\rangle$ and $|k\rangle$, respectively) 
that give rise to the one-phonon VAET feature along the vertical line $\nu_a/\Delta_{31}=1$) in Fig.~\ref{fig:maxPop_N15_kappa_0d01_kBT_1d5_1d5_amplify}(a). 
Similarly, the states $\cong|121\rangle\pm|301\rangle$ at the avoided crossing in Fig.~\ref{fig:vibronicSpectrum}(d) are associated with the single-mode two-phonon VAET indicated by the vertical line, i.e., $\nu_a/\Delta_{31}=0.5$, in Fig.~\ref{fig:maxPop_N15_kappa_0d01_kBT_1d5_1d5_amplify}.
We also see vibronic states associated with the cooperative VAET features. 
In Fig.~\ref{fig:vibronicSpectrum}(e), the avoided crossing of states $\cong|120\rangle\pm|301\rangle$ corresponds to an intersection of the horizontal line $\nu_b/\Delta_{31}=0.5$ and the off-diagonal line $2\nu_a-\nu_b=\Delta_{31}$ in Fig.~\ref{fig:maxPop_N15_kappa_0d01_kBT_1d5_1d5_amplify} that indicates the double-mode cooperative VAET. %

We note that the avoided crossings in the vibronic spectrum become more pronounced as the coupling strength $\kappa_a$ or $\kappa_b$ increases, consistent with the perturbative analysis. This means that not only does the gap between the two adjacent levels increase, but also the shift from the excitonic resonant transition frequencies (e.g., $\nu_a/\Delta_{31}=1$, $0.5$, $0.25$) will be larger. This trend is also visible in the cross-sectional slices in Fig.~\ref{fig:maxPop_N15_kBT_1d5_kappa}(a) and (b). 
Consequently, for given frequencies $\nu_a, \nu_b$, increasing either $\kappa_a$ or $\kappa_b$ to values so large that they are comparable with the excitonic energy differences will be expected to suppress energy transfer processes below values seen for smaller coupling. Indeed, this is consistent with the decrease in $P_3(t)$ seen for large $\kappa_a=\kappa_b$ values in Fig.~\ref{fig:maxPop_N15_kBT_1d5_kappa}(c).

\subsection{Effect of cross couplings in single excitation subspace}

The VAET features presented above are based on the consideration of the effective Hamiltonian Eq.~(\ref{eq:effec_trimerH}) derived as the single electronic excitation restriction of the model in Eq.~(\ref{eq:totalH}) for the trimeric chromophore system.
This  trimeric model, generalized from an experimentally investigated dimer for an artifical excitonic system realized in a trapped ion system~\cite{Gorman18prx}, contains interaction of the vibrations with both excited and the ground states, i.e., $\kappa_a \sigma_z^{(2)}(a^{\dagger}+a)
 + \kappa_b \sigma_z^{(3)}(b^{\dagger}+b)$. 
We saw that the resulting effective model in the single electronic excitation manifold has cross coupling terms, i.e., an interaction of a vibration with the excited states of unconnected sites.

Here we analyze the effects of the cross coupling terms on the excitation energy transfer.
To isolate the effects resulting from these terms, we rewrite the effective Hamiltonian in Eq.~(\ref{eq:effec_trimerH}) as
\begin{eqnarray} \label{eq:effec_H_zeta}
\bar{H}_{tr}(\zeta) &=& \tilde{\omega}_1 |1\rangle\langle 1| 
+\tilde{\omega}_2 |2\rangle\langle 2|
+\tilde{\omega}_3 |3\rangle\langle 3| \\ \notag
&&+ J_{12}(|1\rangle\langle 2| + |2\rangle\langle 1|)
+ J_{23}(|2\rangle\langle 3| + |3\rangle\langle 2|) \\ \notag
&&+\kappa_a(a^{\dagger}+a)(-\zeta|1\rangle\langle 1|+|2\rangle\langle 2| -\zeta|3\rangle\langle 3|) \\ \notag
&&+\kappa_b(b^{\dagger}+b)(-\zeta|1\rangle\langle 1|-\zeta|2\rangle\langle 2| +|3\rangle\langle 3|) \\ \notag
&&+\nu_a a^{\dagger}a +\nu_b b^{\dagger}b. 
\end{eqnarray}
with variable parameter $\zeta$ which interpolates between  Eq.~(\ref{eq:effec_trimerH}) for $\zeta=1$ and the usual single excitation manifold effective Hamiltonian for molecular excitons without cross coupling terms for $\zeta=0$. 
The latter case corresponds to the full Hamiltonian, Eq.~(\ref{eq:totalH}), with the site-vibration coupling in Eq.~(\ref{eq:H_int}) replaced by $\kappa_a \sigma_+^{(2)}\sigma_-^{(2)}(a^{\dagger}+a)
 + \kappa_b \sigma_+^{(3)}\sigma_-^{(3)}(b^{\dagger}+b)$.

\begin{figure}
\centering
  \includegraphics[width=0.9\columnwidth]{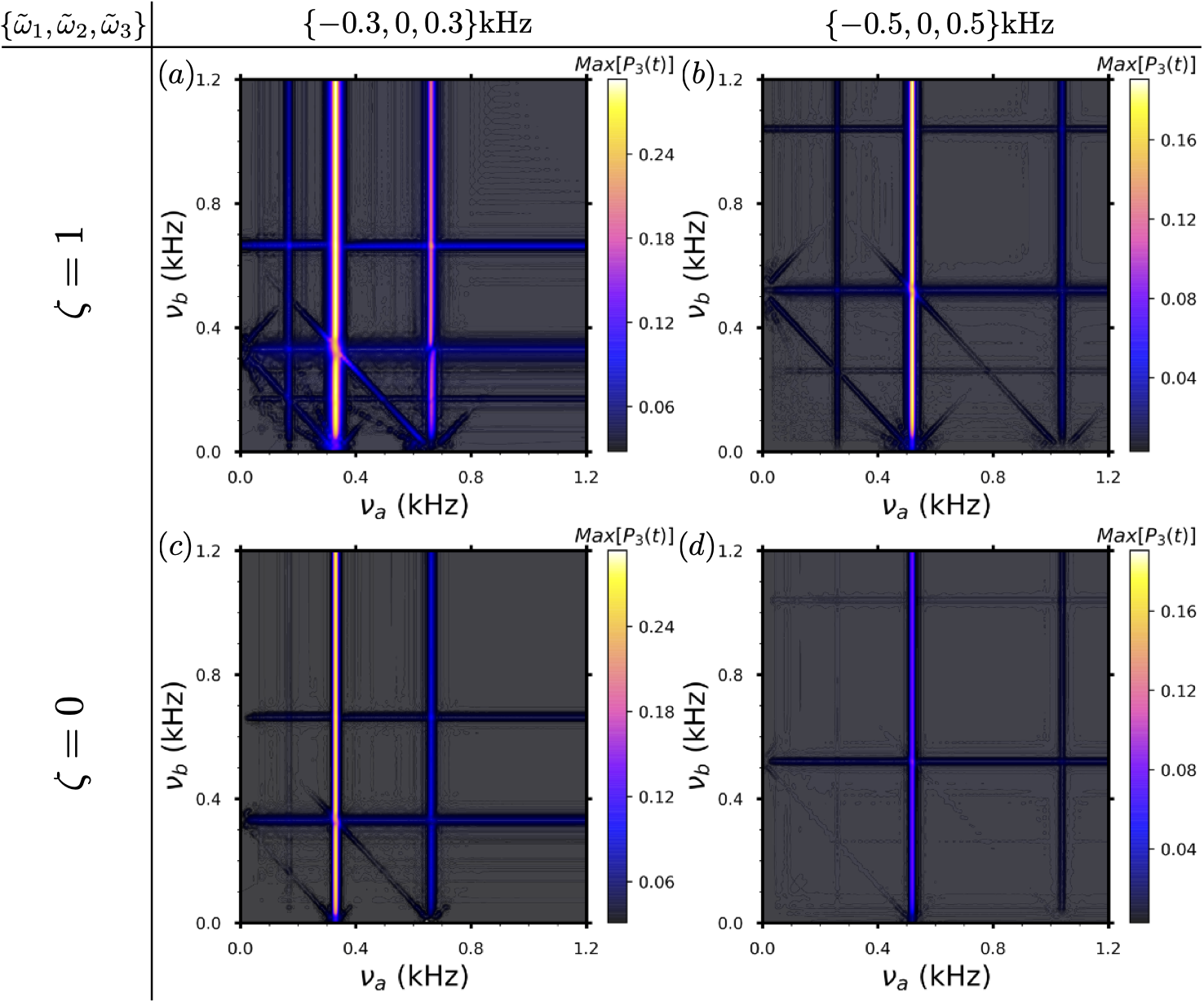} 
\caption{(color online) 
Comparison of 2D VAET spectra of symmetric trimeric chromophore systems described by the effective Hamiltonian $\bar{H}(\zeta)$, Eq.~\eqref{eq:effec_trimerH}
with $\zeta=1$ (upper row) and $\zeta=0$ (lower row). 
The parameter values $\{\tilde{\omega}_1,\tilde{\omega}_2,\tilde{\omega}_3\}$ specified above the panels give excitonic energy differences $\{\Delta_{21},\Delta_{32}\}= \{0.332,0.332\}$kHz in (a), (c) and $\{0.52,0.52\}$kHz in (b), (d)  with $\Delta_{31}=\Delta_{21}+\Delta_{32}$.
The other parameters are $J_{12}=J_{23}=0.1$kHz, $\kappa_a=\kappa_b=0.01$kHz, $k_BT_a=k_BT_b=0.749$kHz, and $N=10$.}
\label{fig:max_popSys3_crossingTerms}
\end{figure}

Fig.~\ref{fig:max_popSys3_crossingTerms} shows the two-dimensional VAET spectra for two symmetric trimeric systems with identical energy gaps (i.e., $\Delta_{21}=\Delta_{32}$) that allow interference VAET to appear. 
Comparison of either the two left panels (a) and (c) with $\Delta_{21}=\Delta_{32}=0.332$kHz, or the two right panels with  $\Delta_{21}=\Delta_{32}=0.52$kHz, shows that the cross coupling terms in $\tilde{H}=\bar{H}_{tr}(\zeta=1)$ significantly enhance the energy transfer. 
For example, the one-phonon VAET at $\nu_b=\Delta_{31}=1.04$kHz, which shows high intensity for $\bar{H}_{tr}(\zeta=1)$ (panel (b)), is considerably less intense $\bar{H}_{tr}(\zeta=0)$ (panel (d)). 

Comparing now the left and right panels of either the upper ($\zeta=1$) or lower ($\zeta=0$) row shows the effect of modifying the energy barrier for both Hamiltonians. Thus the higher probabilities for the two-phonon VAET processes seen in panel (a) are due to the lower excitonic energy barrier $\tilde{\omega}_3-\tilde{\omega}_{2}=\tilde{\omega}_{2}-\tilde{\omega}_{1}=0.3$kHz which is more similar to the excitonic coupling $J=0.1$kHz,  than that of panel (b) for which $\tilde{\omega}_3-\tilde{\omega}_{2}=\tilde{\omega}_{2}-\tilde{\omega}_{1}=0.5$kHz.

\begin{figure}
\centering
  \includegraphics[width=0.9\columnwidth]{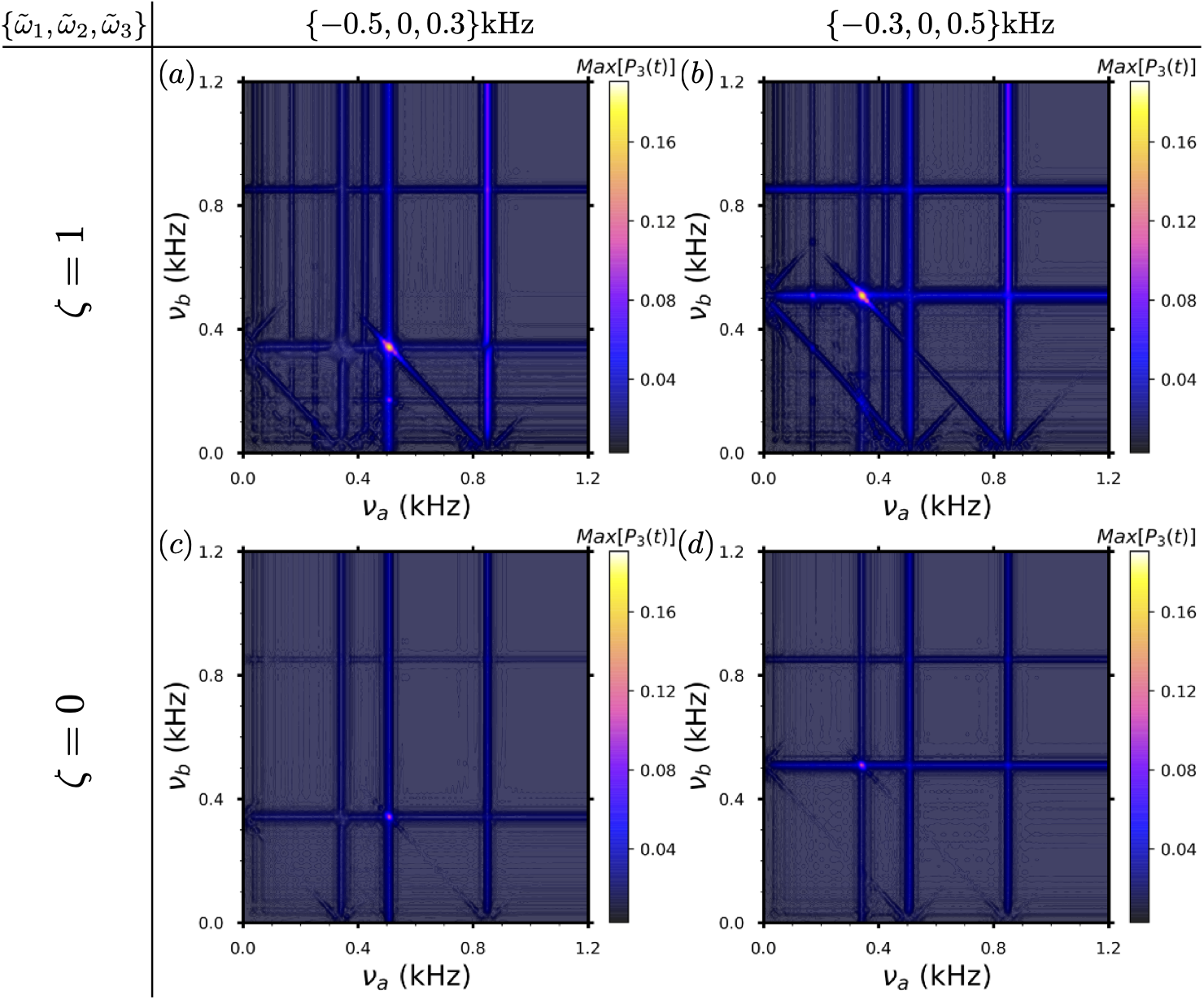} 
\caption{(color online) 
2D VAET spectra of an asymmetric trimeric chromophore system described the effective Hamiltonian $\bar{H}(\zeta)$,  Eq.~\eqref{eq:effec_trimerH} with $\zeta=1$ (upper row) and $\zeta=0$ (lower row).  
The parameter values $\{\tilde{\omega}_1,\tilde{\omega}_2,\tilde{\omega}_3\}$ specified above the panels give excitonic energy differences $\{\Delta_{21},\Delta_{32}\}=\{0.508,0.343\}$ (panels (a), (c)),  $\{0.343,0.508\}$kHz (panels (b), (f)), with $\Delta_{31}=\Delta_{21}+\Delta_{32}$ in all cases.
The other parameters are $J_{12}=J_{23}=0.1$kHz, $\kappa_a=\kappa_b=0.01$kHz, $k_BT_a=k_BT_b=0.749$kHz, and $N=10$. }
\label{fig:max_popSys3_distinguishInterfer}
\end{figure}

\begin{figure}
\centering
  \includegraphics[width=.85\columnwidth]{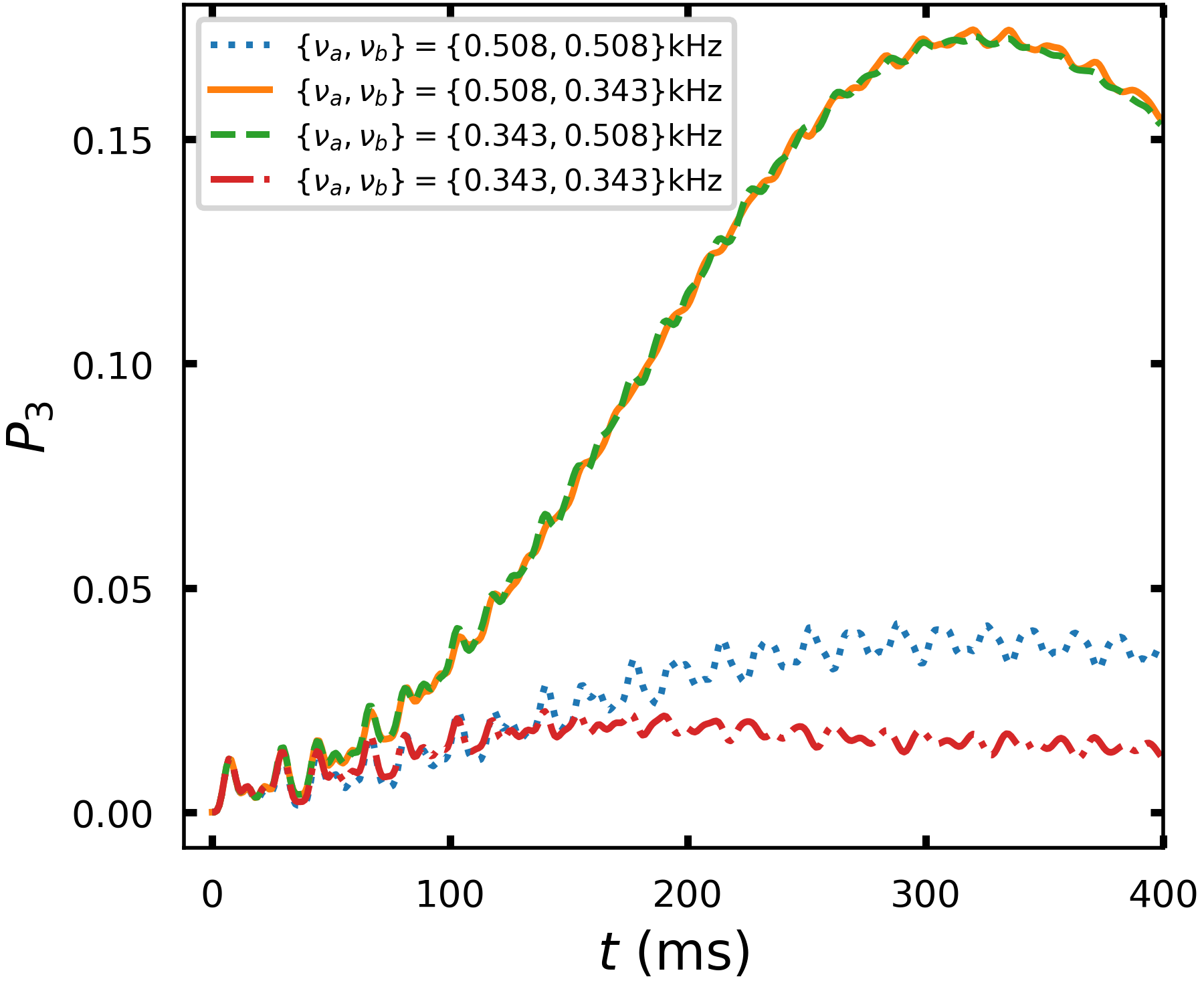}
\caption{(color online) Time evolution of the probability $P_3$ of the acceptor at the asymmetric point $\{\nu_a,\nu_b\}=\{0.508,0.343\}$kHz  in Fig.~\ref{fig:max_popSys3_distinguishInterfer}(a) and  $\{\nu_a,\nu_b\}=\{0.343,0.508\}$kHz in Fig.~\ref{fig:max_popSys3_distinguishInterfer}(b), together with nearby symmetric points $\{\nu_a,\nu_b\}=\{0.508,0.508\}$$kHz, $$\{0.343,0.343\}$kHz. All other parameters are the same as in Fig.~\ref{fig:max_popSys3_distinguishInterfer}.}
\label{fig:nointerference}
\end{figure}

The appearance of an interference VAET requires a specific condition, i.e., $\Delta_{21}=\Delta_{32}$.
To isolate the interference VAET features we therefore present in Fig.~\ref{fig:max_popSys3_distinguishInterfer} 2D VAET spectra for two asymmetric systems that do not host any interferences. 
The upper row of Fig.~\ref{fig:max_popSys3_distinguishInterfer} shows spectra with $\{\tilde{\omega}_1,\tilde{\omega}_2,\tilde{\omega}_3\}$ equal to $\{-0.5,0,0.3\}$kHz (a) and $\{-0.3,0,0.5\}$kHz (b), implying different energy gaps of  $\{\Delta_{21},\Delta_{32}\}=\{0.508,0.343\}$kHz. 
We see that the double-mode two-phonon cooperative VAET, located at the lattice point [$\{\nu_a,\nu_b\}=\{\Delta_{21},\Delta_{32}\}$ in panel (a) and at $\{\nu_a,\nu_b\}=\{\Delta_{32},\Delta_{21}\}$ in panel (b) is now the only dominant process. 
The corresponding time traces are shown in Fig.~\ref{fig:nointerference} where it is evident that they have a maximal values intermediate between those of symmetrically distributed neighboring points, implying an absence of interference at the symmetric points $\nu_a=\nu_b=\Delta_{21}=\Delta_{32}$.
This is in contrast to the 2D VAET spectra for systems with identical energy gaps $\Delta_{21}=\Delta_{32}$ in Fig.~\ref{fig:max_popSys3_crossingTerms}, where 
the interferences at the crossing point of vertical ($\nu_a=0.332$kHz) and horizontal ($\nu_b=0.332$kHz) lines in panel (a) and at the crossing point of $\nu_a=0.52$kHz and $\nu_b=0.52$kHz in panel (b) are clearly visible. The corresponding time traces (not shown), show 
destructive interference along the vertical lines [$\nu_a=0.332$kHz (Fig.10 (a)) and $\nu_a=0.52$kHz (Fig.10 (b)], and constructive interference along the horizontal lines [$\nu_b=0.332$kHz (Fig.10 (a)) and $\nu_b=0.52$kHz (Fig.10 (b)].
We also present the corresponding results for the effective Hamiltonian $\bar{H}_{tr}(\zeta=0)$ in the lower row of Fig.~\ref{fig:max_popSys3_distinguishInterfer}, to emphasize the key role of the cross coupling terms in amplifying these cooperative VAET processes. The interference features are no longer visible here, confirming the critical role of the cross-correlated vibrations in enabling these quantum features.

\begin{figure}
\centering
  \includegraphics[width=.9\columnwidth]{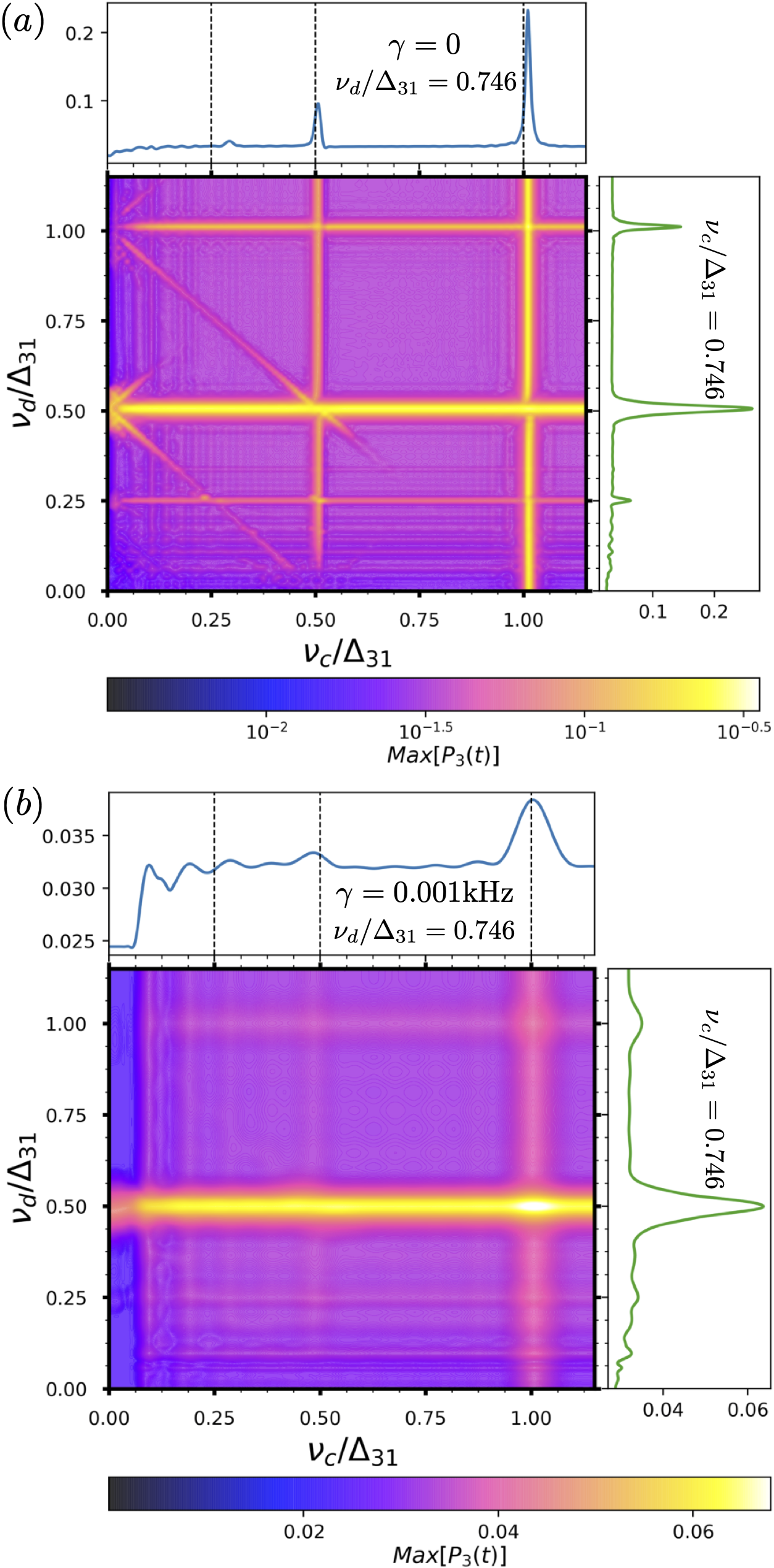}
\caption{(color online) The VAET features of the trimeric chromophore system weakly coupled to correlated and anti-correlated vibrational models described by the effective Hamiltonian Eq.~(\ref{eq:correl_effective}) for the cases of (a) without ($\gamma=0$) and (b) with ($\gamma=0.001$kHz) dissipation. The color bar in (a) is same as in Fig.~\ref{fig:maxPop_N15_kappa_0d01_kBT_1d5_1d5_amplify}(a) for comparison, but note the change in scale of panel (b) relative to panel (a).
The parameters are $\{\tilde{\omega}'_1,\tilde{\omega}'_2,\tilde{\omega}'_3,J'_{12},J'_{23},\kappa'_c,\kappa'_d,k_BT_c,k_BT_d\}=\{-0.5,0,0.5,0.1,0.1,0.01,0.01,1.5,1.5\}$kHz which is same as in Fig.~\ref{fig:maxPop_N15_kappa_0d01_kBT_1d5_1d5_amplify}(a) or Fig.~\ref{fig:dissipation}. }
\label{fig:correVMs}
\end{figure}

\section{VAET with explicitly correlated vibrational modes \label{sec:correl}} 

The energy transfer in the trimeric chromophore system discussed in Sections~\ref{sec:model}-\ref{sec:others} relies on the assistance of two independent vibrational modes that are coupled to the bridge and acceptor sites, respectively [see Eq.~(\ref{eq:H_int})]. 
Experimental realization of this ideal Hamiltonian with the local and independent control of the site-vibration interaction for a trapped-ion quantum simulator requires that the ionic states be coupled to transverse vibrational modes.  This is more challenging than coupling to the longitudinal vibrations, requiring stabilization with regard to the trapping frequency. Coupling to longitudinal modes allows instead access to normal modes that are linear combinations of local vibrations, exemplified for a three-ion system by the following Hamiltonian~\cite{BrozHaffner}: 
\begin{eqnarray} \label{eq:H_lng}
H_{\rm lng} &=& \frac{\omega'_1}{2}\sigma_z^{(1)} +\frac{\omega'_2}{2}\sigma_z^{(2)} +\frac{\omega'_3}{2}\sigma_z^{(3)} \notag\\
&&
+ J'_{12} \sigma_x^{(1)}\sigma_x^{(2)} 
+ J'_{23} \sigma_x^{(2)}\sigma_x^{(3)} 
+ J'_{13} \sigma_x^{(1)}\sigma_x^{(3)} \notag\\
&&
+ \nu_c c^{\dagger}c +\nu_d d^{\dagger}d 
+\kappa'_c (\sigma_z^{(1)} -\sigma_z^{(3)}) (c + c^{\dagger}) \notag\\
&&+\kappa'_d (\sigma_z^{(1)} -2\sigma_z^{(2)} +\sigma_z^{(3)}) (d+d^{\dagger}).
\end{eqnarray}
Note that here we have also included a direct excitonic coupling $J_{13}$ between donor and acceptor sites. 
This Hamiltonian describes the coupling of the trimeric system to the symmetric and asymmetric normal modes of vibration along the longitudinal axis of a linear chain of three ions.
In contrast to the transverse coupling Hamiltonian of Eq.~(\ref{eq:totalH}), we now have one vibrational mode, $\nu_c$ (the symmetric stretch) that shows anti-correlated coupling to the donor and the acceptor and a second vibrational mode $\nu_d$ (the asymmetric stretch) that shows a more complex correlated coupling with all three sites. 
Anti-correlated vibrations have been claimed to drive non-adiabatic electronic energy transfer in photosynthetic light-harvesting systems~\cite{Tiwari12pnas}. 
It is thus of interest to analyze the possibility of VAET processes in such a Hamiltonian possessing both correlated and anti-correlated site-vibration couplings.

Following the procedure outlined in Section~\ref{sec:model} to project the Hamiltonian onto the single electronic excitation subspace, we obtain the following effective Hamiltonian:
\begin{eqnarray}
\tilde{H}_{\rm lng} &=& \frac{\tilde{\omega}'_1}{2} |1\rangle\langle 1| 
+ \frac{\tilde{\omega}'_2}{2} |2\rangle\langle 2| 
+ \frac{\tilde{\omega}'_3}{2} |3\rangle\langle 3| 
+ J'_{12}(|1\rangle\langle 2| + |2\rangle\langle 1|) \notag\\
&&
+ J'_{23}(|2\rangle\langle 3| + |3\rangle\langle 2|)
+ J'_{13}(|1\rangle\langle 3| + |3\rangle\langle 1|) \notag\\
&&
+ \nu_c c^{\dagger}c +\nu_d d^{\dagger}d 
+ 2\kappa'_c (c + c^{\dagger}) (|1\rangle\langle 1|-|3\rangle\langle 3|) \notag\\
&&
+ 2 \kappa'_d (d+d^{\dagger}) (|1\rangle\langle 1| - 2|2\rangle\langle 2| +|3\rangle\langle 3|) ,
\label{eq:correl_effective}
\end{eqnarray}
with $\tilde{\omega}'_1 = \omega'_1 -\omega'_2 - \omega'_3$, $\tilde{\omega}'_2 = \omega'_2 -\omega'_1 - \omega'_3$, $\tilde{\omega}'_3 = \omega'_3 -\omega'_1 - \omega'_2$.
We then perform the numerical simulations to evaluate the maximum energy transfer probability ${\rm Max}[P_3(t)]$ for the effective model 
$\tilde{H}'_{\rm eff}=\tilde{H}_{\rm lng}-\frac{i\gamma}{2}(|1\rangle\langle 1| +|2\rangle\langle 2| +|3\rangle\langle 3|)$, where we also include the effects of dissipation via non-Hermitian decay of chromophore excitations with parameter $\gamma$.  
The resulting 2D VAET spectra for the dynamics with and without dissipation in the weak coupling regime are presented in Fig.~\ref{fig:correVMs}(a) and (b), respectively. 

Fig.~\ref{fig:correVMs}(a) shows that this system with explicitly correlated  couplings to the longitudinal vibrational modes generates a very different relative impact of the symmetric stretch $\nu_c$ and the asymmetric stretch vibration $\nu_d$ on the VAET spectra from that seen for the local transverse couplings in Sections~\ref{sec:model}-\ref{sec:others}. This is evident both in the ratios of single mode two-phonon to one-phonon VAET processes (horizontal and vertical lines), 
and in the variable intensities of the two-mode two-phonon VAET processes (antidiagonal lines). 
As we explain in detail below, this different impact reflects the fact that with the longitudinal modes, the asymmetric stretch $\nu_d$ couples more strongly to the bridge site than to the donor and acceptor sites, while the symmetric stretch $\nu_c$ couples only to the latter. This is true both in the full Hamiltonian $\tilde H_{\rm lng}$ and in the corresponding effective Hamiltonian, Eq.~\eqref{eq:correl_effective}.

The single mode one-phonon VAET line for the asymmetric stretch (vertical line $\nu_c/\Delta_{31} = 1$) is more intense than that for the symmetric 
stretch (horizontal line $\nu_d/\Delta_{31} = 1$).
The two-phonon lines $\nu_d/\Delta_{31}=0.5$ and  $\nu_c/\Delta_{31} = 0.5$ show an even greater disparity, to the extent that the  two-phonon VAET for the symmetric stretch ($\nu_c/\Delta_{31}=0.5$) is not dominant over the corresponding one-phonon process.
The strongest feature in the VAET spectrum is now the single mode two-phonon absorption in $\nu_d$, i.e., the horizontal line $\nu_d/\Delta_{31} = 0.5$.  This parallels the analogous dominance of the single mode two-phonon absorption for the mode coupled to the bridge site in the VAET spectrum of Fig.~\ref{fig:maxPop_N15_kappa_0d01_kBT_1d5_1d5_amplify}(a) (vertical line at $\nu_a/\Delta{31}=0.5$).  
The anomalous observation of the one-phonon symmetric stretch VAET for $\nu_c$ being more intense than the two-phonon process results from the fact that this mode is completely decoupled from 
the excited state of the bridge site in both Eq.~\ref{eq:H_lng} and Eq.~\ref{eq:correl_effective}. This is quite different from not only the interaction of mode $\nu_d$ in Eq.~\eqref{eq:correl_effective}, but also that of modes $\nu_a$ and $\nu_b$ in Eq.~\ref{eq:effec_trimerH}, all of which include some coupling to the excited state of the bridge site in the effective Hamiltonian for the single excitation subspace.

More marked is the behavior in the anti-diagonal lines representing multi-mode VAET (corresponding to TPhonA in Fig.~\ref{fig:maxPop_N15_kappa_0d01_kBT_1d5_1d5_amplify}(b)).   For example, 
the anti-diagonal line $\nu_c/\Delta_{31}+\nu_d/\Delta_{31}=1$ shows significant intensity in the sector $\nu_d > \nu_c$ but negligible intensity in the sector $\nu_d < \nu_c$.  
When dissipation is taken into account, Fig.~\ref{fig:correVMs}(b) shows that all  transfer processes 
are suppressed, similar to what is seen for the uncorrelated trimeric system above (see Fig.~\ref{fig:maxPop_N10_gamma_0d001}).

\section{Discussion and conclusions \label{sec:diss}} 

We have systematically studied the phenomenon of vibrationally assisted energy transfer, VAET, in a donor-bridge-acceptor trimeric chromophore system coupled to two vibrations over a range of coupling strengths. In this work we focused on two types of systems. The first derives from a Hamiltonian with uncorrelated local coupling, as would be obtained by coupling to transverse modes in a trapped ion quantum emulator. The second derives from explicitly correlated non-local coupling, as would be obtained by coupling to normal modes of longitudinal motion in a trapped ion quantum emulator. 
The parameters considered in this work are within the regime of current trapped-ion experiments~\cite{Gorman18prx} and in all cases we considered a parameter set ensuring energetically uphill transitions from both donor to bridge chromophores and bridge to acceptor chromophores.

In the case of local site-vibration couplings, we found a rich array of VAET phenomena going beyond the one-phonon VAET observed previously with a trapped ion quantum emulator~\cite{Gorman18prx}. In particular, we also find clearly resolved signatures of two- and even four-phonon absorption processes in the 2D VAET spectrum at weak site-vibration coupling strength, while
increasing the coupling strength introduces up to to six-phonon VAET processes.
The two-phonon VAET processes constitute a phononic analogue of the well-known two-photon absorption~\cite{Mayer1931} and we refer to them as TPhonA.  They are found to be dominant for all coupling strengths, although the relative contributions of both one- and greater than two-phonon VAET processes do increase with coupling strength,  
gaining intensity from off-resonant contributions in the strong coupling regime.
At all values of coupling strength, we find that for every VAET process the vibration coupled to the bridge has a significantly stronger impact than the terminal vibration, consistent with its central spatial location for energy transport across the chain of sites.

We also found that the two vibrations can give rise to multi-mode VAET processes in which they behave collectively, specifically via cooperation and interference that enhance the efficiency of energy transfer relative to that obtained from VAET with a single vibrational mode. 
This includes cooperative TPhonA in which the two phonons derive from different modes, possibly with different frequencies. We also observe an interesting phenomenon that is formally related to the reverse of this, namely processes in which a phonon from one vibrational mode  simultaneously excites both the excitonic states of the trimeric chromophore system and the other vibrational mode. We term this process ``hetero-excitation".
A vibronic spectral analysis of the VAET features allowed detailed assignment and rationalization of the spectra, revealing the constructive effects of cross coupling terms in our derived effective model. Detailed analysis of transfer processes showing quantum interference was arrived at by considering a generalized asymmetric trimeric analog of the symmetric model for which the bulk of the numerical calculations were made. 

The collective VAET features were found to be reduced but not completely suppressed by dissipative effects.  We showed that they can however be enhanced by raising the temperature of the vibrational modes, as well as by increasing the strength of the site-vibration coupling.  

In the case of explicitly correlated non-local site-vibration couplings, as would be obtained by coupling ions to longitudinal modes, we found generically similar VAET features but with quite different relative strengths. The most important parameter determining the integrated strength of excitonic energy transfer was seen to be the vibrational coupling to the bridge site of the chromophore system.

Our projection of the full Hamiltonian onto a single excitation subspace generates an effective Hamiltonian with induced cross-correlations in the effective site-vibration coupling that   
can  be mapped onto excitonic energy transport for molecular chromophores coupled to correlated vibrational modes.
The richness of the VAET spectra found here raise the intriguing question as to whether some of these VAET processes may be operating in natural systems.  
In particular, the results show the important role played by resonant vibrations in enhancing uphill energy transport. Such modes provide sharp features in the 2D VAET spectra, indicating a significant enhancement of the generic quantum ratcheting of energy transport that is derived from coherent coupling to a quantum vibrational bath~\cite{Hoyer2012spatial}. 
As illustrated in Table~\ref{table:pars_scaleup}, the parameters considered in our study are scaled versions of parameters found in natural photosynthetic systems.
This motivates further analysis of whether examples of the more complex VAET phenomena such as two-phonon absorption and heteroexcitations are present in any  natural systems.

Finally, we emphasize that the cooperative behavior of multiple vibrations seen in this work act not only to enhance the excitation energy transfer but also demonstrate a rich set of VAET phenomena. 
Following the experimental observation of single-mode one-phonon VAET for a dimeric system in a trapped-ion quantum emulator~\cite{Gorman18prx}, generalization of such experiments to three and more ions~\cite{MaierBlattRoos19prl}, as well as to other emulation platforms~\cite{PotocnikWallraff18ncomm} appears feasible.  We look forward to experimental verification of the predictions of VAET signatures for two-phonon absorption and for heteroexcitations in emulations of a trimeric chromophore system.

\begin{acknowledgements}
We thank Joseph Broz and Hartmut H\"{a}ffner for helpful discussions. 

Work at the University of California, Berkeley was supported by the U.S. Department of Energy (DOE), Office of Science, Basic Energy Sciences (BES) under Award \# DE-SC0019376. 
Work at Sandia National Laboratories was supported by the U.S. Department of Energy, Office of Science, Basic Energy Sciences, Chemical Sciences, Geosciences, and Biosciences Division.

Sandia National Laboratories is a multimission laboratory managed and operated by NTESS, LLC., a wholly owned subsidiary of Honeywell International, Inc., for the U.S. DOE's NNSA under contract DE-NA-0003525. This paper describes objective technical results and analysis. Any subjective views or opinions that might be expressed in the paper do not necessarily represent the views of the U.S. Department of Energy or the United States Government.
\end{acknowledgements}

\appendix

\section{Key results of perturbative analysis of energy transfer \label{app:coeff}}

Here we summarize some key results from the analytical perturbative calculations provided in the Supplementary Material~\onlinecite{suppl} that are used in the main text. In the interaction picture, the coefficients $A_{jk}, B_{jk}$  
that determine the perturbative factors discussed in Section~\ref{sec:weak_kappa} are given explicitly as functions of $J$, $\Delta$ and $\Omega$ as
\begin{eqnarray}
A_{12}&=&A_{21}=\frac{2\Delta J(\Delta+\Omega)}{\Omega^2\sqrt{2[J^2+\Delta(\Delta+\Omega)]}}, \label{eq:A_12}\\ 
A_{23}&=&A_{32}=\frac{2\Delta J(\Delta-\Omega)}{\Omega^2\sqrt{2[J^2+\Delta(\Delta-\Omega)]}}, \label{eq:A_23}\\ 
B_{12}&=&B_{21}=\frac{2J^3}{\Omega^2\sqrt{2[J^2+\Delta(\Delta+\Omega)]}}, \label{eq:B_12}\\ 
B_{23}&=&B_{32}=\frac{2J^3}{\Omega^2\sqrt{2[J^2+\Delta(\Delta-\Omega)]}}, \label{eq:B_23}\\ 
A_{13}&=&A_{31}=-\frac{2J^2}{\Omega^2}, \label{eq:A_13} \\
B_{13}&=&B_{31}=\frac{J^2}{\Omega^2} , \label{eq:B_13}
\end{eqnarray}
with $\Delta=\tilde{\omega}_2-\tilde{\omega}_1=\tilde{\omega}_3-\tilde{\omega}_2$ and $\Omega=\sqrt{\Delta^2+2J^2}$. 
The energy transfer probability at the acceptor can be written as
$P_3(t)=Tr[U_0^{\dagger}|3\rangle\langle 3|U_0 U_I(t) |1\rangle\langle 1| \rho_a\rho_b U_I^{\dagger}(t)]$, where $U_0=e^{-i(H_0^{(e)} + H_0^{(\nu)})t}$ and the evolution operator in the interaction picture is $U_I(t)=\mathcal{T} e^{-i\int_0^t ds H_I(s)}$, with $\mathcal{T}$ the time-ordering operator.
To calculate the probability $P_3(t)$, we write the donor and acceptor states in the site basis in terms of eigenstates, giving
$|1\rangle=\alpha |e_1\rangle -\beta |e_2\rangle +\gamma|e_3\rangle$ and $|3\rangle=\gamma |e_1\rangle +\beta |e_2\rangle +\alpha |e_3\rangle$, where
$\alpha=\frac{\sqrt{J^2+\Delta(\Delta+\Omega)}}{\sqrt{2}\Omega}$, 
$\beta = \frac{J}{\Omega}$, and
$\gamma = \frac{\sqrt{J^2+\Delta(\Delta-\Omega)}}{\sqrt{2}\Omega}$. 
Note that in the weak site-site coupling limit, i.e.,  $J\ll\Delta$, we have $\alpha\rightarrow1$ and $\beta,\gamma\rightarrow0$  and the excitonic eigenstates $|e_3\rangle$ and $|e_1\rangle$ can then be approximated by $|3\rangle$ and $|1\rangle$, respectively~\cite{suppl}. 

\begin{figure}
\centering
  \includegraphics[width=.99\columnwidth]{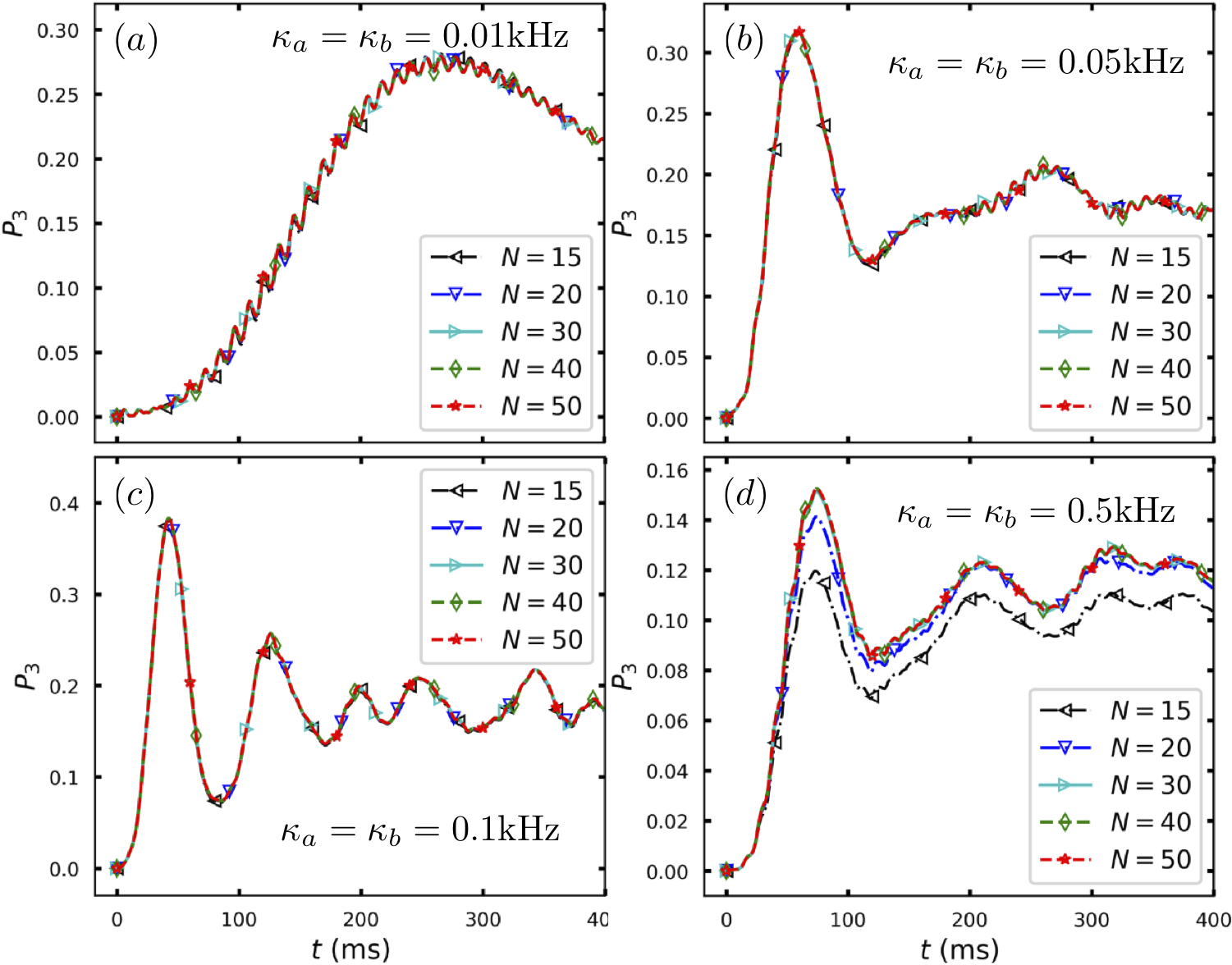} 
\caption{(color online) Convergence of the transfer probability for several values of site-vibration coupling (a) $\kappa_a=\kappa_b=0.01$kHz,  (b) $\kappa_a=\kappa_b=0.05$kHz,  (c) $\kappa_a=\kappa_b=0.1$kHz, and (d)  $\kappa_a=\kappa_b=0.5$kHz. Here we consider $k_BT_a=k_BT_b=1.5$kHz, $\nu_a/\Delta_{31}=\nu_b/\Delta_{31}=0.5$, and other parameters are same as in Fig.~\ref{fig:maxPop_N15_kappa_0d01_kBT_1d5_1d5_amplify}.}
\label{fig:convergence_kBT_1d5}
\end{figure}

\section{Convergence}
\label{app:converge}

As demonstrated in the main text, a large transfer probability can be realized by increasing either the vibrational temperatures $k_BT_a$ and $k_BT_b$, or the site-vibration coupling strengths $\kappa_a,\kappa_b$. However higher values of temperature or site-vibration coupling might cause convergence issues if the truncation number $N$ of each vibrational Fock space is not large enough. 
To address this issue and confirm convergence of the numerical results, we show in Fig.~\ref{fig:convergence_kBT_1d5} the transfer probability for various values of $N$, at vibrational temperatures $k_BT_a=k_BT_b=1.5$kHz for several values of the site-vibration coupling. 
When the coupling strength is weak, e.g., $\kappa_a=\kappa_b=0.01$kHz in Fig.~\ref{fig:convergence_kBT_1d5}(a), and when it is increased to $0.05$kHz and $0.1$kHz as in Fig.~\ref{fig:convergence_kBT_1d5}(b) and (c), respectively, the results are already convergent for $N=15$ . 
This provides evidence for the accuracy of our results in Fig.~\ref{fig:maxPop_N15_kappa_0d01_kBT_1d5_1d5_amplify}(a).
Increasing the coupling strength to an ultra-strong regime~\cite{KockumNori19natrevphys}, e.g., to $\kappa_a=\kappa_b=0.5$kHz, which is comparable to the excitonic transition frequency, requires a value as large as $N=30$ to achieve a good convergence, as shown in Fig.~\ref{fig:convergence_kBT_1d5}(d).
Similarly, convergence of the spectra in the very low vibrational frequency regime (i.e., smaller $\nu_{a(b)}/\Delta_{31}$) might be expected to require a larger value of $N$.  However, the general form of the spectra in this region is already converged at $N=15$, as shown explicitly in the Supplemental Material~\cite{suppl}.

\bibliography{vaet}

\begin{thebibliography}{43}
\bibitem [{Bro()}]{BrozHaffner}%
  \BibitemOpen
  \href@noop {} {}\bibinfo {howpublished} {The three-ion system realizable in
  the current experimental setup is from the communication with Joseph Broz and
  Hartmut H\"{a}ffner.}\BibitemShut {Stop}%
\end{thebibliography}

\pagebreak

\onecolumngrid

\section*{Supporting material for ``Unraveling excitation energy transfer assisted by collective behaviors of vibrations"}

In this supporting material we provide details of our perturbation-based analysis that helps to interpret our findings in the main text. We also present additional numerical results that demonstrate features of vibrationally assisted energy transfer (VAET) in the trimeric chromophore system, including plots of integrated probability, $\int_0^{t_f} P_3(t) dt$, as an alternative measure of the transfer efficiency. 
Finally, we also illustrate the equivalence between the non-Hermitian approach that we employ to study dissipative effects, and a Lindblad master equation approach.

\section*{Perturbation-based analytical results}

In the main text we focus on the symmetric version of the effective model because analytical results for this model can be developed. 
In this section we provide details of the perturbation-based analysis of this model. 

\subsection{The symmetric donor-bridge-acceptor trimeric chromophore model}

Relative to the effective three-level model derived in the main text, its symmetric version under our consideration means identical site-site couplings and effective energy gaps, i.e., $J_{12}=J_{23} = J$ and $\tilde{\omega}_{3}-\tilde{\omega}_2 = \tilde{\omega}_2 - \tilde{\omega}_1=\Delta$. 
This constraint on the effective level energies also implies $\omega_3-\omega_2 = \omega_2-\omega_1 =\Delta/2$, namely, identical difference between frequencies of two adjacent sites in the original model. 
Under these conditions, the effective Hamiltonian given by Eq.~(5) in the main text becomes 
\begin{eqnarray}
H&=&H_0+H_I, \label{eq:H_symm}
\end{eqnarray}
where the unperturbed part is
\begin{eqnarray}
H_0&=&H_0^{(e)}+H_0^{(v)}, \label{eq:free_H_symm} \\ 
H_0^{(e)}&=&-\Delta |1\rangle\langle 1| 
+\Delta|3\rangle\langle 3| 
+ J(|1\rangle\langle 2| +|2\rangle\langle 1|) + J(|2\rangle\langle 3| +|3\rangle\langle 2|) , \label{eq:H_elec}\\
H_0^{(v)}&=&\nu_a a^{\dagger}a +\nu_b b^{\dagger}b,
\end{eqnarray}
and the interaction Hamiltonian is given by
\begin{eqnarray}
H_I&=&\kappa_a (a^{\dagger}+a) (|2\rangle\langle 2| -|1\rangle\langle 1|-|3\rangle\langle 3| ) 
+\kappa_b (b^{\dagger}+b) (|3\rangle\langle 3| -|1\rangle\langle 1|-|2\rangle\langle 2|  ).
\end{eqnarray}

In an interaction picture with respect to the unperturbed Hamiltonian $H_0$ given by Eq.~(\ref{eq:free_H_symm}), the whole Hamiltonian in Eq.~(\ref{eq:H_symm}) becomes 
\begin{eqnarray}
H_I(t)&=&
U_0^{\dagger}H_IU_0
=\kappa_a (a^{\dagger} e^{i\nu_a t}+a e^{-i\nu_a t}) \xi(t) 
+\kappa_b (b^{\dagger} e^{i\nu_b t}+b e^{-i\nu_b t}) \chi(t) , 
\end{eqnarray}
where $U_0^{\dagger}=e^{iH_0t}$ and 
\begin{eqnarray}
\xi(t) &=& e^{iH_0^{(e)}t} (|2\rangle\langle 2|-|1\rangle\langle 1|-|3\rangle\langle 3|) e^{-iH_0^{(e)}t}, \label{eq:xi_def} \\
\chi(t)&=& e^{iH_0^{(e)}t} (|3\rangle\langle 3|-|1\rangle\langle 1|-|2\rangle\langle 2|) e^{-iH_0^{(e)}t} . \label{eq:zeta_def}
\end{eqnarray}
An alternative way of deriving $\xi(t)$ and $\chi(t)$, in addition to the Baker-Campbell-Hausdorff (BCH) formula, is to exploit the operator decomposition in the eigenstate basis $\{|e_j\rangle\}$, e.g., 
\begin{eqnarray}
&&e^{iH_0^{(e)}t} |i\rangle\langle i| e^{-iH_0^{(e)}t} 
= \sum_{j,k} e^{i(\lambda_j - \lambda_k) t} \underbrace{\langle e_j | i\rangle\langle i| e_k\rangle} | e_j\rangle\langle e_k| ,
\end{eqnarray}
where eigenstate states $|e_j\rangle$ satisfy $H_0^{(e)}|e_j\rangle = \lambda_j|e_j\rangle$, and 
\begin{eqnarray}
H_0^{(e)} &=& \sum_{j=1}^3\lambda_j |e_j\rangle\langle e_j| .
\end{eqnarray}
The eigenstates $|e_j\rangle$ are given explicitly by 
\begin{eqnarray}
|e_1\rangle &=& \frac{\sqrt{J^2+\Delta(\Delta+\Omega)}}{\sqrt{2}\Omega} |1\rangle
-\frac{J(\Delta+\Omega)}{\sqrt{2}\Omega\sqrt{J^2+\Delta(\Delta+\Omega)}} |2\rangle 
+\frac{J^2}{\sqrt{2}\Omega\sqrt{J^2+\Delta(\Delta+\Omega)}} |3\rangle , \label{eq:eigenstate1}\\
|e_2\rangle &=& -\frac{J}{\Omega} |1\rangle - \frac{\Delta}{\Omega} |2\rangle +\frac{J}{\Omega} |3\rangle, \label{eq:eigenstate2}\\
|e_3\rangle &=& \frac{\sqrt{J^2+\Delta(\Delta-\Omega)}}{\sqrt{2}\Omega} |1\rangle
-\frac{J(\Delta-\Omega)}{\sqrt{2}\Omega\sqrt{J^2+\Delta(\Delta-\Omega)}} |2\rangle 
+\frac{J^2}{\sqrt{2}\Omega\sqrt{J^2+\Delta(\Delta-\Omega)}} |3\rangle, \label{eq:eigenstate3}
\end{eqnarray}
and the eigenvalues $\lambda_j$ are 
\begin{eqnarray}
\lambda_1 &=& -\Omega, \lambda_2 = 0, \lambda_3 = \Omega,
\end{eqnarray} 
with $\Omega=\sqrt{2J^2+\Delta^2}$. 
The transition frequencies, i.e., $\Delta_{jk}=\lambda_j-\lambda_k$, between eigenstates are   
\begin{eqnarray}
\Delta_{21}&=&\Delta_{32}=-\Delta_{12}=-\Delta_{23}=\Omega, \label{eq:freq_diff_21_32}\\
\Delta_{31}&=&-\Delta_{13}=2\Omega.
\end{eqnarray}

We then obtain the following expression for the electronic operator $\xi(t)$ [defined in Eq.~(\ref{eq:xi_def})], in the interaction picture,
\begin{eqnarray}
\xi(t) &=& \sum_{j,k} A_{jk} e^{i\Delta_{jk}t} |e_j\rangle\langle e_k| ,
\end{eqnarray}
where $A_{jk} e^{i\Delta_{jk}t}$ are the matrix elements of the electronic operator in the interaction Hamiltonian, i.e., $(|2\rangle\langle 2|-|1\rangle\langle 1|-|3\rangle\langle 3|)$, in the eigenstate basis. The $A_{jk}$ are explicitly, 
\begin{eqnarray}
A_{11}&=&-\frac{\Delta^2 (\Delta+\Omega)^2}{2\Omega^2[J^2+\Delta(\Delta+\Omega)]}, 
A_{22} = \frac{\Delta^2-2J^2}{\Omega^2}, 
A_{33} = -\frac{\Delta^2 (\Delta-\Omega)^2}{2\Omega^2[J^2+\Delta(\Delta-\Omega)]}, \label{eq:A_diag} \\
A_{12}&=&A_{21}=\frac{2\Delta J(\Delta+\Omega)}{\Omega^2\sqrt{2[J^2+\Delta(\Delta+\Omega)]}}, 
A_{13}=A_{31}=-\frac{2J^2}{\Omega^2}, 
A_{23}=A_{32}=\frac{2\Delta J(\Delta-\Omega)}{\Omega^2\sqrt{2[J^2+\Delta(\Delta-\Omega)]}} . \label{eq:A_offdiag}
\end{eqnarray}
Similarly, the electronic operator $\chi$ [defined in Eq.~(\ref{eq:zeta_def})], in the interaction picture, becomes
\begin{eqnarray}
\chi(t) &=& \sum_{j,k} B_{jk} e^{i\Delta_{jk}t} |e_j\rangle\langle e_k| ,
\end{eqnarray}
where
\begin{eqnarray}
B_{11}&=&-\frac{(\Delta\Omega+J^2)(\Delta+\Omega)^2}{2\Omega^2[J^2+\Delta(\Delta+\Omega)]}, 
B_{22}=-\frac{\Delta^2}{\Omega^2}, 
B_{33}=\frac{(\Delta\Omega-J^2)(\Delta-\Omega)^2}{2\Omega^2[J^2+\Delta(\Delta-\Omega)]}, \label{eq:B_diag} \\
B_{12}&=&B_{21}=\frac{2J^3}{\Omega^2\sqrt{2[J^2+\Delta(\Delta+\Omega)]}}, 
B_{13}=B_{31}=\frac{J^2}{\Omega^2},
B_{23}=B_{32}=\frac{2J^3}{\Omega^2\sqrt{2[J^2+\Delta(\Delta-\Omega)]}}. \label{eq:B_offdiag}
\end{eqnarray}
Therefore we have the full expression for the Hamiltonian in the interaction picture:
\begin{eqnarray}
H_I(t)&=&\sum_{j,k} \kappa_a A_{jk} [a^{\dagger} e^{i(\Delta_{jk}+\nu_a)t} +a e^{i(\Delta_{jk}-\nu_a)t}] |e_j\rangle\langle e_k|
+\sum_{j,k} \kappa_b B_{jk} [b^{\dagger} e^{i(\Delta_{jk}+\nu_b)t} +b e^{i(\Delta_{jk}-\nu_b)t}] |e_j\rangle\langle e_k| . 
\label{eq:H_I_interPicture}
\end{eqnarray}

\subsection{Perturbed evolution operator}

In the regime of weak site-vibration coupling, e.g.,  $\{\kappa_a,\kappa_b \} < \max\{J,\Delta\}$, the evolution operator in the interaction picture reads 
\begin{eqnarray}
U_I(t)&=&\mathcal{T} e^{-i\int_0^t ds H_I(s)} 
=\sum_{n=0}^{\infty} (-i)^n \int_0^t dt_1\cdots \int_0^{t_{n-1}} dt_n H_I(t_1)H_I(t_2)\cdots H_I(t_n) \\
&=& \underbrace{1}_{U_I^{(0)}} \underbrace{-i\int_0^t dt_1 H_I(t_1) }_{U_I^{(1)}} 
\underbrace{-\int_0^tdt_1\int_0^{t_1}dt_2 H_I(t_1)H_I(t_2) }_{U_I^{(2)}} \notag\\
&&\underbrace{+i\int_0^tdt_1\int_0^{t_1}dt_2\int_0^{t_2}dt_3 H_I(t_1)H_I(t_2)H_I(t_3) }_{U_I^{(3)}}\notag\\
&&\underbrace{+\int_0^tdt_1\int_0^{t_1}dt_2\int_0^{t_2}dt_3\int_0^{t_3}dt_4 H_I(t_1)H_I(t_2)H_I(t_3)H_I(t_4) }_{U_I^{(4)}} +\cdots , 
\label{eq:U_expansion}
\end{eqnarray}
with $\mathcal{T}$ being time-ordering operator and $U_I^{(0)}$, $U_I^{(1)}$, $U_I^{(2)}$, $U_I^{(3)}$, and $U_I^{(4)}$ are the zeroth-, first-, second-, third-, and fourth-order evolution operators, respectively, with respect to the coupling strength $\kappa_a$ or $\kappa_b$. 
Given that the fourth-order perturbative theory [i.e., $U_I(t)\approx1+U_I^{(1)}+U_I^{(2)}+U_I^{(3)}$] already allows us to uncover collective behavior, e.g., not only cooperative behavior but also interference of two vibrational modes, higher-order terms are neglected in the following.

\subsection{Transfer probability  \label{sec:perturb_probablity}}

The probability of finding an excitation at the acceptor (i.e., the third electronic site) of the trimeric chromophore system is 
\begin{eqnarray}
P_3(t)&=&Tr[ U_0^{\dagger}|3\rangle\langle 3|U_0 U_I(t) |1\rangle\langle 1| \rho_a\rho_b U_I^{\dagger}(t)] \notag\\
&=&Tr_{a,b}[ \langle 1| U_I^{\dagger}(t) U_0^{\dagger}|3\rangle \langle 3|U_0 U_I(t) |1\rangle \rho_a\rho_b ] \notag\\
&=& Tr_{a,b} \{ [(\mathcal{A}^{(0)})^{\dagger} +(\mathcal{A}^{(1)})^{\dagger} +(\mathcal{A}^{(2)})^{\dagger} +(\mathcal{A}^{(3)})^{\dagger} +(\mathcal{A}^{(4)})^{\dagger} ]
[ \mathcal{A}^{(0)} +\mathcal{A}^{(1)} +\mathcal{A}^{(2)} +\mathcal{A}^{(3)} +\mathcal{A}^{(4)}] \rho_a\rho_b\} \notag\\
&\approx&
 Tr_{a,b}\{ [(\mathcal{A}^{(0)})^{\dagger} \mathcal{A}^{(0)} +(\mathcal{A}^{(1)})^{\dagger} \mathcal{A}^{(1)} 
 + (\mathcal{A}^{(0)})^{\dagger} \mathcal{A}^{(2)} + (\mathcal{A}^{(2)})^{\dagger} \mathcal{A}^{(0)} \notag\\
&&+(\mathcal{A}^{(2)})^{\dagger} \mathcal{A}^{(2)} 
+ (\mathcal{A}^{(1)})^{\dagger} \mathcal{A}^{(3)} + (\mathcal{A}^{(3)})^{\dagger} \mathcal{A}^{(1)} 
+ (\mathcal{A}^{(0)})^{\dagger} \mathcal{A}^{(4)} + (\mathcal{A}^{(4)})^{\dagger} \mathcal{A}^{(0)}
] \rho_a\rho_b\} ,
\label{eq:transProb_general_def}
\end{eqnarray}
where transition amplitudes at different orders are defined as 
\begin{eqnarray}
\mathcal{A}^{(0)} &=& \langle 3|U_0 |1\rangle, \label{eq:A0_def} \\
\mathcal{A}^{(1)} &=& \langle 3|U_0 U_I^{(1)} |1\rangle, \label{eq:A1_def} \\
\mathcal{A}^{(2)} &=& \langle 3|U_0 U_I^{(2)} |1\rangle, \label{eq:A2_def} \\
\mathcal{A}^{(3)} &=& \langle 3|U_0 U_I^{(3)} |1\rangle, \label{eq:A3_def} \\
\mathcal{A}^{(4)} &=& \langle 3|U_0 U_I^{(4)} |1\rangle. \label{eq:A4_def}
\end{eqnarray}
In the first line of the equation that defines $P_3(t)$, $U_0^{\dagger}|3\rangle\langle 3|U_0$ is the projection operator $|3\rangle\langle 3|$ in the interaction picture. The initial state of the whole system is $|1\rangle\langle 1| \rho_a\rho_b$, independent of  the chosen picture. 
We mention that in the last line of Eq.~(\ref{eq:transProb_general_def}) there are no cross terms like $(\mathcal{A}^{(0)})^{\dagger} \mathcal{A}^{(1)}$ that contain an odd number of vibrational operators and therefore cannot survive the trace (e.g., $Tr_{a,b}[\cdots \rho_a\rho_b]$).

The transfer probability can be written as a sum of probabilities at different orders, namely, 
\begin{eqnarray}
P_3(t)&=&P_3^{(0)}+P_3^{(1)}+P_3^{(2)} ,
\label{eq:transProb}
\end{eqnarray}
where
\begin{eqnarray}
P_3^{(0)} &=& Tr_{a,b}[(\mathcal{A}^{(0)})^{\dagger} \mathcal{A}^{(0)} \rho_a\rho_b], \\
P_3^{(1)} &=& Tr_{a,b} \{ [(\mathcal{A}^{(1)})^{\dagger} \mathcal{A}^{(1)} + (\mathcal{A}^{(0)})^{\dagger} \mathcal{A}^{(2)} + (\mathcal{A}^{(2)})^{\dagger} \mathcal{A}^{(0)} ] \rho_a\rho_b\} , \label{eq:P_3_1_def}\\
P_3^{(2)} &=& Tr_{a,b} \{[ (\mathcal{A}^{(2)})^{\dagger} \mathcal{A}^{(2)} + (\mathcal{A}^{(1)})^{\dagger} \mathcal{A}^{(3)} +(\mathcal{A}^{(3)})^{\dagger} \mathcal{A}^{(1)} 
+(\mathcal{A}^{(0)})^{\dagger} \mathcal{A}^{(4)} +(\mathcal{A}^{(4)})^{\dagger} \mathcal{A}^{(0)}]\rho_a\rho_b \}. \label{eq:P_3_2_def}
\end{eqnarray}
These expressions can be further simplified when the weak site-site coupling regime ($J<\Delta$) is considered, as shown below.

\subsection{Interaction amplitudes}

To gain insight from the perturbative expressions of transfer probability above, we define the $n$th-order interaction amplitude (due to the coupling between the system and the vibrational modes), as
\begin{eqnarray}
W^{q_1x_1,\cdots,q_nx_n}_{jk_1,\cdots,k_{n-1}k_n} &=& \underbrace{\kappa_{x_1}\cdots\kappa_{x_n}} \underbrace{ (X_1)_{jk_1}\cdots(X_n)_{k_{n-1}k_n} } \notag\\
&&\times  \int_0^tdt_1\cdots\int_0^{t_{n-1}}dt_n
e^{i(\Delta_{jk_1}+q_1\nu_{x_1})t_1} \cdots e^{i(\Delta_{k_{n-1}k_n}+q_n\nu_{x_n})t_n} ,
\end{eqnarray}
where $x_i\in\{ a,b\}$ and $X_i\in\{A,B\}$. $q_i\in\{+,-\}$ represents an emission or absorption of a phonon and $j,k_i\in\{1,2,3\}$ denote electronic eigenstates ($|e_1\rangle$, $|e_2\rangle$, or $|e_3\rangle$), respsectively. 
In addition to the site-vibration coupling strength $\kappa_a,\kappa_b$ for vibrational mode $\nu_a$ and $\nu_b$ respectively, the interaction amplitude further depends on $A_{jk}, B_{jk}$, which are matrix elements of the electronic interaction operator in the eigenstate basis and given above. 

In the framework of the fourth-order perturbation theory considered in our work, the first four interaction amplitudes are needed for a full expression of the transfer probability. Here we explicitly write down the first three of them that are relevant in this supplemental material, namely
\begin{eqnarray}
W^{q_1x_1}_{jk} &=& \kappa_{x_1} (X_1)_{jk} \int_0^tdt_1 e^{i(\Delta_{jk}+q_1\nu_{x_1})t_1}, 
\end{eqnarray}
\begin{eqnarray}
W^{q_1x_1,q_2x_2}_{jk,kk'} &=& \kappa_{x_1}\kappa_{x_2} (X_1)_{jk}(X_2)_{kk'} \int_0^tdt_1\int_0^{t_1}dt_2 
e^{i(\Delta_{jk}+q_1\nu_{x_1})t_1} e^{i(\Delta_{kk'}+q_2\nu_{x_2})t_2} ,
\label{eq:w2}
\end{eqnarray}
\begin{eqnarray}
W^{q_1x_1,q_2x_2,q_3x_3}_{jk,kk',k'k''} &=& \kappa_{x_1}\kappa_{x_2}\kappa_{x_3} (X_1)_{jk}(X_2)_{kk'}(X_3)_{k'k''} \notag\\
&&\times  \int_0^tdt_1\int_0^{t_1}dt_2\int_0^{t_2}dt_3
e^{i(\Delta_{jk}+q_1\nu_{x_1})t_1} e^{i(\Delta_{kk'}+q_2\nu_{x_2})t_2} e^{i(\Delta_{k'k''}+q_3\nu_{x_3})t_3} .
\end{eqnarray}

\subsection{Weak site-site coupling \label{sec:weakJ}}

In this subsection we specialize to a weak site-site coupling $J<\Delta$ (or  $J<\Omega$) regime. 
This assumption allows us to approximate initial and final states in the site basis roughly by the eigenstates, i.e.,  
\begin{eqnarray}
|1\rangle &\approx& |e_1\rangle, |3\rangle \approx |e_3\rangle ,
\end{eqnarray}
and the transition amplitude appearing in Eq.~(\ref{eq:transProb_general_def}) becomes
\begin{eqnarray}
\langle 3| U_0 U_I |1\rangle &\approx& \langle e_3| U_0^{(\nu)} U_0^{(e)} U_I |e_1\rangle 
= U_0^{(\nu)} e^{-i\lambda_3 t} \langle e_3| U_I |e_1\rangle \notag\\
&=& U_0^{(\nu)} e^{-i\lambda_3 t} \langle e_3| 1+ U_I^{(1)} + U_I^{(2)} + U_I^{(3)} + U_I^{(4)} |e_1\rangle
= \mathcal{A}^{(0)} + \mathcal{A}^{(1)} + \mathcal{A}^{(2)} + \mathcal{A}^{(3)} + \mathcal{A}^{(4)}
\end{eqnarray}
with 
\begin{eqnarray}
\mathcal{A}^{(0)} &=& 0, \label{eq:eq:A0_def_weakJ} \\
\mathcal{A}^{(1)} &=& U_0^{(\nu)} e^{-i\lambda_3 t} \langle e_3| U_I^{(1)} |e_1\rangle, \label{eq:A1_def_weakJ} \\
\mathcal{A}^{(2)} &=& U_0^{(\nu)} e^{-i\lambda_3 t} \langle e_3| U_I^{(2)} |e_1\rangle, \label{eq:A2_def_weakJ} \\
\mathcal{A}^{(3)} &=& U_0^{(\nu)} e^{-i\lambda_3 t} \langle e_3| U_I^{(3)} |e_1\rangle, \label{eq:A3_def_weakJ} \\
\mathcal{A}^{(4)} &=& U_0^{(\nu)} e^{-i\lambda_3 t} \langle e_3| U_I^{(4)} |e_1\rangle . \label{eq:A4_def_weakJ}
\end{eqnarray}
Therefore the probability in Eq.~(\ref{eq:transProb}) becomes
\begin{eqnarray}
P_3(t)
&=&
P_3^{(1)} + P_3^{(2)}
= Tr_{a,b} \{ [(\mathcal{A}^{(1)})^{\dagger} \mathcal{A}^{(1)} +(\mathcal{A}^{(2)})^{\dagger} \mathcal{A}^{(2)}
+ (\mathcal{A}^{(1)})^{\dagger} \mathcal{A}^{(3)} + (\mathcal{A}^{(3)})^{\dagger} \mathcal{A}^{(1)} ] \rho_a\rho_b \} .
\end{eqnarray}
The amplitude $\mathcal{A}^{(0)}$ is zero and does not contribute to the probability, which is however, not true when $J$ is not small as shown in the next subsection. This implies that, without assistance from the vibrational modes, it is impossible for an excitation to transfer from the initial donor state $|1\rangle$ to the target acceptor state $|3\rangle$.

\subsubsection{One-phonon processes \label{sec:weakJ_1phonon}}

By using the amplitude $\mathcal{A}^{(1)}$ in Eq.~(\ref{eq:A1_def_weakJ}), we obtain the first-order probability as
\begin{eqnarray}
P_3^{(1)} &=& n_a W_{13}^{+a} W_{31}^{-a} + n_b W_{13}^{+b } W_{31}^{-b} = n_a (W_{31}^{-a})^* W_{31}^{-a} + n_b (W_{31}^{-b})^* W_{31}^{-b} .
\label{eq:P_3_1}
\end{eqnarray}
Here we used $(W_{ij}^{+a})^*=W_{ji}^{-a}$ and $(W_{ij}^{+b})^*=W_{ji}^{-b}$
and the average phonon numbers of two vibrational modes $\nu_a$ and $\nu_b$ are given by
\begin{eqnarray}
n_a&=&Tr_a[a^{\dagger}a\rho_a] =\langle a^{\dagger}a \rangle =\frac{1}{e^{\nu_a/K_BT_a}-1}, \label{eq:n_a}\\
n_b&=&Tr_b[b^{\dagger}b\rho_b] =\langle b^{\dagger}b \rangle =\frac{1}{e^{\nu_b/K_BT_b}-1}, \label{eq:n_b}
\end{eqnarray}
respectively. 
We mention that terms like $(n_a+1) (W_{31}^{+a})^*W_{31}^{+a} $ that do not conserve energy have been neglected in $P_3^{(1)}$ above. 
This direct transition from $|1\rangle$ to $|3\rangle$ results from the assumption of weak $J$ under which we have used $|1\rangle\approx|e_1\rangle$ and $|3\rangle\approx|e_3\rangle$. It is obvious the transition becomes forbidden when $J\rightarrow0$ due to $A_{13}=A_{31}=-\frac{2J^2}{\Omega^2}$ [Eq.~(\ref{eq:A_offdiag})] or $B_{13}=B_{31}=\frac{J^2}{\Omega^2}$ [Eq.~(\ref{eq:B_offdiag})]. 

When the transition becomes resonant, i.e., $\nu_a=\Delta_{31}$ or $\nu_b=\Delta_{31}$, the terms in $P_3^{(1)} $, corresponding to a one-phonon process assisted by either mode $\nu_a$ or $\nu_b$, can be further simplified as $(W_{31}^{-a})^* W_{31}^{-a} =t^2 \kappa_a^2 A_{13}^2$ and $(W_{31}^{-b})^* W_{31}^{-b}=t^2 \kappa_b^2 B_{13}^2$.

\begin{figure}[h]
\centering
  \includegraphics[width=.75\columnwidth]{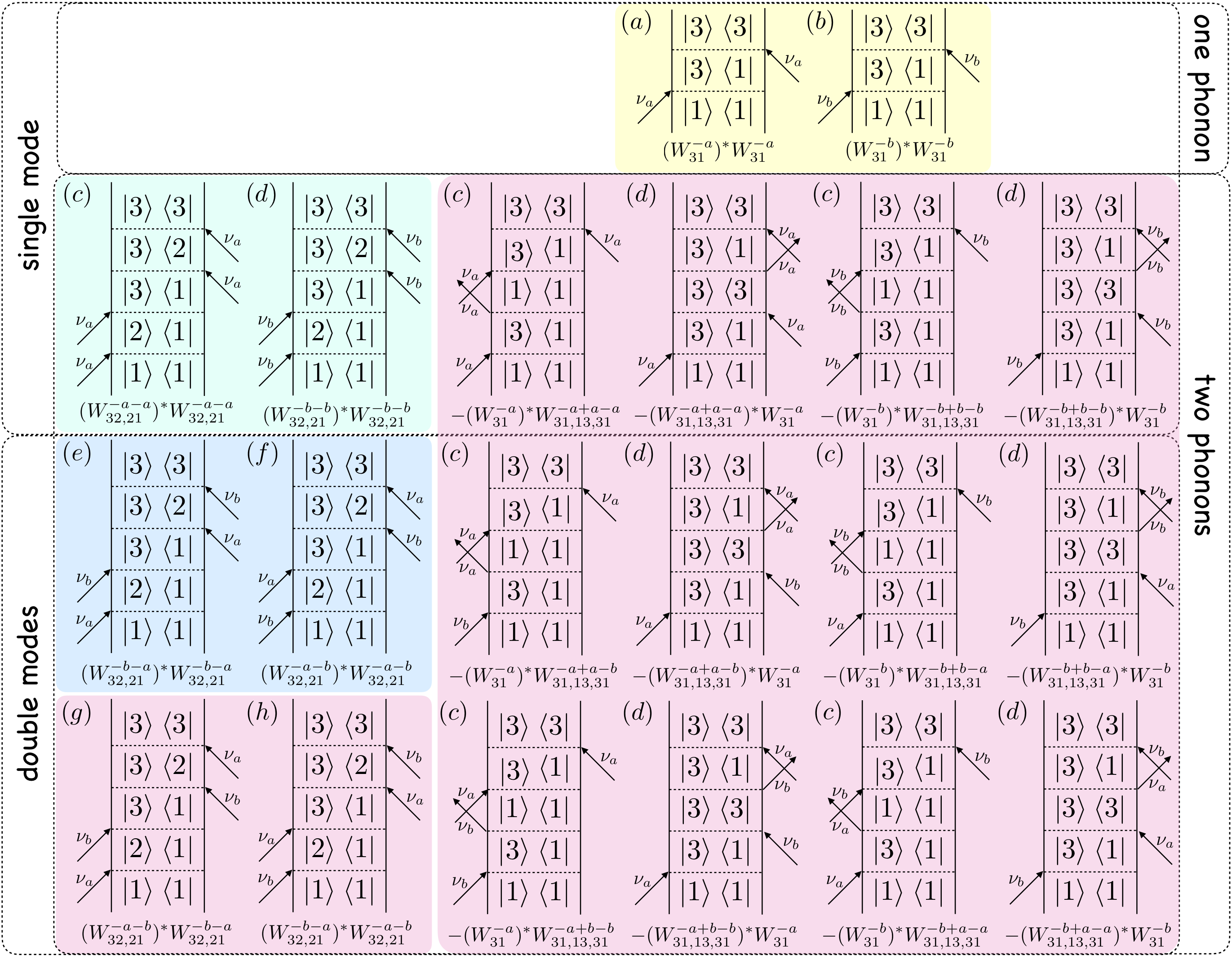} 
\caption{(color online) Double-sided Feynman diagrams at weak $J$ and around resonance conditions. }
\label{fig:feynman_weakJ}
\end{figure}

\subsubsection{Two-phonon processes}

The two-phonon processes refer to transitions from the donor $|1\rangle$ to the acceptor $|3\rangle$ via an intermediate state $|2\rangle$. 
The corresponding second-order probability is 
\begin{eqnarray}
P_3^{(2)} &=& P_3^{(2,1)} + P_3^{(2,2)} ,
\end{eqnarray}
where 
\begin{eqnarray}
P_3^{(2,1)} &=&
2n_a^2 (W_{32,21}^{-a-a})^* W_{32,21}^{-a-a} + 2n_b^2 (W_{32,21}^{-b-b})^* W_{32,21}^{-b-b} \notag\\
&&+ n_a n_b [(W_{32,21}^{-b-a})^* W_{32,21}^{-b-a} +  (W_{32,21}^{-a-b})^* W_{32,21}^{-a-b} 
+ (W_{32,21}^{-a-b})^* W_{32,21}^{-b-a} +(W_{32,21}^{-b-a})^* W_{32,21}^{-a-b} ] ,
\label{eq:P_3_2_1}
\end{eqnarray}
and 
\begin{eqnarray}
P_3^{(2,2)} &=& 
-  n_a(2n_a+1) [ W_{13}^{+a}W_{31,13,31}^{-a+a-a} + W_{31}^{-a}W_{13,31,13}^{+a-a+a} ]
-  n_b(2n_b+1) [W_{13}^{+b}W_{31,13,31}^{-b+b-b} + W_{31}^{-b}W_{13,31,13}^{+b-b+b} ] \notag\\
&&
- n_a(n_b+1) [W_{13}^{+a} W_{31,13,31}^{-a+a-b} + W_{31}^{-a} W_{13,31,13}^{+a-a+b} ]
- (n_a+1)n_b [W_{13}^{+b}  W_{31,13,31}^{-b+b-a}  + W_{31}^{-b} W_{13,31,13}^{+b-b+a} ] \notag\\
&& 
- n_a n_b [ W_{13}^{+a} W_{31,13,31}^{-a+b-b} + W_{13}^{+b} W_{31,13,31}^{-b+a-a}  
+W_{31}^{-a} W_{13,31,13}^{+a-b+b} + W_{31}^{-b} W_{13,31,13}^{+b-a+a} ] \} .
\label{eq:P_3_2_2}
\end{eqnarray}
The coefficients in front of interaction amplitudes accounting for the effect of temperature come from averages of products of four vibrational operators, evaluated via Wick's theorem~\cite{FetterWalecka1971}, 
\begin{eqnarray}
\langle aaa^{\dagger}a^{\dagger}\rangle &=& 2(n_a+1)^2, \langle a^{\dagger}a^{\dagger}aa\rangle=2n_a^2, 
\langle a^{\dagger}aa^{\dagger}a\rangle = n_a(2n_a+1), \label{eq:example_Wick_a} \\
\langle aa^{\dagger}aa^{\dagger}\rangle &=& (n_a+1)(2n_a+1), 
\langle a^{\dagger}aaa^{\dagger}\rangle = \langle aa^{\dagger}a^{\dagger}a\rangle = 2n_a(n_a+1), \label{eq:example_Wick}
\end{eqnarray}
\begin{eqnarray}
\langle bbb^{\dagger}b^{\dagger}\rangle &=& 2(n_b+1)^2, \langle b^{\dagger}b^{\dagger}bb\rangle=2n_b^2, 
\langle b^{\dagger}bb^{\dagger}b\rangle = n_b(2n_b+1), \label{eq:example_Wick_b1} \\
\langle bb^{\dagger}bb^{\dagger}\rangle &=& (n_b+1)(2n_b+1), 
\langle b^{\dagger}bbb^{\dagger}\rangle = \langle bb^{\dagger}b^{\dagger}b\rangle = 2n_b(n_b+1), \label{eq:example_Wick_b2}
\end{eqnarray}
\begin{eqnarray}
\langle aa^{\dagger}bb^{\dagger}\rangle &=& (n_a+1)(n_b+1), \langle a^{\dagger}ab^{\dagger}b\rangle = n_a n_b,
\langle aa^{\dagger}b^{\dagger}b\rangle = (n_a+1)n_b, \langle a^{\dagger}abb^{\dagger}\rangle = n_a(n_b+1) . \label{eq:example_Wick_ab}
\end{eqnarray}

Substituting the expressions for $W_{ik,kl}$ from Eqs. \eqref{eq:w2} into the expressions for $P_3^{(2,1)}$ and $P_3^{(2,2)}$, we find that the resonance conditions for the two probabilities are different. For $P_3^{(2,2)}$ the resonance condition is $\nu_a = \Delta_{31}$ and/or $\nu_b = \Delta_{31}$, whereas for $P_3^{(2,1)}$ the resonance condition is $\nu_a = \Delta_{21}=\Delta_{32}$ and/or $\nu_b = \Delta_{21}=\Delta_{32}$.

The double-sided Feynman diagrams corresponding to terms in Eqs.~(\ref{eq:P_3_1}), (\ref{eq:P_3_2_1}), and (\ref{eq:P_3_2_2}) are presented in Fig.~\ref{fig:feynman_weakJ}.

Since $P_3^{(2,1)}$ is much more relevant to the two-phonon VAET processes shown in the main text, we examine it further. We simplify each term in $P_3^{(2,1)}$ by considering resonant transitions, i.e., $\nu_a,\nu_b\rightarrow \Delta_{21}=\Delta_{32}$, in which case,
\begin{eqnarray}
2n_a^2 (W_{32,21}^{-a-a})^* W_{32,21}^{-a-a} &=& 2n_a^2 t^4 \kappa_a^4 A_{12}A_{23}A_{32}A_{21}, \\
2n_b^2 (W_{32,21}^{-b-b})^* W_{32,21}^{-b-b} &=& 2n_b^2  t^4 \kappa_b^4 B_{12}B_{23}B_{32}B_{21}, \\
n_a n_b (W_{32,21}^{-b-a})^* W_{32,21}^{-b-a} &=& n_an_b  t^4 \kappa_a^2\kappa_b^2 A_{12}B_{23}B_{32}A_{21}, \\
n_a n_b (W_{32,21}^{-a-b})^* W_{32,21}^{-a-b} &=& n_an_b  t^4 \kappa_a^2\kappa_b^2 B_{12}A_{23}A_{32}B_{21}, \\
n_a n_b (W_{32,21}^{-a-b})^* W_{32,21}^{-b-a} &=& n_an_b  t^4 \kappa_a^2\kappa_b^2 A_{12}B_{23}A_{32}B_{21}, \\
n_a n_b (W_{32,21}^{-b-a})^* W_{32,21}^{-a-b} &=& n_an_b  t^4 \kappa_a^2\kappa_b^2  B_{12}A_{23}B_{32}A_{21}. 
\end{eqnarray}

\subsection{Strong site-site coupling}

The weak site-site coupling $J<\Delta$ considered above does simply the problem, but it is, however, actually not necessary for obtaining  perturbation-based analytical results, which only require a small $\kappa_a$ and $\kappa_b$. 
In this section we therefore present results that go beyond weak site-site coupling. 

To calculate the transition amplitudes, e.g., $\mathcal{A}^{(i)}$, and therefore probability $P_3(t)=P_3^{(0)}+P_3^{(1)}+P_3^{(2)}$ (see Sec.~\ref{sec:perturb_probablity}), we need to first transform the initial and target states from the site basis into the eigenstate basis. By using eigenstates defined in Eqs.~(\ref{eq:eigenstate1}), (\ref{eq:eigenstate2}), and (\ref{eq:eigenstate3}), we have
\begin{eqnarray}
|1\rangle=\sum_i |e_i\rangle\langle e_i|1\rangle &=&\alpha |e_1\rangle -\beta |e_2\rangle +\gamma|e_3\rangle, \\
|3\rangle=\sum_i |e_i\rangle\langle e_i|3\rangle &=&\gamma |e_1\rangle +\beta |e_2\rangle +\alpha |e_3\rangle, 
\end{eqnarray}
where
\begin{eqnarray}
\alpha&=&\frac{\sqrt{J^2+\Delta(\Delta+\Omega)}}{\sqrt{2}\Omega}, 
\beta = \frac{J}{\Omega},
\gamma = \frac{\sqrt{J^2+\Delta(\Delta-\Omega)}}{\sqrt{2}\Omega} .
\end{eqnarray}
We then obtain an analytical expression of the probability of finding an excitation in the final state $|3\rangle$ (acceptor) transferred from the initial state $|1\rangle$ (donor). 

\subsubsection{$P_3^{(0)} $}

By first deriving the transition amplitude $\mathcal{A}^{(0)}=\langle 3|U_0|1\rangle$ in Eq.~(\ref{eq:A0_def}), we obtain the zeroth-order probability
\begin{eqnarray}
P_3^{(0)}&=&Tr_{a,b}[(\mathcal{A}^{(0)})^{\dagger} \mathcal{A}^{(0)} \rho_a\rho_b]
= \frac{J^4}{\Omega^4} [\cos(\Omega t)-1]^2 .
\end{eqnarray}
As expected, the transition amplitude is independent of interaction with vibrational modes and therefore is consistent with unitary evolution. 

\subsubsection{$P_3^{(1)} $ }

The first-order probability  $P_3^{(1)}$ defined in Eq.~(\ref{eq:P_3_1_def}) is written as a sum of two terms that have different resonant transition frequencies, i.e., 
\begin{eqnarray}
P_3^{(1)}&=&P_3^{(1,1)}+P_3^{(1,2)}.
\end{eqnarray}
The first term on the right-hand-side is obtained as 
\begin{eqnarray}
P_3^{(1,1)} &=&Tr_{a,b}\{ (\mathcal{A}^{(1)})^{\dagger} \mathcal{A}^{(1)} \rho_a\rho_b\}
=  \langle aa^{\dagger}\rangle  |A_{a+}^{(1)}|^2 + \langle a^{\dagger}a \rangle  |A_{a-}^{(1)}|^2 ,
+ \langle bb^{\dagger}\rangle  |B_{b+}^{(1)}|^2 +  \langle b^{\dagger}b \rangle  |B_{b-}^{(1)}|^2 ,
\end{eqnarray}
where
\begin{eqnarray}
|A_{a+}^{(1)}|^2 &=&  \beta^2\gamma^2 (W_{12}^{+a})^* W_{12}^{+a} + \gamma^4 (W_{13}^{+a})^* W_{13}^{+a} + \beta^2\gamma^2 (W_{23}^{+a})^* W_{23}^{+a} \notag\\
&& - \beta^2 \gamma^2 e^{-i\Delta_{21}t} (W_{12}^{+a})^* W_{23}^{+a} - \beta^2 \gamma^2 e^{i\Delta_{21}t} (W_{23}^{+a})^* W_{12}^{+a} , \\
|A_{a-}^{(1)}|^2  &=& \alpha^2 \beta^2 (W_{21}^{-a})^* W_{21}^{-a} + \alpha^4 (W_{31}^{-a})^* W_{31}^{-a} + \alpha^2 \beta^2 (W_{32}^{-a})^* W_{32}^{-a} \notag\\
&& - \alpha^2 \beta^2 e^{-i\Delta_{32}t} (W_{21}^{-a})^* W_{32}^{-a} -\alpha^2 \beta^2 e^{i\Delta_{32}t} (W_{32}^{-a})^* W_{21}^{-a}  , \\
|B_{b+}^{(1)}|^2 &=& \beta^2\gamma^2 (W_{12}^{+b})^* W_{12}^{+b} + \gamma^4 (W_{13}^{+})^* W_{13}^{+b} + \beta^2\gamma^2 (W_{23}^{+b})^* W_{23}^{+b} \notag\\
&& - \beta^2 \gamma^2 e^{-i\Delta_{21}t} (W_{12}^{+b})^* W_{23}^{+b} - \beta^2 \gamma^2 e^{i\Delta_{21}t} (W_{23}^{+b})^* W_{12}^{+b}  , \\
|B_{b-}^{(1)}|^2 &=& -\alpha^2 \beta^2 (W_{21}^{-b})^* W_{21}^{-b} + \alpha^4 (W_{31}^{-b})^* W_{31}^{-b} + \alpha^2 \beta^2 (W_{32}^{-b})^* W_{32}^{-b} \notag\\
&& -\alpha^2 \beta^2 e^{-i\Delta_{32}t} (W_{21}^{-b})^* W_{32}^{-b} -\alpha^2 \beta^2 e^{i\Delta_{32}t} (W_{32}^{-b})^* W_{21}^{-b} .
\end{eqnarray}

The other term is given by
\begin{eqnarray}
P_3^{(1,2)} &=& Tr_{a,b} \{ [(\mathcal{A}^{(0)})^{\dagger} \mathcal{A}^{(2)} + (\mathcal{A}^{(2)})^{\dagger} \mathcal{A}^{(0)} ] \rho_a\rho_b \} \notag\\
&=&-\frac{J^2}{2\Omega^2}[\cos(\Omega t)-1] \{
\langle aa^{\dagger}\rangle [A_{a-a+}^{(2)}+(A_{a-a+}^{(2)})^* ] + \langle a^{\dagger}a \rangle [A_{a+a-}^{(2)}+(A_{a+a-}^{(2)})^*]  \notag\\
&& + \langle bb^{\dagger}\rangle [B_{b-b+}^{(2)}+(B_{b-b+}^{(2)})^*] + \langle b^{\dagger}b \rangle [B_{b+b-}^{(2)}+(B_{b+b-}^{(2)})^*]  \}, 
\end{eqnarray}
where
\begin{eqnarray}
A_{a-a+}^{(2)}+(A_{a-a+}^{(2)})^*  &=&
-\beta^2 [ e^{-i\lambda_2t} W_{21,12}^{-a+a} + e^{i\lambda_2t} (W_{21,12}^{-a+a})^* ] 
+\alpha\gamma [ e^{-i\lambda_3t} W_{31,13}^{-a+a}  + e^{i\lambda_3t} (W_{31,13}^{-a+a})^* ] \notag\\
&&+\alpha\gamma [ e^{-i\lambda_3t} W_{32,23}^{-a+a}  + e^{i\lambda_3t} (W_{32,23}^{-a+a})^* ] , \\
A_{a+a-}^{(2)}+(A_{a+a-}^{(2)})^* &=& 
\alpha\gamma [e^{-i\lambda_1t} W_{12,21}^{+a-a} + e^{i\lambda_1t} (W_{12,21}^{+a-a})^* ]
+ \alpha\gamma [ e^{-i\lambda_1t} W_{13,31}^{+a-a} + e^{i\lambda_1t} (W_{13,31}^{+a-a})^* ]  \notag\\
&& - \beta^2 [ e^{-i\lambda_2t} W_{23,32}^{+a-a} + e^{i\lambda_2t} (W_{23,32}^{+a-a})^* ] , \\
B_{b-b+}^{(2)}+(B_{b-b+}^{(2)})^* &=& 
-\beta^2 [e^{-i\lambda_2t} W_{21,12}^{-b+b}  +  e^{i\lambda_2t} (W_{21,12}^{-b+b})^* ]
+\alpha\gamma [ e^{-i\lambda_3t} W_{31,13}^{-b+b}  + e^{i\lambda_3t} (W_{31,13}^{-b+b})^* ] \notag\\
&&+\alpha\gamma [ e^{-i\lambda_3t} W_{32,23}^{-b+b}  + e^{i\lambda_3t} (W_{32,23}^{-b+b})^* ] , \\
B_{b+b-}^{(2)}+(B_{b+b-}^{(2)})^* &=& 
\alpha\gamma [ e^{-i\lambda_1t} W_{12,21}^{+b-b} + e^{i\lambda_1t} (W_{12,21}^{+b-b})^* ]
+ \alpha\gamma [ e^{-i\lambda_1t} W_{13,31}^{+b-b} + e^{i\lambda_1t} (W_{13,31}^{+b-b})^* ]  \notag\\
&& - \beta^2 [ e^{-i\lambda_2t} W_{23,32}^{+b-b} + e^{i\lambda_2t} (W_{23,32}^{+b-b})^* ] .
\end{eqnarray}
For thermal states, we would set $\langle a^{\dagger}a \rangle=n_a$, $\langle aa^{\dagger} \rangle=n_a+1$,  $\langle b^{\dagger}b \rangle=n_b$, and $\langle bb^{\dagger} \rangle=n_b+1$.

\subsubsection{$P_3^{(2)} $}

The second-order probability $P_3^{(2)}$ defined in Eq.~(\ref{eq:P_3_2_def}) is
\begin{eqnarray} 
P_3^{(2)} &=& P_3^{(2,1)} + P_3^{(2,2)} + P_3^{(2,3)} . 
\end{eqnarray}
The first term on the right hand side is given by
\begin{eqnarray} 
P_3^{(2,1)} &=& Tr_{a,b} \{[ (\mathcal{A}^{(2)})^{\dagger} \mathcal{A}^{(2)} ]\rho_a\rho_b  \} , \notag\\
&=&
\langle aaa^{\dagger}a^{\dagger}\rangle  |A_{a+a+}^{(2)}|^2 
+\langle a^{\dagger}a^{\dagger}aa\rangle |A_{a-a-}^{(2)}|^2
+\langle a^{\dagger}aa^{\dagger}a\rangle |A_{a+a-}^{(2)}|^2 
+\langle aa^{\dagger}aa^{\dagger}\rangle |A_{a-a+}^{(2)}|^2 \notag\\
&&+\langle a^{\dagger}aaa^{\dagger}\rangle (A_{a+a-}^{(2)})^* A_{a-a+}^{(2)} 
+ \langle aa^{\dagger}a^{\dagger}a \rangle (A_{a-a+}^{(2)})^*A_{a+a-}^{(2)} \notag\\
&&+ \langle bbb^{\dagger}b^{\dagger}\rangle |B_{b+b+}^{(2)}|^2 
+ \langle b^{\dagger}b^{\dagger}bb\rangle |B_{b-b-}^{(2)}|^2 
+ \langle b^{\dagger}bb^{\dagger}b\rangle |B_{b+b-}^{(2)}|^2 
+ \langle bb^{\dagger}bb^{\dagger}\rangle |B_{b-b+}^{(2)}|^2 \notag\\
&&+ \langle b^{\dagger}bbb^{\dagger}\rangle (B_{b+b-}^{(2)})^* B_{b-b+}^{(2)} 
+ \langle bb^{\dagger}b^{\dagger}b \rangle (B_{b-b+}^{(2)})^*B_{b+b-}^{(2)} \notag\\
&&+ \langle aa^{\dagger}bb^{\dagger}\rangle |C_{a+b+}^{(2)}|^2 
+ \langle a^{\dagger}ab^{\dagger}b\rangle   |C_{a-b-}^{(2)}|^2 
+ \langle aa^{\dagger}b^{\dagger}b\rangle   |C_{a+b-}^{(2)}|^2 
+ \langle a^{\dagger}abb^{\dagger}\rangle   |C_{a-b+}^{(2)}|^2 ,
\end{eqnarray}
where
\begin{eqnarray} 
|A_{a+a+}^{(2)}|^2 &=& \gamma^4 ( W_{12,23}^{+a+a} )^* W_{12,23}^{+a+a} 
(\approx t^4 \gamma^4 \kappa_a^4 A_{12}^2 A_{23}^2) , \\
|A_{a-a-}^{(2)}|^2 &=& \alpha^4 ( W_{32,21}^{-a-a})^* W_{32,21}^{-a-a} 
(\approx \frac{t^4}{4} \alpha^4 \kappa_a^4 A_{32}^2 A_{21}^2 ) , \\
|A_{a+a-}^{(2)}|^2 &=& \alpha^2\gamma^2 (W_{12,21}^{+a-a})^* W_{12,21}^{+a-a} 
 +  \alpha^2\gamma^2 (W_{13,31}^{+a-a})^* W_{13,31}^{+a-a} 
 +\beta^4 (W_{23,32}^{+a-a})^* W_{23,32}^{+a-a}  \notag\\
&&  -\alpha\beta^2\gamma e^{-i\Delta_{21}t} (W_{12,21}^{+a-a})^* W_{23,32}^{+a-a}
 - \alpha\beta^2\gamma e^{i\Delta_{21}t} (W_{23,32}^{+a-a})^* W_{12,21}^{+a-a} , \\
|A_{a-a+}^{(2)}|^2 &=& \beta^4 (W_{21,12}^{-a+a})^* W_{21,12}^{-a+a} 
+ \alpha^2\gamma^2 (W_{31,13}^{-a+a})^* W_{31,13}^{-a+a} 
+ \alpha^2\gamma^2 (W_{32,23}^{-a+a})^* W_{32,23}^{-a+a} \notag\\
&&- \alpha\beta^2\gamma e^{-i\Delta_{32}t} (W_{21,12}^{-a+a})^* W_{32,23}^{-a+a}
-\alpha\beta^2\gamma e^{i\Delta_{32}t} (W_{32,23}^{-a+a})^* W_{21,12}^{-a+a} , \\
(A_{a+a-}^{(2)})^* A_{a-a+}^{(2)} 
&=& -\alpha\gamma e^{-i\Delta_{21}t} (W_{12,21}^{+a-a})^* W_{21,12}^{-a+a} 
 + \beta^4 (W_{23,32}^{+a-a})^* W_{21,12}^{-a+a} 
+\alpha^2\gamma^2 e^{-i\Delta_{31}t} (W_{12,21}^{+a-a})^* W_{32,23}^{-a+a} \notag\\
&&  -\alpha\beta^2\gamma e^{-i\Delta_{32}t} (W_{23,32}^{+a-a})^* W_{32,23}^{-a+a} 
+ \alpha^2\gamma^2 e^{-i\Delta_{31}t} (W_{13,31}^{+a-a})^* W_{31,13}^{-a+a} , \\
(A_{a-a+}^{(2)})^*A_{a+a-}^{(2)} 
&=&
 -\alpha\beta^2\gamma e^{i\Delta_{21}t} (W_{21,12}^{-a+a})^* W_{12,21}^{+a-a}
 +\beta^4 (W_{21,12}^{-a+a})^* W_{23,32}^{+a-a}
 +\alpha^2\gamma^2 e^{i\Delta_{31}t} (W_{32,23}^{-a+a})^* W_{12,21}^{+a-a} \notag\\
&&-\alpha\beta^2\gamma e^{i\Delta_{32}t} (W_{32,23}^{-a+a})^* W_{23,32}^{+a-a} 
 +\alpha^2\gamma^2 e^{i\Delta_{31}t} (W_{31,13}^{-a+a})^* W_{13,31}^{+a-a}  ,
\end{eqnarray}

\begin{eqnarray} 
|B_{b+b+}^{(2)}|^2 &\approx& = \gamma^4 (W_{12,23}^{+b+b})^* W_{12,23}^{+b+b} 
(\approx \frac{t^4}{4} \gamma^4\kappa_b^4 B_{12}^2 B_{23}^2 ) ,\\
|B_{b-b-}^{(2)}|^2 &\approx& = \alpha^4 (W_{32,21}^{-b-b})^* W_{32,21}^{-b-b} 
(\approx  \frac{t^4}{4} \alpha^4 \kappa_b^4 B_{32}^2 B_{21}^2 )  ,\\
|B_{b+b-}^{(2)}|^2 &\approx& \alpha^2\gamma^2 (W_{12,21}^{+b-b})^* W_{12,21}^{+b-b} 
 +  \alpha^2\gamma^2 (W_{13,31}^{+b-b})^* W_{13,31}^{+b-b} 
 +\beta^4 (W_{23,32}^{+b-b} )^* W_{23,32}^{+b-b} \notag\\
&&  -\alpha\beta^2\gamma e^{-i\Delta_{21}t} (W_{12,21}^{+b-b})^* W_{23,32}^{+b-b}
 - \alpha\beta^2\gamma e^{i\Delta_{21}t} (W_{23,32}^{+b-b})^* W_{12,21}^{+b-b}  ,\\
|B_{b-b+}^{(2)}|^2 &\approx& \beta^4 (W_{21,12}^{-b+b})^* W_{21,12}^{-b+b}
+ \alpha^2\gamma^2 (W_{31,13}^{-b+b})^* W_{31,13}^{-b+b} 
+ \alpha^2\gamma^2 (W_{32,23}^{-b+b})^* W_{32,23}^{-b+b}  \notag\\
&& - \alpha\beta^2\gamma e^{-i\Delta_{32}t} (W_{21,12}^{-b+b})^* W_{32,23}^{-b+b}
-\alpha\beta^2\gamma e^{i\Delta_{32}t} (W_{32,23}^{-b+b})^* W_{21,12}^{-b+b}  ,\\
(B_{b+b-}^{(2)})^* B_{b-b+}^{(2)} 
&\approx&  
-\alpha\gamma e^{-i\Delta_{21}t} (W_{12,21}^{+b-b})^* W_{21,12}^{-b+b} 
 + \beta^4 (W_{23,32}^{+b-b})^* W_{21,12}^{-b+b} 
+\alpha^2\gamma^2 e^{-i\Delta_{31}t} (W_{12,21}^{+b-b})^*  W_{32,23}^{-b+b} \notag\\
&&  -\alpha\beta^2\gamma e^{-i\Delta_{32}t} (W_{23,32}^{+b-b})^*  W_{32,23}^{-b+b} 
+ \alpha^2\gamma^2 e^{-i\Delta_{31}t} (W_{13,31}^{+b-b})^* W_{31,13}^{-b+b} ,\\
(B_{b-b+}^{(2)})^*B_{b+b-}^{(2)} 
&\approx&
 -\alpha\beta^2\gamma e^{i\Delta_{21}t} (W_{21,12}^{-b+b})^* W_{12,21}^{+b-b}
 +\beta^4 (W_{21,12}^{-b+b})^* W_{23,32}^{+b-b}
 +\alpha^2\gamma^2 e^{i\Delta_{31}t} (W_{32,23}^{-b+b})^*W_{12,21}^{+b-b} \notag\\
&&-\alpha\beta^2\gamma e^{i\Delta_{32}t} (W_{32,23}^{-b+b})^* W_{23,32}^{+b-b} 
 +\alpha^2\gamma^2 e^{i\Delta_{31}t} (W_{31,13}^{-b+b})^* W_{13,31}^{+b-b} ,
\end{eqnarray}

\begin{eqnarray} 
|C_{a+b+}^{(2)}|^2 &\approx& \gamma^4 [(W_{12,23}^{+a+b})^* W_{12,23}^{+a+b}  
+ (W_{12,23}^{+b+a})^* W_{12,23}^{+b+a}
+ (W_{12,23}^{+a+b})^* W_{12,23}^{+b+a} 
+ (W_{12,23}^{+b+a})^* W_{12,23}^{+a+b} ] \label{eq:proba_C_a+b+_summary}  \notag\\
(&\approx& \frac{t^4}{4} \gamma^4 \kappa_a^2 \kappa_b^2 (B_{12} A_{23} + A_{12} B_{23})^2 ) ,\\
|C_{a-b-}^{(2)}|^2 &\approx& 
 \alpha^4 [ (W_{32,21}^{-a-b})^* W_{32,21}^{-a-b} 
\underline{+ (W_{32,21}^{-b-a})^* W_{32,21}^{-b-a} }
+ (W_{32,21}^{-a-b})^* W_{32,21}^{-b-a} 
+ (W_{32,21}^{-b-a} )^* W_{32,21}^{-a-b} ]  \notag \\
(&\approx& \frac{t^4}{4} \alpha^4 \kappa_a^2  \kappa_b^2 ( A_{23} B_{21} + A_{32} B_{21})^2 ) ,\\
|C_{a+b-}^{(2)}|^2 &\approx&  
\alpha^2\gamma^2 (W_{12,21}^{+a-b} )^* W_{12,21}^{+a-b}
+ \alpha^2\gamma^2 (W_{13,31}^{+a-b})^* \mathcal{G}_{13,31}^{+a-b} 
+ \beta^4 (W_{23,32}^{+a-b})^* W_{23,32}^{+a-b} \notag\\
&&+ \beta^4 (W_{21,12}^{-b+a})^* W_{21,12}^{-b+a} 
 +\alpha^2\gamma^2 (W_{31,13}^{-b+a})^* W_{31,13}^{-b+a} 
 +\alpha^2\gamma^2 (W_{32,23}^{-b+a})^* W_{32,23}^{-b+a}  ,\\
&& 
 -\alpha\beta^2\gamma [ e^{-i\Delta_{21}t} (W_{12,21}^{+a-b})^* W_{23,32}^{+a-b} 
+ e^{i\Delta_{21}t} (W_{23,32}^{+a-b})^* W_{12,21}^{+a-b} ] \notag\\
&&%
- \alpha\beta^2\gamma [ e^{-i\Delta_{32}t} (W_{21,12}^{-b+a})^* W_{32,23}^{-b+a} 
+ e^{i\Delta_{32}t} (W_{32,23}^{-b+a})^* W_{21,12}^{-b+a} ]  \notag\\
&&
 -\alpha\beta^2\gamma [ e^{-i\Delta_{21}t} (W_{12,21}^{+a-b})^* W_{21,12}^{-b+a} 
+ e^{i\Delta_{21}t} (W_{21,12}^{-b+a} )^* W_{12,21}^{+a-b} ] \notag\\
&&%
 -\alpha\beta^2\gamma [ e^{-i\Delta_{32}t} (W_{23,32}^{+a-b})^* W_{32,23}^{-b+a} 
 + e^{i\Delta_{32}t} (W_{32,23}^{-b+a})^* W_{23,32}^{+a-b} ] \notag\\
&&
 + \alpha^2\gamma^2 [ e^{-i\Delta_{31}t} (W_{13,31}^{+a-b})^* W_{31,13}^{-b+a} 
 + e^{i\Delta_{31}t} (W_{31,13}^{-b+a})^* W_{13,31}^{+a-b} ] \notag\\
&&%
 +\alpha^2\gamma^2 [ e^{-i\Delta_{31}t} (W_{12,21}^{+a-b})^* W_{32,23}^{-b+a} 
+ e^{i\Delta_{31}t} (W_{32,23}^{-b+a})^*W_{12,21}^{+a-b} ] \notag\\
&&+\beta^4 [ (W_{21,12}^{-b+a})^* W_{23,32}^{+a-b} 
+ (W_{23,32}^{+a-b})^* W_{21,12}^{-b+a} ]  ,\\
|C_{a-b+}^{(2)}|^2 &\approx& 
\alpha^2\gamma^2 (W_{12,21}^{+b-a})^* W_{12,21}^{+b-a} 
+ \alpha^2\gamma^2 (W_{13,31}^{+b-a})^* W_{13,31}^{+b-a} 
+ \beta^4 (W_{23,32}^{+b-a})^* W_{23,32}^{+b-a} \notag\\
&& +\beta^4 (W_{21,12}^{-a+b})^* W_{21,12}^{-a+b} 
+\alpha^2\gamma^2 (W_{31,13}^{-a+b})^* W_{31,13}^{-a+b} 
+\alpha^2\gamma^2 (W_{32,23}^{-a+b})^* W_{32,23}^{-a+b}  \notag\\
&&
-\alpha\beta^2\gamma [ e^{-i\Delta_{21}t} (W_{12,21}^{+b-a})^* W_{23,32}^{+b-a} 
+ e^{i\Delta_{21}t} (W_{23,32}^{+b-a})^* W_{12,21}^{+b-a} ] \notag\\
&&%
- \alpha\beta^2\gamma [e^{-i\Delta_{32}t} (W_{21,12}^{-a+b})^* W_{32,23}^{-a+b} 
+ e^{i\Delta_{32}t} (W_{32,23}^{-a+b})^* W_{21,12}^{-a+b} ] \notag\\
&&
-\alpha\beta^2\gamma [ e^{-i\Delta_{21}t} (W_{12,21}^{+b-a})^* W_{21,12}^{-a+b} 
+ e^{i\Delta_{21}t} (W_{21,12}^{-a+b})^* W_{12,21}^{+b-a} ] \notag\\
&&%
- \alpha\beta^2\gamma [ e^{-i\Delta_{32}t} (W_{23,32}^{+b-a})^* W_{32,23}^{-a+b} 
+ e^{i\Delta_{32}t} (W_{32,23}^{-a+b})^* W_{23,32}^{+b-a} ] \notag\\
&&
+\alpha^2\gamma^2 [ e^{-i\Delta_{31}t} (W_{13,31}^{+b-a})^* W_{31,13}^{-a+b} 
+ e^{i\Delta_{31}t} (W_{31,13}^{-a+b})^* W_{13,31}^{+b-a} ] \notag\\
&&%
+\alpha^2\gamma^2 [ e^{-i\Delta_{31}t} (W_{12,21}^{+b-a})^* W_{32,23}^{-a+b} 
+ e^{i\Delta_{31}t} (W_{32,23}^{-a+b})^* W_{12,21}^{+b-a} ] \notag\\
&&
+ \beta^4 [ (W_{21,12}^{-a+b})^* W_{23,32}^{+b-a} 
+ (W_{23,32}^{+b-a})^* W_{21,12}^{-a+b} ] . 
\end{eqnarray}

Since analytical expressions for the two other terms, $P_3^{(2,2)} = Tr_{a,b} \{[ (\mathcal{A}^{(1)})^{\dagger} \mathcal{A}^{(3)} +(\mathcal{A}^{(3)})^{\dagger} \mathcal{A}^{(1)} ]\rho_a\rho_b \}$ and $P_3^{(2,3)} = Tr_{a,b} \{[ (\mathcal{A}^{(0)})^{\dagger} \mathcal{A}^{(4)} +(\mathcal{A}^{(4)})^{\dagger} \mathcal{A}^{(0)} ]\rho_a\rho_b \} $, are very lengthy and not directly relevant to an explanation of our findings in the main text, we therefore do not include them in this supplemental material. 

To see, for example,  cooperation and interference effects to lowest order in $J$ -- namely, the fourth order in $J/\Delta$ -- notice that
\begin{eqnarray}
&& \alpha \sim 1, \beta\sim \frac{J}{\Delta}, \gamma\sim \frac{J}{\Delta}, \\
&& A_{12}\sim \frac{J}{\Delta},  A_{23}\sim \frac{J}{\Delta}, A_{13}\sim\big(\frac{J}{\Delta}\big)^2 ,  \\
 &&B_{12}\sim \big(\frac{J}{\Delta}\big)^3, B_{23}\sim \frac{J}{\Delta}, B_{13}\sim\big(\frac{J}{\Delta}\big)^2,
\end{eqnarray}
we find that
\begin{eqnarray}
&& P_3[\big(\frac{J}{\Delta}\big)^4 ] \notag\\
&\approx& n_a[
\alpha^2 \beta^2 (W_{21}^{-a})^* W_{21}^{-a}
+ \alpha^2 \beta^2 (W_{32}^{-a})^* W_{32}^{-a} 
 -\alpha^2 \beta^2 e^{-i\Delta_{32}t} (W_{21}^{-a})^* W_{32}^{-a} 
-\alpha^2 \beta^2 e^{i\Delta_{32}t} (W_{32}^{-a})^* W_{21}^{-a} 
+ \alpha^4 (W_{31}^{-a})^* W_{31}^{-a} ] \notag\\
&&
+ n_b \alpha^2 \beta^2 (W_{32}^{-b})^* W_{32}^{-b}
+2n_a^2 \alpha^4 ( W_{32,21}^{-a-a})^* W_{32,21}^{-a-a}
+ n_a n_b \alpha^4 (W_{32,21}^{-b-a})^* W_{32,21}^{-b-a}  \notag\\
&=& n_a[
- \alpha^2 \beta^2 \kappa_a^2 t^2 A_{12}^2
 + \alpha^2 \beta^2 \kappa_a^2 t^2 A_{23}^2 
 -\alpha^2 \beta^2 \kappa_a^2 t^2 A_{32}A_{12} e^{-i\Delta_{32}t} 
- \alpha^2 \beta^2 \kappa_a^2 t^2 A_{21}A_{23} e^{i\Delta_{32}t} 
+ \alpha^4 \kappa_a^2 t^2 A_{13}^2 ] \notag\\
&&
+n_b \alpha^2 \beta^2 \kappa_b^2 t^2 B_{23}^2
+n_a^2 \frac{t^4}{2} \alpha^4 \kappa_a^4 A_{32}^2 A_{21}^2
+ n_a n_b \frac{t^4}{4} \alpha^4 \kappa_a^2  \kappa_b^2  B_{23} A_{21} \notag\\
&=& n_a \alpha^2 \beta^2 \kappa_a^2 t^2 \{
  A_{12}^2  +  A_{23}^2 - A_{32}A_{12} e^{-i\Delta_{32}t} -  A_{21}A_{23} e^{i\Delta_{32}t} + n_b \kappa_b^2/(n_a\kappa_a^2) B_{23}^2 \} \notag\\
 &&
+ \frac{t^4}{2} n_a^2 \alpha^4 \kappa_a^4 A_{32}^2 A_{21}^2
+ \frac{t^4}{4} n_an_b \alpha^4 \kappa_a^2  \kappa_b^2  B_{23} A_{21} 
+ n_a\alpha^4 \kappa_a^2 t^2 A_{13}^2 .
\end{eqnarray}
We note that although this expression is different from the corresponding one for weak $J$ presented in the previous subsection, as expected, this does not affect our explanation and understanding of the results in the main text.

\begin{figure}
\centering
  \includegraphics[width=.55\columnwidth]{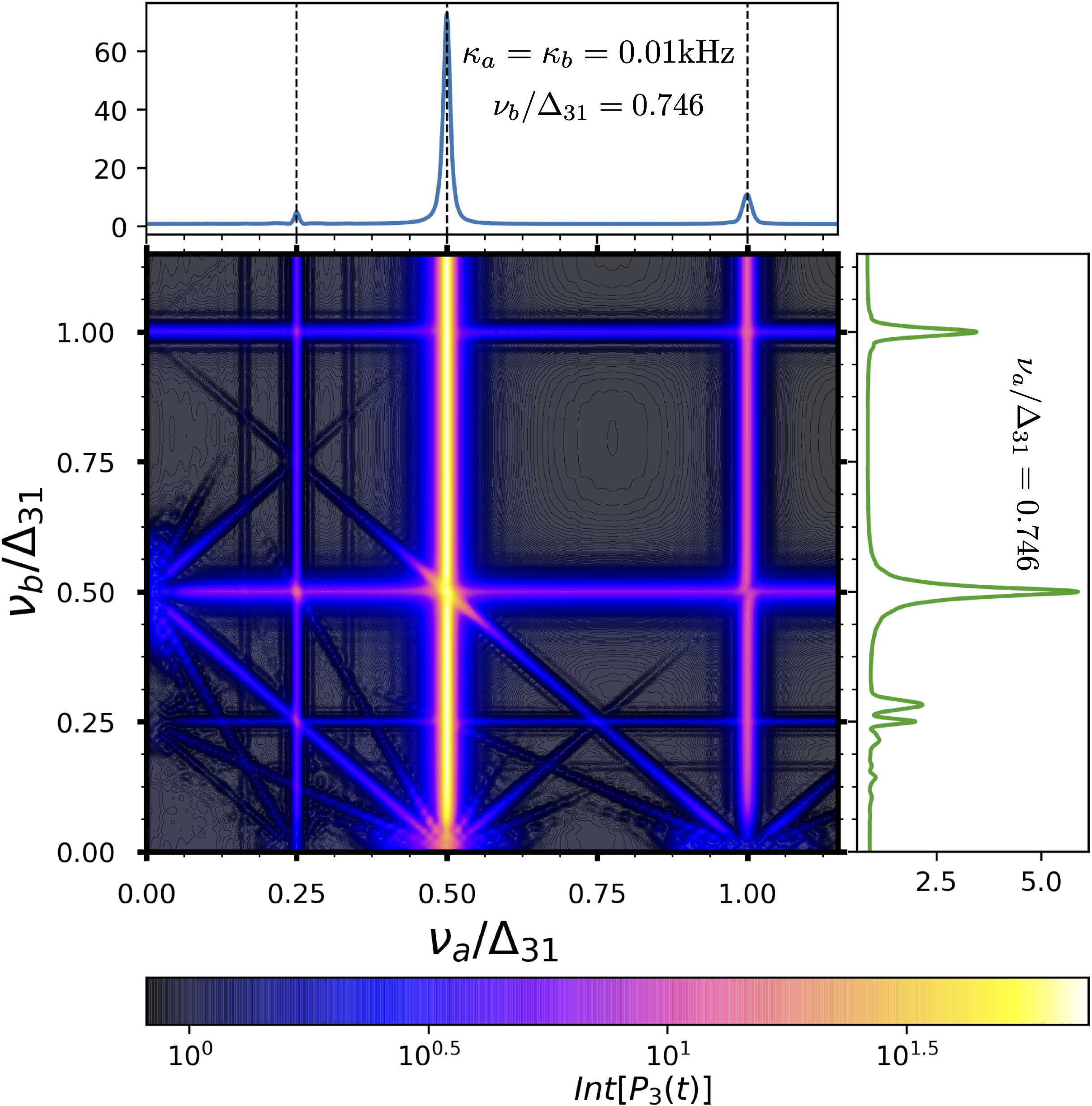} 
\caption{(color online) (a) 2D VAET spectrum of a symmetric trimeric chromophore system coupled to two non-interacting vibrations in the weak site-vibration coupling regime $\kappa_a=\kappa_b=0.01$kHz. The integration transfer probability ${\rm Int}[P_3(t)]$ is taken during a time period $t\in[0,400]$ms. 
$\Delta_{ij}$ is the energy difference between eigenstates $|e_i\rangle$ and $|e_j\rangle$ of the electronic part in Eq.~(5) in the main text 
with $\{\tilde{\omega}_1,\tilde{\omega}_2,\tilde{\omega}_3,J_{12},J_{23}\}=\{-0.5,0,0.5,0.1,0.1\}$kHz, satisfying  $\Delta_{31}=2\Delta_{21}$~\cite{suppl}. 
We consider two vibrations with identical temperatures $k_BT_a=k_BT_b=1.5$kHz. The truncation number of each vibrational Fock space is $N=15$. The 1-dimensional slices on the top and right of the 2-dimensional contour plot are taken at $\nu_b/\Delta_{31}=0.746$ and $\nu_a/\Delta_{31}=0.746$, respectively. }
\label{fig:intPop_N15_kappa_0d01_kBT_1d5_1d5_amplify}
\end{figure}

\begin{figure}
\centering
  \includegraphics[width=.85\columnwidth]{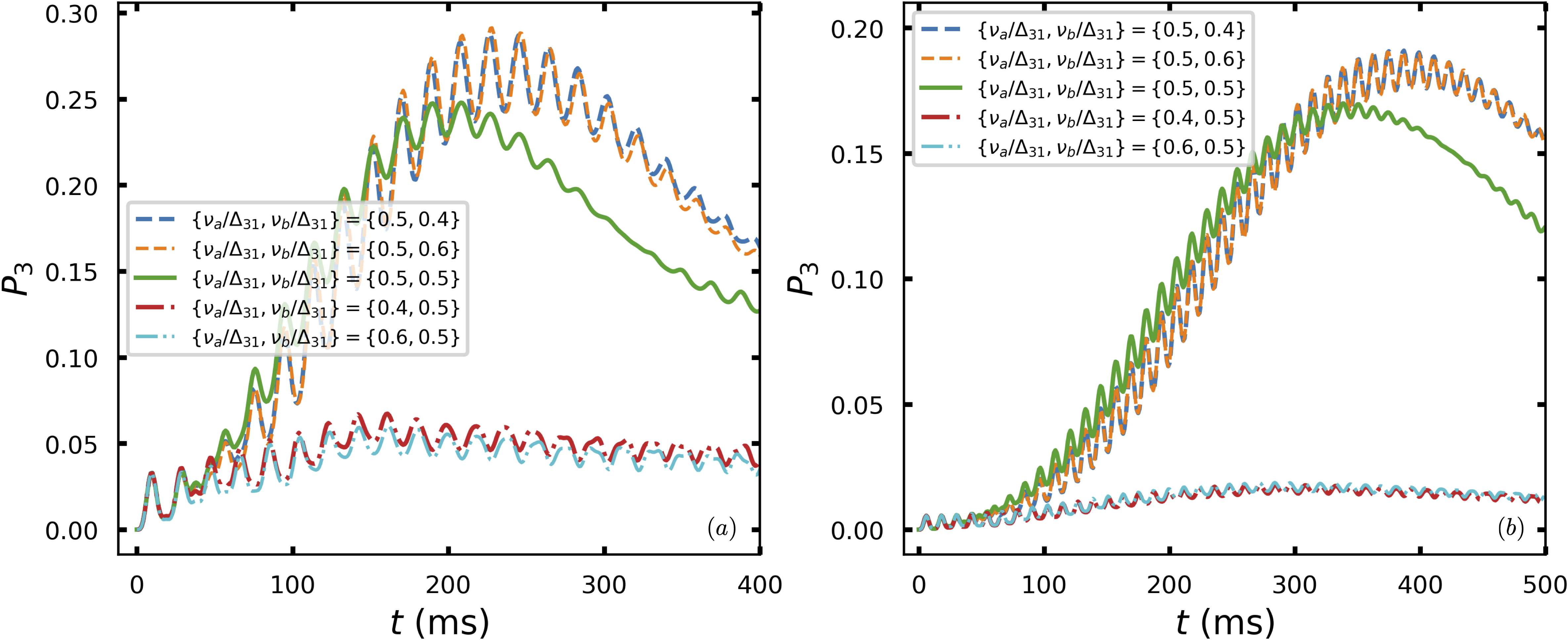}
\caption{(color online) (a) and (b) Time evolution of the probability $P_3$ of the acceptor at the symmetric point $\{\nu_a/\Delta_{31},\nu_b/\Delta_{31}\}=\{0.5,0.5\}$ and nearby points along the $\nu_a/\Delta{31}=0.5$ and $\nu_a/\Delta{31}=0.5$ lines in Fig.10 (a) and (b) of the main text, respectively. All other parameters are the same as in Fig.10 (a) and (b) of the main text.}
\label{fig:symmetric_offresonance}
\end{figure}

\section*{Additional simulation results}

\subsection{Accumulated population and time traces for Fig.10}

In addition to the maximum population considered in the main text, in Fig.~\ref{fig:intPop_N15_kappa_0d01_kBT_1d5_1d5_amplify} we employ the accumulated population ${\rm Int}[P_3(t)]=\int_0^{t_f}P_3(t)dt$ during a given time period $t_f$ as an alternative quantitative measure of the excitonic energy transfer efficiency. 
It is shown that VAET features in Fig.~\ref{fig:intPop_N15_kappa_0d01_kBT_1d5_1d5_amplify} are very similar to those in Fig.~2(a) in the main text and therefore support the VAET transfer processes presented therein. 
In particular, the 1-dimensional slice to the right of Fig.~\ref{fig:intPop_N15_kappa_0d01_kBT_1d5_1d5_amplify} shows that the integrated strength of TPhona VAET is stronger, supporting the dominance of TPhonA VAET by the  terminal site vibration.

In Fig.~\ref{fig:symmetric_offresonance}, we present the time traces for Fig.~10 of the main text to support the statement that destructive interference appears along the vertical line [$\nu_a=0.332$kHz (Fig.10 (a)) or $\nu_a=0.52$kHz (Fig.10 (b)] and constructive interference appears along the horizontal line [$\nu_b=0.332$kHz (Fig.10 (a)) or $\nu_b=0.52$kHz (Fig.10 (b)].

\subsection{The convergence in a 2D spectra}

\begin{figure}
\centering
  \includegraphics[width=.55\columnwidth]{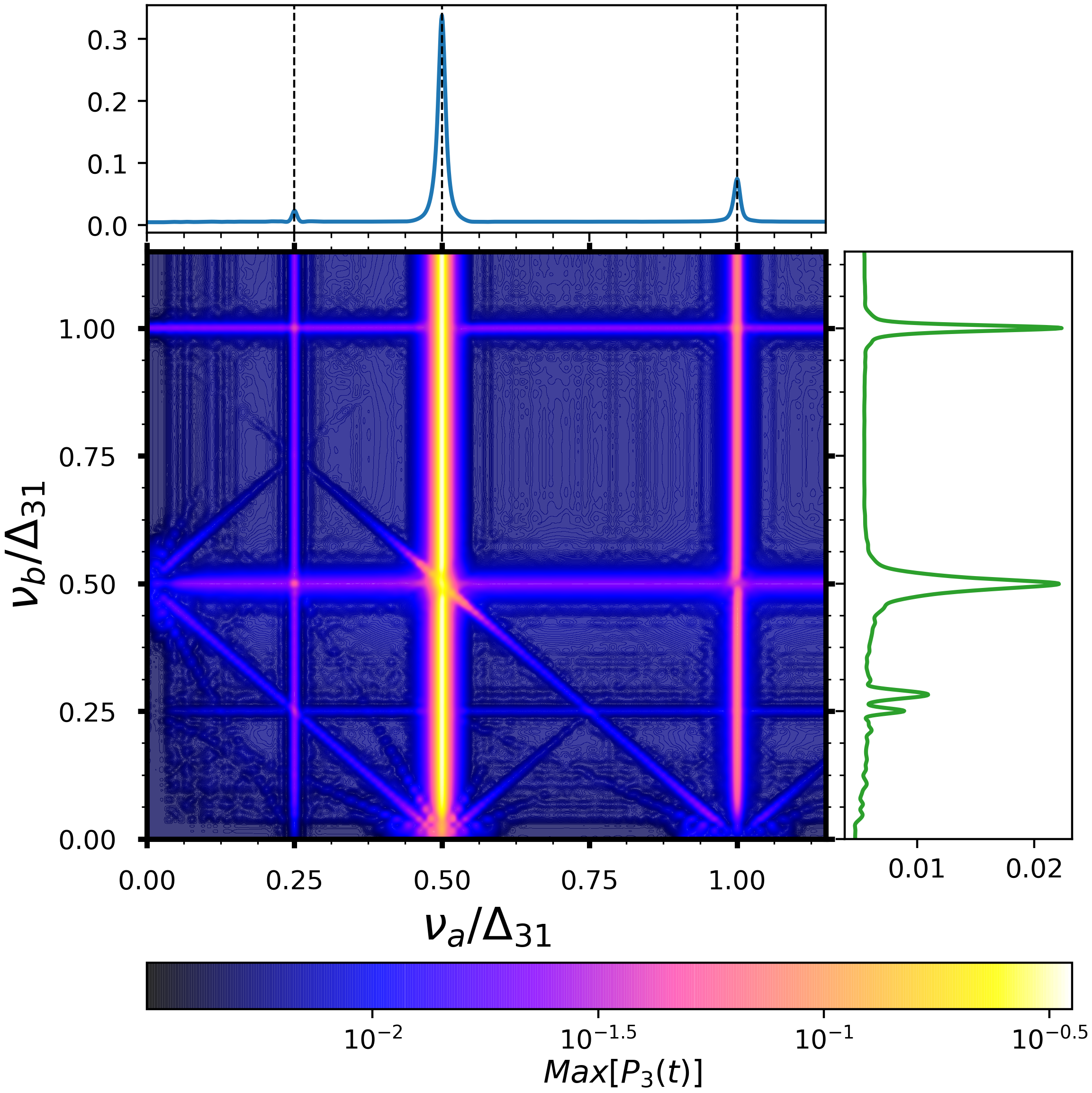}
\caption{(color online) Same as Fig.~2 in the main text but with $N=10$.}
\label{fig:N10}
\end{figure}

In our simulations we model the vibrational mode with a truncated Fock basis. The truncation number $N$ must be large enough so that there is negligible population in the high-lying Fock states in order to minimize the effects of this truncation approximation. For all simulations presented in the main text we use $N=15$. It is reasonable to question whether this truncation is sufficient for accurate simulation at low vibrational frequencies, especially in the regime $k_B T > \nu$. For example, in Fig.~2 in the main text the 2D spectra have points in the lower left with very small vibrational frequencies (the vibrational frequency step we used to generate these figures was $0.01$kHz (which gives $\sim 0.019$ for $\nu_{a(b)}/\Delta_{31})$ in the 2D spectra), together with temperature $k_BT_a=k_BT_b=1.5$KHz. Since the vibrational modes are off-resonant with the electronic energy gaps in this regime we expect that the effect on energy transfer dynamics in the electronic degrees of freedom will not be strongly effected by the truncation error in this regime. To confirm this, in Fig.~\ref{fig:N10}, we plot the same 2D spectrum with a smaller truncation, $N=10$, and verify that the main features observed in Fig.~2 of main text, with $N=15$, are still present. The transfer probability is underestimated with decreasing $\nu_{a(b)}/\Delta_{31}$, but the positions of the multiphonon VAET processes are not affected even in the strong site-vibration coupling.

\subsection{A remark on the FMO parameters}

\begin{figure}
\centering
  \includegraphics[width=.55\columnwidth]{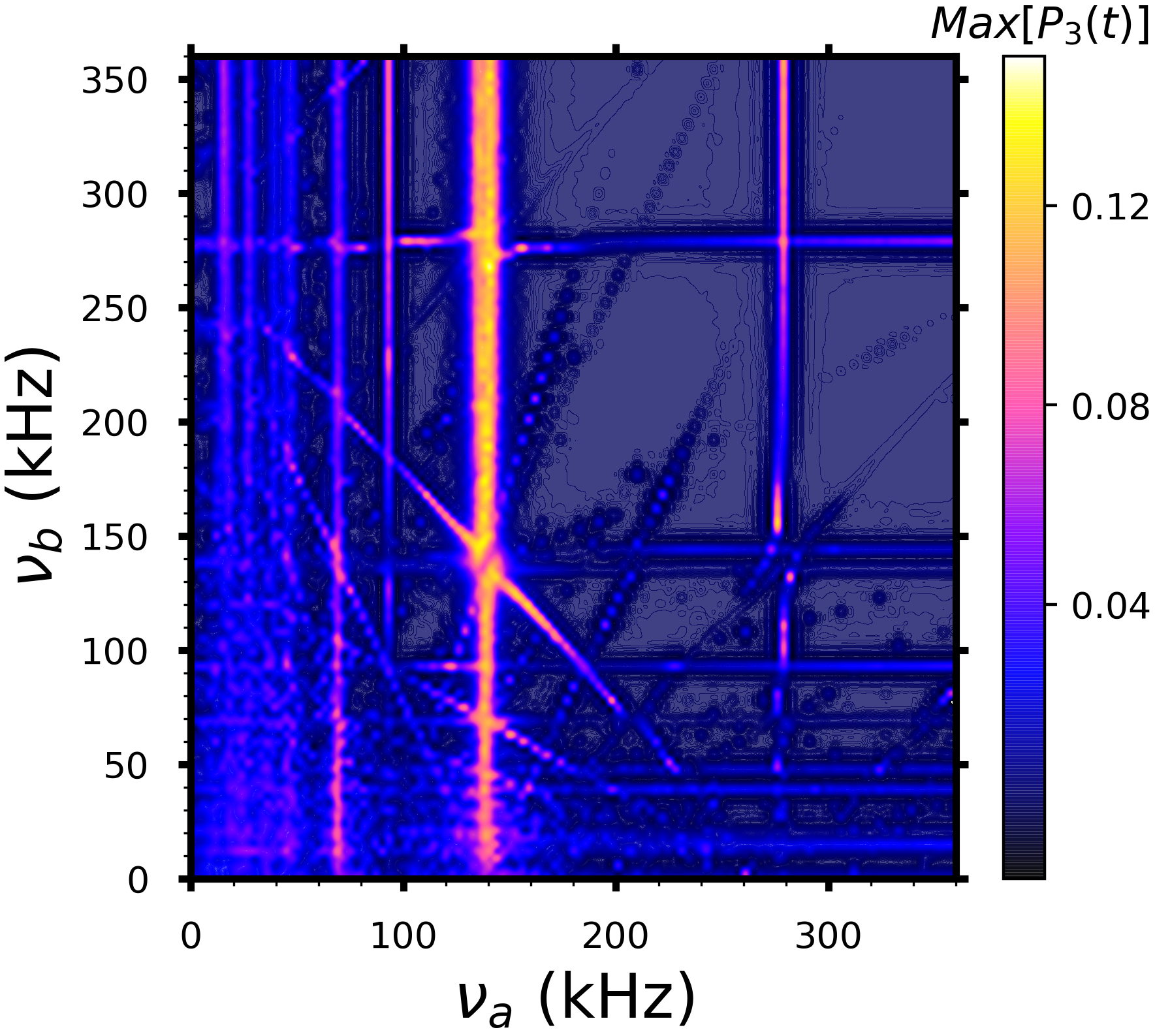}
\caption{(color online) The 2D VAET spectra with the FMO parameters in Table 1 of the main text.}
\label{fig:naturalsystems}
\end{figure}

The parameters we use for simulations in the main text are those relevant to trapped-ion simulations. We show that if these are scaled, they coincide reasonably well with the relevant parameters in typical photosynthetic complexes, specifically the FMO complex. However, in Table 1 in the main text, it can be seen that the site-site couplings ($J$) in FMO are stronger than the scaled values by a factor of $2$ to $5$. Therefore, to verify that our observations are valid even in these regimes of stronger site-site couplings, we perform the simulations with the FMO parameters in the line 3 of Table 1 and present the result in Fig.~\ref{fig:naturalsystems}. Indeed, the results look similar to Fig.~2 of the main text, with prominent VAET and multi-phonon VAET features.

\section*{Non-Hermitian approach for dissipation}

In this section we justify the non-Hermitian approach that we use to study effects of the dissipation.

To include the effect of dissipation in the effective Hamiltonian, we go back to the full Hamiltonian [i.e., $H=H_{\rm s}+H_{\rm v}+H_{\rm int}$ with $H_{\rm s}$, $H_{\rm v}$, and $H_{\rm int}$ given by Eqs.~ (1), (2), and (3), respectively, in the main text]. 
Suppose the $j$th site excited state decays with a rate $\gamma_j$ into the ground state (e.g. spontaneous emission), independent of other sites. Then the master equation for the density matrix $\rho$ of the coupled exciton-vibration system reads 
\begin{equation}
    \frac{d}{dt}\rho = -\frac{i}{\hbar} [H, \rho] + \sum_{j=1}^3 \gamma_j D[\sigma_-^{(j)}](\rho) ,
\end{equation}
where $D[A](\rho) = A\rho A^\dagger - \frac{1}{2} A^\dagger A \rho -\frac{1}{2}\rho A^\dagger A$.
Rewriting the dissipator as a non-Hermitian Hamiltonian and quantum jump term, we have
\begin{equation}
    \frac{d}{dt}\rho = -\frac{i}{\hbar}(H_{\text{eff}} \rho - \rho H_{\text{eff}}^\dagger) + \sum_{j=1}^3 \gamma_j \sigma_-^{(j)}\rho\sigma_+^{(j)} ,
    \label{Lindblad1}
\end{equation}
with $H_{\text{eff}} = H -\frac{i\hbar}{2} \sum_{j=1}^3 \gamma_j |e\rangle_j\langle e|$.
In the main text, we restrict attention to the single excitation subspace, and so projecting this equation into the single excitation yields:

\begin{equation}
    \frac{d}{dt}\rho_{1} = -\frac{i}{\hbar}(H_{\text{eff}}^{1} \rho_{1} - \rho_{1} H_{\text{eff}}^{1\dagger}) ,
\end{equation}
where $\rho_1$ is the density matrix projected into the single excitation subspace, and 
\begin{align}
\begin{split}
    H_{\text{eff}}^{1} = \Tilde{H} -i\hbar\big(&\frac{\gamma_1}{2} |1\rangle\langle 1| + \frac{\gamma_2}{2}  |2\rangle\langle 2|+\frac{\gamma_3}{2}  |3\rangle\langle 3| \big)
\end{split} ,
\end{align}
with $\tilde{H}$ defined in the main text.

\end{document}